%% file: Journal.tex
\documentclass{tlp}

\usepackage{verbatim}
\usepackage{subfigure}
\usepackage{color}		% Need the color package
\usepackage{epsfig}
\usepackage{graphicx}
\usepackage{ifpdf}
\usepackage{multirow}
\usepackage[all]{xy}
\usepackage{amssymb}
\usepackage{times}
\usepackage{rotating}
\usepackage{enumerate}

\makeatletter
\newif\if@restonecol
\makeatother

\usepackage[linesnumbered,ruled]{algorithm2e}

\newif\ifremarks
\remarksfalse %makes REMARKs invisible

\ifremarks
\advance\oddsidemargin       0mm
\advance\evensidemargin    50mm
\advance\marginparwidth 34mm
\fi

\newcommand{\ms}[1]{\ifmmode%
\mathord{\mathcode`-="702D\it #1\mathcode`\-="2200}\else%
$\mathord{\mathcode`-="702D\it #1\mathcode`\-="2200}$\fi}

\newtheorem{definition}{Definition}
\newtheorem{theorem}{Theorem}
\newtheorem{proposition}{Proposition}

\newtheorem{corol}{Corollary}

\newcommand{\queryM}{$request$}
\newcommand{\IDM}{$\ms{id}$}
\newcommand{\ReqM}{$req$}

\newcommand{\query}{\emph{request}}
\newcommand{\ID}{\emph{\ms{id}}}
\newcommand{\Req}{\emph{req}}

\def\qTuple#1#2#3{\ensuremath{\langle #1,#2,#3\rangle}}

\newcommand{\answerM}{$response$}
\newcommand{\AnssM}{$Ans$}
\newcommand{\ASM}{$S_{ans}$}
\newcommand{\LoopsM}{$Loops$}
\newcommand{\answer}{\emph{response}}

\newcommand{\Anss}{\emph{Ans}}
\newcommand{\AS}{\emph{$S_{ans}$}}
\newcommand{\Loops}{\emph{Loops}}

\newcommand{\newSM}{$new$}
\newcommand{\actSM}{$active$}
\newcommand{\ansSM}{$answer$}

\newcommand{\dispSM}{$disposed$}
\newcommand{\loopPSM}{$loop(\ID_1)$}
\newcommand{\loopSSM}{$loop(\IDs)$}

\newcommand{\loopLoopsSM}{$loop(\Loops)$}
\newcommand{\newS}{\emph{new}}
\newcommand{\actS}{\emph{active}}
\newcommand{\ansS}{\emph{answer}}

\newcommand{\loopSS}{\emph{loop(\IDs)}}

\newcommand{\dispS}{\emph{disposed}}

\newcommand{\IDsM}{$\ms{ID}$}
\newcommand{\node}{\emph{node}}

\newcommand{\IDs}{\emph{\ms{ID}}}

\def\nTuple#1#2#3{\ensuremath{\langle #1,#2,#3\rangle}}

\newcommand{\tabM}{$Table$}

\newcommand{\QM}{$\ms{HR}$}
\newcommand{\LoopQM}{$\ms{LR}$}
\newcommand{\ActiveLoopsM}{$ActiveGoals$}
\newcommand{\AnsSetM}{$AnsSet$}
\newcommand{\TreeM}{$Tree$}

\newcommand{\tab}{\emph{Table}}

\newcommand{\Q}{\emph{HR}}
\newcommand{\LoopQ}{\emph{LR}}
\newcommand{\ActiveLoops}{\emph{ActiveGoals}}
\newcommand{\AnsSet}{\emph{AnsSet}}
\newcommand{\Tree}{\emph{Tree}}

\newcommand{\Ans}{\emph{ans}}
\def\ALTuple#1#2{\ensuremath{\langle #1,#2\rangle}}
\def\ASTuple#1#2{\ensuremath{\langle #1,#2\rangle}}
\def\tTuple#1#2#3#4#5{\ensuremath{\langle #1,#2,#3,#4,#5\rangle}}
\newcommand{\newsubgoal}{\emph{\sc Process Request}}
\newcommand{\createtable}{\emph{\sc Create Table}}
\newcommand{\positivereturn}{\emph{\sc Process Response}}
\newcommand{\newactive}{\emph{\sc Activate Node}}
\newcommand{\sendanswer}{\emph{\sc Send Response}}
\newcommand{\answerreturn}{\emph{\sc Generate Response}}
\newcommand{\completion}{\emph{\sc Terminate}}

\newcommand{\orderingIn}{\sqsubset}
\newcommand{\orderingSide}{\hookrightarrow}

\title[GEM: a Distributed Goal Evaluation Algorithm]
			{GEM: a Distributed Goal Evaluation Algorithm for Trust Management}

\author[D. Trivellato, N. Zannone, S. Etalle]
         {DANIEL TRIVELLATO$^*$, NICOLA ZANNONE$^*$, SANDRO ETALLE$^{*,+}$\\
         $^*$Eindhoven University of Technology, Eindhoven, The Netherlands\\
         $^+$University of Twente, Enschede, The Netherlands\\
         \email{\{d.trivellato,n.zannone,s.etalle\}@tue.nl}}
         
 \submitted {30 May 2011}
 \revised {13 March 2012; 17 July 2012}
 \accepted{24 September 2012}

\pagerange{\pageref{firstpage}--\pageref{lastpage}}

\begin{document}

\maketitle

\label{firstpage}

\vspace{-0.2cm}
\textbf{Note:} To appear in \emph{Theory and Practice of Logic Programming} (TPLP).\\

\begin{abstract}

Trust management is an approach to access control in distributed systems where access decisions are based on policy statements issued by multiple principals and stored in a distributed manner.
In trust management, the policy statements of a principal can refer to other principals' statements; thus, the process of evaluating an access request (i.e., a goal) consists of finding a ``chain'' of policy statements that allows the access to the requested resource.
Most existing goal evaluation algorithms for trust management either rely on a centralized evaluation strategy, which consists of collecting all the relevant policy statements in a single location (and therefore they do not guarantee the confidentiality of intensional policies), or do not detect the termination of the computation (i.e., when all the answers of a goal are computed).
In this paper we present GEM, a distributed goal evaluation algorithm for trust management systems that relies on function-free logic programming for the specification of policy statements.
GEM detects termination in a completely distributed way without disclosing intensional policies, thereby preserving their confidentiality.
We demonstrate that the algorithm terminates and is sound and complete with respect to the standard semantics for logic programs. 

\end{abstract}


%\newpage

\input{introduction}
\input{preliminaries}

\input{gem}

\input{properties}

%
\input{evaluation}

\input{advanced}

\input{rw}

\input{conclusions}

%%%%%%%%%%%% ACKNOWLEDGEMENTS %%%%%%%%%%%

%\vspace{0.2cm}{\begin{center}
% {\bf Acknowledgements}
%\end{center}}
\vspace{0.1cm}
\paragraph{\textbf{Acknoledgements.}}
\noindent This work has been carried out as part of the POSEIDON project under the responsibility of the Embedded Systems Institute (ESI). This project is partially supported by the Dutch Ministry of Economic Affairs under the BSIK03021 program.
%This work has been also funded by the EU-IST-IP-216287 TAS$^3$ project.

%%%%%%%%%%%%% BIBLIOGRAPHY %%%%%%%%%%%%%

%\newpage
\bibliographystyle{acmtrans}
\bibliography{biblio}

%\appendix
%
\input{appendix}

\end{document}

%% file: introduction.tex
\section{Introduction}
\label{sec:intro}

The widespread availability of the Internet has led to a significant increase in the number of collaborations, services and transactions carried out over networks spanning multiple administrative domains (e.g., web services).
%Examples of these services include location-based services~\cite{K-B-05}, e-commerce, and e-learning.
Such collaborations are frequently characterized by the interaction of users and institutions (hereafter indistinctly referred to as \emph{principals}) who do not know each other beforehand.
For this reason, in such distributed settings, attribute-based approaches to access control are mostly preferred to identity-based solutions~\cite{EFLRTY-RFC-99}.
Consider, for instance, an international medical research project \emph{Alpha} involving several companies worldwide.
Project \emph{Alpha} is funded and coordinated by the multinational pharmaceutical company \emph{mc} which, among its tasks, appoints the partners of the project consortium.
In this scenario, it is likely that company \emph{mc} does not know the project members of each partner company personally, i.e., does not know their identity.
Therefore, rather than on the identity of the project members, the policies regulating the access to project's documents will be based on their attributes (e.g., project membership, specialization) and their relationships with other principals (e.g., partner companies, departments within a company).
%Online car renting services, for example, offer their service to all the principals with a valid driver's license and whose age is above 21.

Trust management is an approach to access control in distributed systems where access decisions are based on the attributes of principals, which are attested by digitally signed certificates called \emph{digital credentials}~\cite{BFL-SP-96}.
Digital credentials (or simply credentials) are the digital counterpart of paper credentials.
%For instance, a machine-readable employment contract signed with the private key of the employing institution represents a digital credential.
%In the remainder of the paper we refer to digital credentials simply as credentials.
Credentials are defined and derived by means of policy statements that specify the conditions upon which a credential is issued, where conditions are in turn represented by credentials.
A distinguishing ingredient of trust management is that all the principals in a distributed system are free to define such policy statements and determine where to store them.
The set of policy statements defined by a principal forms the \emph{policy} of that principal.
In the scenario above, for instance, the rules of company \emph{mc} dictating the conditions for the membership of a user to project \emph{Alpha} (e.g., a Master degree in chemistry) form the policy of \emph{mc}.
These statements can be stored by \emph{mc} or at another principal's location (e.g., by each partner company).

In trust management languages, policy statements are often expressed as Horn clauses \cite{LM-PADL-03} where each atom represents a credential, and is possibly annotated with the storage location of the statements defining the credential.
Depending on the language, the location can be expressed implicitly \cite{CE-ICLP-07,LWM-CS-03} or explicitly \cite{ADNO-POLICY-06,B-TR-05}.
%A distinguishing ingredient of trust management is that all the principals in a distributed system are free to define policy statements and determine where to store them.
While typically principals do not have direct access to each other policies, the statements of a principal can refer to other principals' policies, thereby delegating authority to them. 
For instance, assume that a hospital $h$ authorizes the members of project \emph{Alpha} certified by the local pharmaceutical company \emph{c$1$} to access the (anonymized) medical records of its patients suffering from genetic diseases.
%Project \emph{Alpha} is funded and coordinated by the multinational pharmaceutical company \emph{mc} which, among its tasks, has to select the partners of the project consortium.
The policies governing this scenario can be represented by the following clauses:

\vspace{0.1cm}
{\small 
1. ${\sf mayAccessMedRec}(h\mbox{,}X) \leftarrow {\sf memberOfAlpha}(c1\mbox{,}X).$\\
\indent 2. ${\sf memberOfAlpha}(c1\mbox{,}X) \leftarrow {\sf projectPartner}(mc\mbox{,}Y), {\sf memberOfAlpha}(Y\mbox{,}X).$\\
\indent 3. ${\sf projectPartner}(mc\mbox{,}c2).$\\
\indent 4. ${\sf projectPartner}(mc\mbox{,}c3).$\\
\indent 5. ${\sf projectPartner}(mc\mbox{,}c4).$\\
\indent 6. ${\sf memberOfAlpha}(c2\mbox{,}X) \leftarrow {\sf memberOfAlpha}(c1\mbox{,}X).$ \\
\indent 7. ${\sf memberOfAlpha}(c2\mbox{,}alice).$\\
\indent 8. ${\sf memberOfAlpha}(c3\mbox{,}bob).$\\
\indent 9. ${\sf memberOfAlpha}(c4\mbox{,}charlie).$
}
\vspace{0.1cm} 

\noindent where the first parameter of each atom denotes the location where the credential that the atom represents is defined.
%
%\noindent Clause~1 represents a policy statement of \emph{ehvH}, which states that a principal is authorized to access the hospital's medical records if \emph{c$1$} certifies that the principal is a member of project \emph{Alpha}.
%Clause~2 is specified (and issued) by \emph{c$1$}, and defines the members of project \emph{Alpha}: a principal is certified by \emph{c$1$} as a member of project \emph{Alpha} if she (or he) is a project member at a project partner company appointed by \emph{mc}.
%Clauses~3 to~5 define the partner companies appointed by \emph{mc} to project \emph{Alpha}, namely companies \emph{c$2$}, \emph{c$3$}, and \emph{c$4$}.
%Clause~6 states that \emph{c$2$} certifies that a principal is a member of project \emph{Alpha} if the principal is defined as a project member by \emph{c$1$}.
%Finally, clauses~7 to~9 state that \emph{alice}, \emph{bob}, and \emph{charlie} are certified as project members by companies \emph{c$2$}, \emph{c$3$}, and \emph{c$4$} respectively.
%
%In the example policy, 
Here, hospital $h$ relies on the policy statements of \emph{c$1$} to determine who is authorized to access the hospital's medical records (clause~1); in turn, \emph{c$1$} relies on the policy statements of \emph{mc} and the partner companies appointed by \emph{mc} for the definition of project \emph{Alpha}'s members (clause~2). %; similarly, company \emph{c$2$} relies on the statements of \emph{c$1$} to determine its project members (clause~6).
%
%Suppose that hospital $h$ wants to determine who can access the medical records of its patients.
%Three main existing approaches can be used by $h$ to compute the set of authorized principals: top-down approaches~\cite{K-IP-74}, bottom-up approaches~\cite{P-MI-69}, and ``hybrid'' approaches (e.g., magic templates~\cite{R-JLP-91}).
%Bottom-up and hybrid approaches to the evaluation of a \emph{goal} (i.e., a query) require knowledge of all the policy statements that depend on a given credential to compute the answers of the goal.
%Since in trust management systems policies are stored at different locations, this requirement cannot be guaranteed. CANNOT WRITE IT HERE: if policies were not confidential, it would be enough to collect all of them.
%
Therefore, the process of evaluating a request to access the hospital's records (i.e., a \emph{goal}) consists of deriving a ``chain'' of policy statements delegating the authority from hospital $h$ (i.e., the resource owner) to the members of project \emph{Alpha} (i.e., the authorized principals).
This process, referred to as \emph{credential chain discovery}~\cite{LWM-CS-03}, can be addressed using \emph{goal evaluation algorithms}.

Since in trust management policies are stored at different locations, goal evaluation algorithms require principals to disclose their policy statements to other principals to enable credential chain discovery.
In particular, for a successful computation, principals need to disclose at least (part of) their \emph{extensional policy}, that is, the credentials that can be derived from their policy and are required for an access decision.
For example, suppose that hospital $h$ wants to determine who can access the medical records of its patients.
To compute the answers of this goal, it is clear that \emph{c$1$} has to disclose to $h$ the credentials certifying all the project members.
Most of the existing goal evaluation algorithms (e.g.,~\cite{CE-ICLP-07,LWM-CS-03}), however, rely on a centralized evaluation strategy and require principals to disclose also (part of) their \emph{intensional policy}, i.e., the policy statement used to derive those credentials (e.g., clause~2 in the example policy).

We argue that one of the advanced desiderata of goal evaluation algorithms for trust management is that the amount of information about intensional policies that principals reveal to each other should be minimized.
In fact, intensional policies might contain sensitive information about the relationships among principals, whose disclosure would leak valuable business information that can be exploited by other principals in the domain (e.g., rival companies)~\cite{YW-SP-03}.
For example, if \emph{c$1$}'s policy was disclosed to other principals for evaluation (e.g., to hospital $h$), the involvement of \emph{mc} in project \emph{Alpha} along with the list of all project partners would be exposed.
As a consequence, some competitors of \emph{mc} could start investing on similar projects, or could try to get at the project members to acquire sensitive information and project results.
Furthermore, the loss of confidentiality of intensional policies can result in attempts by other principals to influence the policy evaluation process \cite{SKBLF-TR-08}, and allows adversaries to know what credentials they need to forge to illegitimately get access to a resource \cite{FAL-TC-06}.

To protect the confidentiality of intensional policies, it is necessary to design a completely distributed goal evaluation algorithm that discloses as few information on intensional policies as possible.
Since bottom-up approaches to goal evaluation (e.g., fixpoint semantics~\cite{P-MI-69}, magic templates~\cite{R-JLP-91} and magic sets~\cite{C-IJIS-97}) require knowledge of all the policy statements that depend on a given credential, they do not represent an applicable solution to our problem.
Hence, a top-down approach to goal evaluation needs to be employed.
The design of a distributed top-down algorithm, however, requires addressing two main problems: (a) loop detection, and (b) termination detection.
In addition, to reduce network overhead, the goal evaluation algorithm should attempt to decrease the number of messages that principals exchange.

\begin{figure}[!t]
	\centering
	\fbox{
{\footnotesize
\ms{\xymatrix@R=20pt@C=22pt@M=1pt{
mayAccessMedRec(h1,X) \ar[r] & memberOfAlpha(c1,X) \ar[d] \ar[r] \ar[ld] \ar[rd] & projectPartner(mc,Y) \\
memberOfAlpha(c2,X) \ar@(u,r)[ru] & memberOfAlpha(c3,X) & memberOfAlpha(c4,X) \\ %\ar@/^15pt/[ur]
}}
}}
	\caption{Call Graph of the Evaluation of the Example Policies}
	\label{fig:intro-graph}
\end{figure}
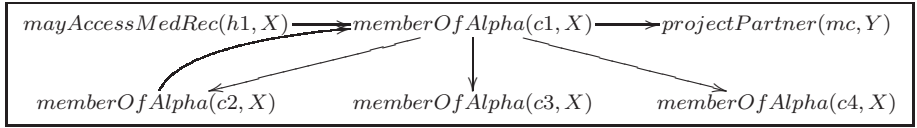

Loops are formed when the evaluation of a goal leads to a new request for the same goal.
In our scenario, for example, to determine the set of project members without disclosing its intensional policy to hospital $h$, \emph{c$1$} should first request the list of project partners to \emph{mc}, and then the list of their project members to \emph{c$2$}, \emph{c$3$} and \emph{c$4$}.
Since \emph{c$2$} in turn relies on the policy statements of \emph{c$1$} to determine its project members, \emph{c$2$} would pose the same request back to \emph{c$1$}, forming a loop. 
Fig.~\ref{fig:intro-graph} shows the ``call graph'' originating from this sequence of requests.
Intuitively, \emph{c$1$} should detect the loop and refrain from evaluating \emph{c$2$}'s request, as doing so could lead to a non-terminating chain of requests.
Even though in the example scenario loops could be avoided, for instance, by requiring a single company (e.g., \emph{mc}) to define the set of project members, this cannot be guaranteed in distributed systems characterized by the absence of a coordinating principal.
Examples of such scenarios include self-organizing networks~\cite{SFHKMRUVA-ESOS-04} and access control policies based on independent information sources (e.g,. the Friend of a Friend - FOAF - project~\cite{FOAF}).
%The detection of loops can be achieved using \emph{tabling} techniques~\cite{CW-JACM-96}.

Existing goal evaluation algorithms employ \emph{tabling} techniques~\cite{B-DOOD-89,CW-JACM-96,TS-ICLP-86,V-ICLP-87} for the detection of loops.
Although some of these algorithms resort to a distributed tabling strategy (e.g.,~\cite{D-TAPD-00,H-THESIS-97}), they rely on centralized data structures to detect termination - i.e., to detect when all the answers have been collected - thus leaking some policy information.
In fact, the real challenge in designing a goal evaluation algorithm that does not disclose intensional policies lies in detecting termination distributedly. 
In the example, we have the following possible answer flow: 
\emph{c$3$} returns \emph{bob} as answer to \emph{c$1$}, which forwards it to $h$ and \emph{c$2$};
\emph{c$4$} returns \emph{charlie} as answer to \emph{c$1$};
\emph{c$1$} sends \emph{charlie} as additional answer to $h$ and \emph{c$2$};
\emph{c$2$} returns \emph{alice}, \emph{bob}, and \emph{charlie} as answer to \emph{c$1$}, which sends \emph{alice} as additional answer to $h$ and \emph{c$2$}.
%Bob gives a (partial?) answer to Alice's request (answers: $man1,man2,man3$); 
%Alice gives $man1$ as an additional answer to Bob's request. 
At this point, all the requests have been fully evaluated, but \emph{c$1$} does not know whether \emph{c$2$} will ever send additional answers. 
In other words, \emph{c$1$} is waiting for \emph{c$2$} to announce that its evaluation has terminated, and in turn \emph{c$2$} is waiting for \emph{c$1$} to announce that its evaluation has terminated. 
This situation is not acceptable in the context of access control, where a decision (positive or negative) always needs to be made.
A few top-down goal evaluation algorithms are able to detect the termination of a computation distributedly~\cite{ADNO-POLICY-06,ZW-ESORICS-08}; however, they do not detect when the single goals within a computation are fully evaluated.
In top-down goal evaluation, detecting when the evaluation of a goal has terminated is necessary to allow (a) for memory deallocation and (b) the use of negation, which is employed by some systems to express non-monotonic constraints (e.g., separation of duty) \cite{CTDEHH-ENTCS-06,CD-IFIPTM-10}.

Finally, another non-trivial issue in designing a distributed goal evaluation algorithm is determining when a principal should send the answers to a request. 
The simplest solution is to force each principal to send an answer as soon as the principal has computed it, as done, for example, in~\cite{ADNO-POLICY-06}. 
This is, however, suboptimal from the viewpoint of network overhead; in the example above, \emph{c$1$} eventually sends three distinct messages to $h$ and \emph{c$2$}, one for each answer. 
A more network-efficient solution would be for \emph{c$1$} to wait for the answers from \emph{c$3$} and \emph{c$4$} before sending its answers to the other principals. 
A na\"{i}ve ``wait'' mechanism, on the other hand, might cause deadlocks.
For instance, if \emph{c$1$} also waits for \emph{c$2$}'s answers, the computation deadlocks. 
In a trust management system, where network latency is likely to be a bigger bottleneck than computational power, it is preferable to have a mechanism that allows principals to wait until they collect the maximum possible set of answers before sending them to the requester, while avoiding deadlocks.
Even though this solution may delay the identification of the answers of ground goals (i.e., goals expecting a single answer), the ``superfluous'' computed answers might become relevant for the evaluation of other goals, reducing the delay in future computations.

In this paper we present GEM, a goal evaluation algorithm for trust management systems that addresses all the above mentioned problems.
In GEM, policy statements are expressed as function-free logic programming clauses; each statement is stored by the principal defining it.
GEM computes the answers of a goal in a completely distributed way without disclosing intensional policies of principals, thereby preserving their confidentiality. %addresses the above mentioned problems. 
The algorithm deals with loops in three steps: (1) detection, (2) processing, and (3) termination.
To enable loop detection, we employ a distributed tabling strategy and associate an identifier to each request for the evaluation of a goal.
%The table of a goal stores all the requests previously received for the goal, the answers computed so far, and the status of its evaluation.
After its detection, a loop is processed by iteratively evaluating the goals in the loop until a fixpoint is reached, i.e., no more answers of the goals in the loop are computed, at which point their evaluation is terminated.
This three-steps approach enables GEM to detect both when the whole computation has terminated, and when the single goals within a computation are fully evaluated, allowing for the use of non-monotonic constraints in policies.
In addition, by exploiting the information stored in the table of a goal, principals are able to delay the response to a request until a ``maximal'' set of answers of the goal has been computed without running the risk of deadlocks.
We demonstrate that GEM terminates and is sound and complete with respect to the standard semantics for logic programs.

The remainder of the paper is structured as follows.
Section~\ref{sec:preliminaries} presents preliminaries on logic programming and SLD resolution. 
Section~\ref{sec:gem} introduces GEM and its implementation.
Section~\ref{sec:gem-properties} demonstrates the soundness, completeness and termination of the algorithm, and discusses what information is disclosed by GEM during the evaluation of a goal.
Section~\ref{sec:evaluation} presents the results of experiments conducted to evaluate the performance of GEM.
A possible extension of GEM to deal with negation is presented in Section~\ref{sec:advanced}.
Section~\ref{sec:rw} discusses related work.
Finally, Section~\ref{sec:conclusions} concludes and gives directions for future work.

%% file: preliminaries.tex
\section{Preliminaries on Logic Programming}
\label{sec:preliminaries}

In this section we revisit the concepts of logic programming \cite{A-HTCS-90} that are relevant to this paper. 
In particular, we review function-free logic programs.

An \emph{atom} is an object of the form $p(t_1,\ldots,t_n)$ where $p$ is an $n$-ary predicate symbol and $t_1,\ldots,t_n$ are terms (i.e., variables or constants).
An atom is \textit{ground} if $t_1,\ldots,t_n$ are constants.
A \emph{clause} is an expression of the form $H\leftarrow B_1,\ldots, B_n$ (with $n\geq 0$), where $H$ is an atom called \textit{head} and $B_1,\ldots,B_n$ (called \textit{body}) are atoms.
If $n=0$, the clause is a \textit{fact}. 
%\footnote{In the examples through the paper, we will adopt the syntactic conventions of Prolog so that each clause ends with the period ``.'' and ``$\leftarrow$'' is omitted in facts.}
A \textit{program} is a finite set of clauses.
We say that an atom $A$ is \textit{defined in the program} $P$ if and only if there is a clause in $P$ that has an atom $A'$ in its head such that $A$ and $A'$ are unifiable.
%Each program $P$ has an associated \emph{dependency graph}, which consists of a set of nodes and edges: all the atoms defined in $P$ are represented by a node in its dependency graph; given two atoms $A_1$ and $A_2$, there is an edge from $A_1$ to $A_2$ in the graph iff $A_2$ appears in the body of a clause which has $A_1$ in its head.
Finally, a \textit{goal} is a clause with no head atom, i.e., a clause of the form $\leftarrow B_1,\ldots,B_n$. %finite, possibly empty sequence of atoms $A_1,\ldots,A_n$ (
%The atoms $B_1,\ldots,B_n$ in the body of a goal are also called a \emph{query}.
%In the remainder of the paper, we omit symbol $\leftarrow$, and denote a goal $\leftarrow B_1,\ldots,B_n$ simply $B_1,\ldots,B_n$.
%In addition, w
Without loss of generality, in this paper we restrict to goals with $0\leq n\leq 1$, that is, consisting of at most one atom.
The empty goal is denoted by $\varnothing$.

SLD resolution (Selective Linear Definite clause resolution)~\cite{K-IP-74} is the standard operational semantics for logic programs. 
In this paper, we refer to SLD resolution with leftmost selection rule (extending the algorithm to an arbitrary selection rule is trivial).
Computations are constructed as sequences of ``basic'' steps. 
Consider a goal $G_0 =$ $\leftarrow B_1,\ldots,B_n$ and a clause $c$ in a program $P$. 
Let $H\leftarrow B'_1,\ldots,B'_m$ be a variant of $c$ variable disjoint from $\leftarrow B_1,\ldots,$ $B_n$. 
Let $B_1$ and $H$ unify with \emph{most general unifier} (\emph{mgu}) $\theta$.  
The goal $G_1 =$ $\leftarrow (B'_1,\ldots,B'_m,B_2,\ldots,B_n)\theta$ is called a \emph{resolvent of} $G_0$ \emph{and} $c$ \emph{with selected atom} $B_1$ \emph{and mgu} $\theta$.  
An SLD \emph{derivation step} is denoted by $G_0 \stackrel{\theta}\rightarrow G_1$.
Clause $H\leftarrow B'_1,\ldots,B'_m$ is called \textit{input clause}, and atom $B_1$ is called the \textit{selected atom} of $G_0$.

An SLD derivation is obtained by iterating derivation steps. The sequence $\delta:=G_0\stackrel{\theta_1}\rightarrow G_1 \stackrel{\theta_2}\rightarrow \cdots \stackrel{\theta_n}\rightarrow G_n \stackrel{\theta_{n+1}}\rightarrow \cdots$
is called a \emph{derivation of} $P \cup \{G_0\}$, where at every step the input clause employed is variable disjoint from the initial goal $G_0$ and from the substitutions and the input clauses used at earlier steps.
Given a program $P$ and a goal $G_0$, SLD resolution builds a search tree for $P \cup \{G_0\}$, called \textit{(derivation) tree} of $G_0$, whose branches are SLD derivations of $P \cup \{G_0\}$.
Any selected atom in the SLD resolution of $P \cup \{G_0\}$ is called a \emph{subgoal}.
SLD derivations can be finite or infinite.  
If $\delta:=G_0\stackrel{\theta_1}\rightarrow \cdots \stackrel{\theta_n}\rightarrow G_n$ is a finite prefix of a derivation, we say that $\theta=\theta_1\cdots\theta_n$ is a \textit{partial derivation} and $\theta$ is a \textit{partial computed answer substitution} of $P \cup \{G_0\}$. 
If $\delta$ ends with the empty goal $\varnothing$, $\theta$ is called
\textit{computed answer substitution} (\textit{c.a.s.}). 
Let $G_0 =$ $\leftarrow B_1$.
Then, we also call $\theta$ a \textit{solution} of $G_0$ and $B_1\theta$ an
\textit{answer} of $G_0$. The length of a (partial) derivation
$\delta$, denoted by $len(\delta)$, is the number of derivation steps
in $\delta$.  

The most commonly employed technique to prevent infinite derivations is \textit{tabling}~\cite{B-DOOD-89,CW-JACM-96,GG-ICLP-01,SYYZ-TPLP-01,TS-ICLP-86,V-ICLP-87,ZS-PPDP-03}.
Given a goal $G_0$ consisting of an atom defined in a program $P$, tabling-based goal evaluation algorithms create a table for each (sub)goal in the SLD resolution of $P \cup \{G_0\}$, to keep track of the previously evaluated goals and thus avoid the reevaluation of a subgoal.
Tabling algorithms differ mainly in the data structures employed for the evaluation of goal $G_0$.
Linear tabling~\cite{SYYZ-TPLP-01,ZS-PPDP-03} and DRA~\cite{GG-ICLP-01}, for instance, evaluate $G_0$ by building a single derivation tree of $G_0$.
In SLG resolution~\cite{CW-JACM-96}, on the other hand, goal $G_0$ is evaluated by producing a forest of (partial) derivation trees, one for each subgoal in the resolution of $P \cup \{G_0\}$. 
%Each tree has an associated table, where the derived answers are stored. 
In SLG, the evaluation of $G_0$ starts by ordinary resolution with the clauses in $P$; as in SLD, a subgoal $G_1$ is selected in a resolvent of $G_0$. 
If a tree for a variant of $G_1$ already exists, $G_1$ is added to the set of \textit{consumers} of the corresponding table. 
Otherwise, a tree for $G_1$ is created. 
When a new answer of a subgoal is found, it is stored in the respective table and it is propagated to its consumer subgoals. 
The evaluation of a goal by means of a forest of derivation trees proposed by SLG resolution is at the basis of the distributed goal evaluation algorithm proposed in this paper.

%% file: gem.tex
\section{The GEM Algorithm}
\label{sec:gem}

In this section we first introduce some definitions and basic assumptions underlying our work. %on the properties of TM system for which GEM is designed.
Then, we present GEM and discuss its implementation.

\input{system}

\input{basic}

\input{implementation}

%% file: system.tex
\subsection{Definitions and Assumptions}
\label{sec:system}

Similarly to other works on trust management (e.g.,~\cite{LWM-CS-03,ADNO-POLICY-06}), we consider policy statements expressed as function-free logic programming clauses. %; trust management policies are sets of clauses (i.e., programs).
As in most trust management systems, policy statements are stored at different locations: each location is controlled by a principal who is responsible for defining and evaluating the policy statements at that location.
We assume a one-to-one correspondence between locations and principals; accordingly, we use a principal's identifier to refer to the location she controls.
To represent the location where a policy statement is stored, we require every atom to have the form $p(loc,t_1,\ldots,t_n)$, where $loc$ is a mandatory term that represents the location where the atom is defined, and $t_1,\ldots,t_n$ are terms. 
For instance, $p(bob, \ldots)$ refers to $p$ as defined by Bob and thus stored at Bob's location. 

Let $a$ be a principal in the trust management system.
We call the set of policy statements defined by $a$ the \emph{local trust management policy} (or simply the \emph{policy}) of principal $a$.
The set of clauses with non-empty body in $a$'s policy is the \emph{intensional policy} of $a$, while the set of facts that can be derived from principal $a$'s policy forms the \emph{extensional policy} of $a$.
The set of all the local policies in the trust management system is called \emph{global policy}.

%Finally, s
Since we consider the confidentiality of intensional policies to be a main concern in trust management systems, we assume that principals do not have access to the policies %and the state of the computation (i.e., the partial derivation trees) 
at other principals' locations.
As a consequence, the answers of a goal cannot be derived by building the derivation tree of the goal as done by SLD resolution, as this might involve input clauses defined by different principals.
Similarly to SLG resolution, in GEM a principal computes the answers of a goal defined in her policy by building the \textit{partial derivation tree} of the goal.
Differently from a derivation tree, in the partial derivation tree of a goal $G$ only the first derivation step is obtained by resolution with the clauses defining $G$; all the subsequent steps are by substitution with the solutions of the subgoals of $G$.

\begin{definition}
\label{def:partial-derivation-tree}
Let $G =$ $\leftarrow A$ be a goal and $P_A$ be the policy in which $A$ is defined.
%Let $\Anss$ be a set of pairs of the form $(B,\theta)$, where $B$ is a goal and $B\theta$ is an answer of goal $B$.
A \emph{partial derivation tree} of $G$ is a derivation tree with the following properties:
\begin{itemize}
	\item the root is the node $(A\leftarrow A)$;
	\item there is a derivation step $(A\leftarrow A) \stackrel{\theta}{\rightarrow} (A\leftarrow B_1,\ldots,B_n)\theta$, where $(A\leftarrow A)$ is the root, iff there exists a clause $H\leftarrow B_1,\ldots,B_n$ in $P_A$ (renamed so that it is variable disjoint from $A$) s.t. $A$ and $H$ unify with $\theta=mgu(A,H)$; 
	\item let $(A\leftarrow B_1,\ldots,B_n)$ be a non-root node, and $\Anss$ be a set of answers of goal $\leftarrow B_1$; for each answer $B'_1\in\Anss$ (renamed so that it is variable disjoint from $B_1$) there is a derivation step $(A\leftarrow B_1,\ldots,B_n) \stackrel{\theta}{\rightarrow} (A\leftarrow B_2,\ldots,B_n)\theta$, where $\theta=mgu(B_1,B'_1)$;
	\item for each branch $(A\leftarrow A) \stackrel{\theta_0}{\rightarrow} (A\leftarrow B_1,\ldots,B_n)\theta_0 \stackrel{\theta_1}{\rightarrow} \ldots \stackrel{\theta_n}{\rightarrow} (A\leftarrow\varnothing)\theta_0\theta_1\cdots\theta_n$, 
	we say that $A\theta$ (with $\theta=\theta_0\theta_1\cdots\theta_n$) is an answer of $G$ \emph{using} clause $H\leftarrow B_1,\ldots,B_n$.\hfill$\Box$
\end{itemize}
\end{definition}

Note that, to enable the evaluation of an atom $B_1$ in the partial derivation tree of goal $G$, the location where $B_1$ is defined must be known by the principal evaluating $G$.
A straightforward solution for guaranteeing that this requirement is satisfied would be to impose the location parameter of each atom in a policy to be ground at policy definition time.
This, however, would limit the constraints that a principal can express.
Consider, for instance, clause~2 on page~2.
In the clause, the location parameter of atom \emph{memberOfAlpha($Y$,$X$)} is determined at runtime based on the answers of \emph{projectPartner(mc,$Y$)}.
Therefore, rather than relying on a ``static'' safety condition, we require the location parameter of an atom to be ground \emph{when the atom is selected for evaluation}. 
If this is not the case, the computation \emph{flounders}. %, as it is not possible to establish the location where the atom is defined.
A discussion on how to write flounder-free programs and queries is orthogonal to the scope of this paper.
Here, we just mention that there exist well-established techniques based on \emph{modes} \cite{AM-FAC-94} which guarantee that certain parameters of an atom are ground when the atom is selected for evaluation. 
%Informally speaking, a mode is an assignment indicating how the parameters of each atom should be used, i.e., which are the ``input'' and which are the ``output'' parameters.
%This allows to derive properties such as absence of floundering for \emph{well-moded} programs and queries \cite{AM-FAC-94}. 

\begin{figure}[!t]
	\centering
	\subfigure[Extensional Policy Confidentiality Levels]{
	{\small
	\begin{oldtabular}{|c|p{4cm}|}
	\cline{1-2}
	{\bf Level} & {\bf Disclosed Information}\\
	\cline{1-2}
	\multirow{2}{*}{E0} & \multirow{2}{*}{None} \\
	& \\
	\cline{1-2}
	\multirow{2}{*}{E1} & \multirow{2}{*}{Answers of a goal} \\
	& \\
	\cline{1-2}
	\end{oldtabular}
	}
	\label{fig:extensional}
	}
	\subfigure[Intensional Policy Confidentiality Levels]{
	{\small
	\begin{oldtabular}{|c|p{4cm}|}
	\cline{1-2}
	{\bf Level} & {\bf Disclosed Information}\\
	\cline{1-2}
	I0 & None \\
	\cline{1-2}
	I1 & Part of the dependency graph \\
	\cline{1-2}
	I2 & Full dependency graph \\
	\cline{1-2}
	I3 & Clauses \\
	\cline{1-2}
	\end{oldtabular}
	}
	\label{fig:intensional}
}
	\caption{Classification of Goal Evaluation Algorithms in Terms of Disclosed Policy Information}
	\label{fig:taxonomy}
\end{figure}

Finally, we define a classification criteria for goal evaluation algorithms based on disclosed policy information.
We will use such criteria to compare GEM with the existing algorithms (see Section~\ref{sec:rw}).
The classification criteria consists of two elements: an extensional and an intensional policy confidentiality level.
Intuitively, the first characterizes algorithms in terms of how much information about extensional policies they disclose during goal evaluation, while the second refers to the disclosure of intensional policies.
Confidentiality levels define an increasing scale used to characterized from the most conservative approaches where no policy information is disclosed, to the least confidentiality-preserving solutions which disclose respectively extensional and intensional policies in full.
%More precisely, the extensional policy confidentiality taxonomy consists of two levels; the intensional policy confidentiality taxonomy of four levels.
The extensional and intensional confidentiality levels are presented in Fig.~\ref{fig:taxonomy}.
In a goal evaluation algorithm classified as E1-I2, for example, principals disclose to a requester all the answers of the requested goals.
In addition, all the dependencies among the goals in the global policy are disclosed to the principals responsible for goal evaluation.

%% file: basic.tex
\subsection{Intuition}
\label{sec:intuition}

In this section we describe how GEM computes the answers of a goal.
Given a goal $G$, GEM computes the answers of $G$ by evaluating one branch of its partial derivation tree at a time; this may involve the generation of evaluation requests for subgoals that are processed by different principals at different locations.
When all the answers from each branch of the tree of $G$ have been computed, they are sent to the principal(s) that requested the evaluation of $G$. 
$G$ is \emph{completely evaluated} when no more answers of $G$ can be computed.

To illustrate how GEM works in detail, we consider the scenario presented in Section~\ref{sec:intro}, where several pharmaceutical companies collaborate in the research project \emph{Alpha}.
However, we slightly modify the global policy to better focus on the algorithm's features. 
In particular, we assume that company \emph{c$1$} already knows which are the partner companies in project \emph{Alpha}, without needing to request them to \emph{mc}, and we reduce the partner companies to \emph{c$2$} and \emph{c$3$} only.
Furthermore, we consider a research institute \emph{ri} that works on project \emph{Alpha} in partnership with company \emph{c$2$}.
As a result, we have the following global policy:

\vspace{0.1cm} 
{\small
%\begin{tabular}{lll}
1. ${\sf memberOfAlpha}(c1\mbox{,}X) \leftarrow {\sf memberOfAlpha}(c2\mbox{,}X).$ \\
\indent 2. ${\sf memberOfAlpha}(c1\mbox{,}X) \leftarrow {\sf memberOfAlpha}(c3\mbox{,}X).$ \\
\indent 3. ${\sf memberOfAlpha}(c2\mbox{,}X) \leftarrow {\sf memberOfAlpha}(ri\mbox{,}X).$ \\
\indent 4. ${\sf memberOfAlpha}(c2\mbox{,}alice).$\\
\indent 5. ${\sf memberOfAlpha}(c3\mbox{,}bob).$\\
%\end{tabular}
}
\vspace{-0.3cm} 

\begin{figure}[!t]
	\centering
	\fbox{
{\footnotesize
$\xymatrix@R=10pt@C=8pt@M=1pt{
\ar[d] & \\
membe\ms{rOfA}lpha(c1\mbox{,}X) \ar[d] \ar@(r,u)[rd] & \\
membe\ms{rOfA}lpha(c2\mbox{,}X) \ar[d] & membe\ms{rOfA}lpha(c3\mbox{,}X) \\
membe\ms{rOfA}lpha(ri\mbox{,}X) &
}$
}}
	\caption{Call Graph of the Example Global Policy}
	\label{fig:simple-graph}
\end{figure}
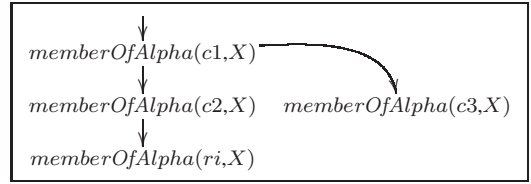

Recall that the first parameter of the atom in the head indicates the principal storing and evaluating a clause: clauses~1 and~2 are evaluated by \emph{c$1$}, clauses~3 and~4 by \emph{c$2$}, and clause~5 by \emph{c$3$}. 
Suppose that hospital $h$ sends to company \emph{c$1$} a request for (the evaluation of) goal $\leftarrow$ {\it memberOfAlpha(c$1$,X)} (for a matter of readability, from here on we omit the $\leftarrow$ symbol when referring to a goal $\leftarrow A$, and we simply refer to it as $A$).
Fig.~\ref{fig:simple-graph} shows the \emph{call graph} of the evaluation of {\it memberOfAlpha(c$1$,X)} with respect to the example global policy. 
A call graph is a directed graph where nodes represents goals and edges connect each goal to its subgoals~\cite{LMS-LOPSTR-98}.
In other words, edges represent (evaluation) requests.

GEM performs a depth-first computation.
When \emph{c$1$} receives the initial goal, it evaluates the first applicable clause in its policy (i.e., clause~1) and sends a request for {\it memberOf-Alpha(c$2$,X)} to \emph{c$2$}.
In turn, \emph{c$2$} sends a request for {\it memberOfAlpha(ri,X)} to \emph{ri}. 
\emph{ri} does not have any clause applicable to {\it memberOfAlpha(ri,X)} and returns an empty set of answers to \emph{c$2$}.
\emph{c$2$} evaluates the next applicable clause (i.e., {\it memberOfAlpha(c$2$,alice)}), which is a fact.
Since \emph{c$2$} does not have any other clause to evaluate, it sends the computed answer to \emph{c$1$}.
\emph{c$1$} applies the next clause (clause~2) and sends a request for {\it memberOfAlpha(c$3$,X)} to \emph{c$3$}, that returns answer {\it memberOfAlpha(c$3$,bob)} to \emph{c$1$} after applying clause~5.
At this point, {\it memberOfAlpha(c$1$,X)} is completely evaluated and \emph{c$1$} sends answers {\it memberOfAlpha(c$1$, alice)} and {\it memberOfAlpha(c$1$,bob)} to hospital $h$.

The evaluation of a subgoal of a goal $G$, however, may lead to new requests for $G$, forming a \emph{loop}. 
In our scenario, this reflects the ``sharing'' of project members among partner companies.
Consider, for instance, the global policy above with the following two additional clauses, stored by \emph{c$2$} and \emph{ri} respectively:

\vspace{0.1cm}
{\small
%\begin{tabular}{ll}
6. ${\sf memberOfAlpha}(c2\mbox{,}X) \leftarrow {\sf memberOfAlpha}(c1\mbox{,}X).$ \\
\indent 7. ${\sf memberOfAlpha}(ri\mbox{,}X) \leftarrow {\sf memberOfAlpha}(c2\mbox{,}X).$ \\
%\end{tabular}
}
\vspace{-0.3cm}

\begin{figure}[!t]
	\centering
		\subfigure[Call Graph with Loops]{
	\fbox{
	{\footnotesize
$\xymatrix@R=10pt@C=10pt@M=1pt{
\ar[d] & \\
membe\ms{rOfA}lpha(c1\mbox{,}X) \ar[d] \ar@(r,r)[rd] & \\
membe\ms{rOfA}lpha(c2\mbox{,}X) \ar[d] \ar@(r,r)[rd]  & membe\ms{rOfA}lpha(c3\mbox{,}X) \\
membe\ms{rOfA}lpha(ri\mbox{,}X) \ar[d] & *+[F]{membe\ms{rOfA}lpha(c1\mbox{,}X)} \\
*+[F]{membe\ms{rOfA}lpha(c2\mbox{,}X)} & 
}$
}}
	\label{fig:graph-loops}	
	}
	\subfigure[Compact Call Graph with Request Identifiers and Loops]{
	\fbox{
	{\footnotesize
$\xymatrix@R=16pt@C=10pt@M=1pt@L=2pt{
\ar[d]^-{h_1} & \\
membe\ms{rOfA}lpha(c1\mbox{,}X) \ar[d]^-{h_1c1_1} \ar@(r,u)[rd]^-{h_1c1_2} & \\
membe\ms{rOfA}lpha(c2\mbox{,}X) \ar[d]^-{h_1c1_1c2_1} \ar@/^15pt/ [u]^{h_1c1_1c2_2} & membe\ms{rOfA}lpha(c3\mbox{,}X) \\
membe\ms{rOfA}lpha(ri\mbox{,}X) \ar@/^15pt/[u]^{h_1c1_1c2_1ri_1} &
}$
}}
	\label{fig:compact-graph}	
	}
	\caption{Call Graph of the Example Global Policy with Loops}
	\label{fig:graph}
\end{figure}
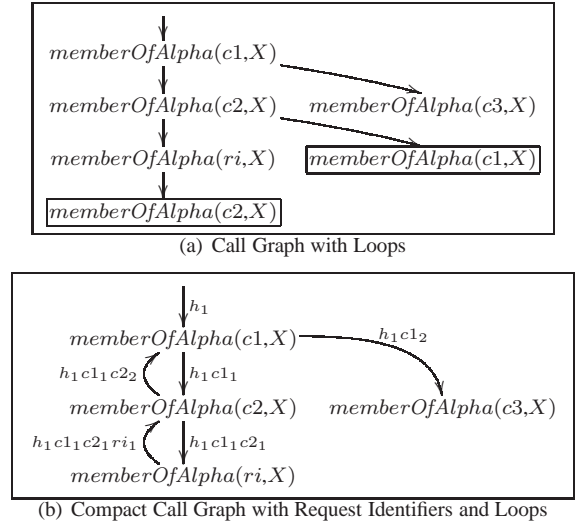

The new call graph is shown in Fig.~\ref{fig:graph-loops}. 
Now, when \emph{ri} receives the request for {\it memberOfAlpha(ri,X)} from \emph{c$2$}, it applies clause~7 and sends a request for {\it memberOfAlpha(c$2$,X)} back to \emph{c$2$}, forming a loop.
Similarly, the evaluation of clause~6 by \emph{c$2$} leads to another loop.
The requests forming a loop are identified by boxed atoms in Fig.~\ref{fig:graph-loops}.
Formally, a loop is defined as follows.

\begin{definition}
\label{def:loop}
Let $C$ be the call graph of the evaluation of a goal $G$ with respect to a global policy $P$.
A \emph{loop} is a maximal subgraph of $C$ consisting of goals $G_1,\ldots,G_k$ such that for each $G_i\in\{G_1,\ldots,G_k\}$ there exists a path that leads from $G_1$ to $G_i$ and from $G_i$ to (a variant of) $G_1$.
Then, we say that goals $G_1,\ldots,G_k$ are \emph{involved} in the loop. %, and goal $G_1$ is called the loop \emph{coordinator}.
\hfill$\Box$
\end{definition}

Intuitively, requests forming a loop should not be further evaluated.
However, in the example above \emph{c$1$} and \emph{c$2$} cannot detect whether a request forms a loop, as in a distributed system several independent requests for the same goal can occur. 
%GEM employs tabling techniques to avoid infinite derivations. 
In most of the existing goal evaluation algorithms (e.g., \cite{CW-JACM-96,D-TAPD-00,LWM-CS-03}), loop detection (and termination) is made possible by the system's ``global view'' on the derivation process. 
For example, centralized goal evaluation algorithms such as SLG~\cite{CW-JACM-96} and RT~\cite{LWM-CS-03} identify loops by observing goal dependencies respectively in the call stack and in the call graph of the global policy.
In a similar way, the distributed algorithm proposed by Damasio~\cite{D-TAPD-00} requires the dependency graph of the global policy to be known to all principals.
Such global view, however, implies the loss of policy confidentiality. 
GEM detects loops and their termination in a completely distributed way without resorting to any centralized data structure.
In GEM, loops are handled in three steps: (1) detection, (2) processing, and (3) termination.

\paragraph{Loop Detection.}
Loops are detected by dynamically identifying Strongly Connected Components (SCCs). 
A SCC is a set of mutually dependent subgoals.
More precisely, a set of goals $G_1,\ldots,G_k$ is part of a SCC if for each $G_i\in\{G_1,\ldots,G_k\}$ there exists a goal $G_j\in\{G_1,\ldots,G_{i-1},G_{i+1},\ldots,G_k\}$ such that $G_i$ and $G_j$ are involved in a common loop.
To enable the identification of SCCs, we assign to each request a unique identifier from an \emph{identifier domain}.

\begin{definition}
\label{def:identifier-domain}
An \emph{identifier domain} is a triple $\langle I,\orderingIn,\orderingSide\rangle$, where:
\begin{itemize}
	\item $I$ is a set of sequences of characters called \emph{identifiers};
	\item $\orderingIn$ is a partial order on the identifiers in $I$. Given two identifiers $\ID_1,\ID_2\in I$ s.t. $\ID_1\orderingIn \ID_2$, we say that $\ID_1$ is \emph{lower} than $\ID_2$, and $\ID_2$ is \emph{higher} than $\ID_1$;
	\item $\orderingSide$ is a partial order on the identifiers in $I$. Given two identifiers $\ID_1,\ID_2\in I$ s.t. $\ID_1\orderingSide \ID_2$, we say that $\ID_2$ is \emph{side} of $\ID_1$. 
	\item The following property holds: $\forall\ID_1, \ID_2, \ID_3, \ID_4 \in I$ if $\ID_1\orderingIn \ID_2$, $\ID_3\orderingIn \ID_4$, and $\ID_2\orderingSide \ID_4$, then $\ID_1\orderingSide \ID_3$.\hfill$\Box$
\end{itemize}
\end{definition}

Intuitively, $\orderingIn$ defines a top-down ordering and $\orderingSide$ defines a left-to-right ordering with respect to the call graph of the global policy.
In other words, $\orderingIn$ reflects the order in which the subgoals in a branch of the graph are evaluated, whereas $\orderingSide$ reflects the order in which the branches are inspected. 
Several identifier domains can be employed whose identifiers respect these partial orders (e.g., based on alphanumeric ordering).
For the sake of simplicity, in the sequel we consider identifiers obtained as follows: given a request for a goal $G$ with identifier $\ID_0$, the identifier of the request for a subgoal $G_1$ of $G$ has the form $\ID_0s_1$, denoting the concatenation of $\ID_0$ with a sequence of characters $s_1$.
Then, $\orderingIn$ is a prefix relation, and we have that $\ID_0s_1\orderingIn\ID_0$.
Ordering $\orderingSide$ is a partial order on the strings composing the identifiers.
For example, consider another subgoal $G_2$ of $G$ with identifier $\ID_0s_2$, which is evaluated after $G_1$. 
Then, we have that $\ID_0s_1\orderingSide\ID_0s_2$.
Even though identifiers from this domain leak some policy information (see Section~\ref{sec:disclosed} for more details), they allow for an easy visualization of the relationships among identifiers.
When applying GEM in practice, more confidentiality-preserving identifier domains can be employed.

A loop is detected when a principal receives a request with identifier $\ID_2$ for a goal $G$ such that a request $\ID_1$ for a variant of $G$ has been previously received and $\ID_2\orderingIn\ID_1$.
Accordingly, we call request $\ID_2$ a \textit{lower request} for $G$, while request $\ID_1$ is a \textit{higher request} for $G$.
We use the identifier of the higher request for $G$, $\ID_1$, as the loop identifier.
Goal $G$ is called the \textit{coordinator} of the loop.
An SCC may contain several loops. 
Given two loops with identifiers $\ID_1$ and $\ID_2$, we say that loop $\ID_2$ is \emph{lower} than loop $\ID_1$ if $\ID_2\orderingIn\ID_1$. 
The coordinator of the highest loop of the SCC (i.e., the loop with the highest identifier) is called the \emph{leader} of the SCC. 

Fig.~\ref{fig:compact-graph} represents a compact version of the call graph in Fig.~\ref{fig:graph-loops}, where loop coordinators are depicted only once.
In addition, in Fig.~\ref{fig:compact-graph} edges are labeled with the corresponding request identifier. 
In the remainder of the paper, we concatenate the identifier of a request for a goal evaluated by company \emph{c$1$} with meta-variables of the form $c1_i$. 
Thus, for instance, $c1_1$ and $c1_2$ are two distinct sequences of characters generated by \emph{c$1$}.
In the figure, identifiers $h_1$, $h_1c1_1$, $h_1c1_2$, and $h_1c1_1c2_1$ identify higher requests for goals {\it memberOfAlpha(c$1$,X)}, {\it memberOfAlpha(c$2$,X)}, {\it memberOfAlpha(c$3$,X)}, and {\it memberOfAlpha(ri,X)} respectively; identifiers $h_1c1_1c2_1ri_1$ and $h_1c1_1c2_2$ identify lower requests for goals {\it memberOfAlpha(c$2$,X)} and {\it memberOfAlpha(c$1$,X)} respectively.
Goals inherit the ordering associated with the identifier of their higher request.
Therefore, in Fig.~\ref{fig:compact-graph} goal {\it memberOfAlpha(c$1$,X)} is higher than {\it memberOfAlpha(c$2$,X)}, {\it memberOf-Alpha(c$3$,X)}, and {\it memberOfAlpha(ri,X)}.
Goals {\it memberOfAlpha(c$2$,X)} and {\it memberOfAl-pha(c$1$,X)} are the coordinators of loops $h_1c1_1$ and $h_1$ respectively.
Loop $h_1c1_1$ is lower than loop $h_1$, which is the highest loop of the SCC; therefore, {\it memberOfAlpha(c$1$,X)} is the leader of the SCC.
The identifier of the lower requests enables \emph{c$1$} and \emph{c$2$} to determine the subgoals involved in the loop, which are {\it memberOfAlpha(c$2$,X)} and {\it memberOfAlpha(ri,X)} respectively. 

\paragraph{Loop Processing.}
When a loop is detected, GEM sends the answers of the coordinator already computed to the requester of the lower request together with a notification about the loop. 
The loop is then processed iteratively as follows:
in turn, each principal (a) processes the received answers, (b) ``freezes'' the evaluation of the subgoal involved in the loop and evaluates other branches of the partial derivation tree of the locally defined goal. 
Then, when all branches have been evaluated, (c) the new answers are sent to the requester of the higher request with a notification about the loop. 
We call the execution of actions (a), (b), and (c) a \emph{loop iteration step}.

\begin{definition}
\label{def:loop-iteration-step}
Let $G$ be a goal, and $G_1,\ldots,G_k$ be the subgoals of $G$ s.t. $G,G_1,\ldots,G_k$ are involved in a loop $\ID$.
A \emph{loop iteration step} for goal $G$ is a three-phases process in which the principal evaluating $G$:
\end{definition}
\begin{enumerate}[(a)]
	\item Receives a set of answers of the subgoals $G_1,\ldots,G_k$ of $G$.
	\item Evaluates all the nodes in the partial derivation tree of $G$ whose selected atom is not involved in a loop.
	\item Sends the newly derived answers of $G$ to the requester of $G$.
	\hfill$\Box$
\end{enumerate}

The definition above applies to all the goals involved in a loop but the loop coordinator.
The loop iteration step for a loop coordinator differs in the order in which the phases occur.
In particular, for the coordinator phase (c) precedes (a) and (b), and the latter two are executed only after a loop iteration step for the other goals in the loop has been performed.
In other words, the processing of the coordinator occurs only after all the other goals in the loop have been processed.
This is because the coordinator, being the ``highest goal'' in the loop, is assigned the task of overseeing its processing.
More precisely, it is in charge of starting (phase (c)) a new \emph{loop iteration} whenever the answers of its subgoals lead to new answers of the coordinator (computed in phases (a) and (b)), i.e., until a fixpoint is reached.
This difference is reflected in the definition below.

\begin{definition}
\label{def:loop-iteration}
Let $G_1,\ldots,G_k$ be the goals involved in a loop $\ID_1$ s.t. goal $G_1$ is the loop coordinator.
A \emph{loop iteration} is a process where:
\end{definition}
\begin{enumerate}
	\item The answers of $G_1$ that have not been previously sent are sent to the requesters of the lower requests for $G_1$ (phase (c) of the loop iteration step for the coordinator).
	\item For each $G_i\in\{G_2,\ldots,G_k\}$ a loop iteration step for $G_i$ is performed, s.t. for each $G_j\in\{G_2,\ldots,G_k\}$, if $G_j$ is lower than $G_i$ then the loop iteration step for $G_j$ is executed before the loop iteration step for $G_i$.
	\item The principal evaluating $G_1$ receives a set of answers of the subgoals of $G_1$ involved in loop $\ID_1$ (phase (a) of the loop iteration step for $G_1$). All the nodes in the partial derivation tree of $G_1$ whose selected atom is not involved in a loop are processed (phase (b) of the loop iteration step). 
\hfill$\Box$
\end{enumerate}

%When the notification gets back to the coordinator (together with a possibly empty set of new answers), all lower goals have been processed.
If the processing of the received answers leads to new answers of the coordinator, these new answers are sent to the requesters of lower requests, starting a new iteration. Otherwise, a fixpoint has been reached (i.e., all possible answers of the goals in the loop have been computed) and the answers of the coordinator are sent to the requester of the higher request.
Notice that a goal in a higher loop may eventually provide new answers to a goal in a lower loop: the fixpoint for a loop must be recalculated every time new answers of its coordinator are computed.

In the example (Fig.~\ref{fig:compact-graph}), when \emph{c$2$} identifies loop $h_1c1_1$, it informs \emph{ri} that they are both involved in loop $h_1c1_1$.
Since \emph{ri} has no more clauses to evaluate, it returns an empty set of answers to \emph{c$2$} notifying it that {\it memberOfAlpha(ri,X)} is in loop $h_1c1_1$. 
The further evaluation of {\it memberOfAlpha(c$2$,X)} leads to the identification of loop $h_1$ and to a new answer {\it memberOfAlpha(c$2$,alice)}, which is sent first to \emph{ri} in the context of loop $h_1c1_1$.
In turn, \emph{ri} computes answer {\it memberOfAlpha(ri,alice)} and sends it to \emph{c$2$}. 
Now, a fixpoint for loop $h_1c1_1$ has been reached and \emph{c$2$} sends {\it memberOfAlpha(c$2$,alice)} to \emph{c$1$} notifying it that {\it memberOfAlpha(c$2$,X)} is in loop $h_1$.
Notice that {\it memberOfAlpha(c$2$,X)} is also in loop $h_1c1_1$, but since this loop does not involve \emph{c$1$}, \emph{c$1$} is not notified of it. 
Next, \emph{c$1$} computes answers {\it memberOfAlpha(c$1$,alice)} and {\it memberOfAlpha(c$1$,bob)} (the latter being found through the evaluation of {\it memberOfAlpha(c$3$,X)}), and sends them to \emph{c$2$} in the context of loop $h_1$.
In turn, \emph{c$2$} computes {\it memberOfAlpha(c$2$,bob)}. 
Now, \emph{c$2$} has to find a fixpoint for loop $h_1c1_1$ given the new answer before proceeding with the evaluation of loop $h_1$.
It is worth noting that \emph{ri} is not aware of loop $h_1$.
This is because loop notifications are only transmitted to higher requests (except for the lower request that has formed the loop).

\paragraph{Loop Termination.}
The termination of the evaluation of all the goals in an SCC is initiated by the principal handling the leader of the SCC when a fixpoint for the loop it coordinates has been reached.
In the example, when the answers of {\it memberOfAlpha(c$2$,X)} do not lead to new answers of {\it memberOfAlpha(c$1$,X)}, \emph{c$1$} informs \emph{c$2$} (which in turn informs \emph{ri}) that the evaluation of {\it memberOfAlpha(c$1$,X)} is terminated and sends answers {\it memberOfAlpha(c$1$,alice)} and {\it memberOfAlpha(c$1$,bob)} to \emph{h}.

\paragraph{Side Requests.}
So far we have only considered ``linear'' loops, i.e., loops formed by lower requests.
%More complex situations arise when the evaluation of a goal $G_1$ in an SCC leads to a higher request for a goal $G_2$ in the same SCC.
However, higher requests can also lead to a loop.
Consider, for instance, the following additional clause stored by company \emph{c$3$}:

\vspace{0.1cm}
{\small
8. ${\sf memberOfAlpha}(c3\mbox{,}X) \leftarrow {\sf memberOfAlpha}(ri\mbox{,}X).$ \\
}
\vspace{-0.3cm}

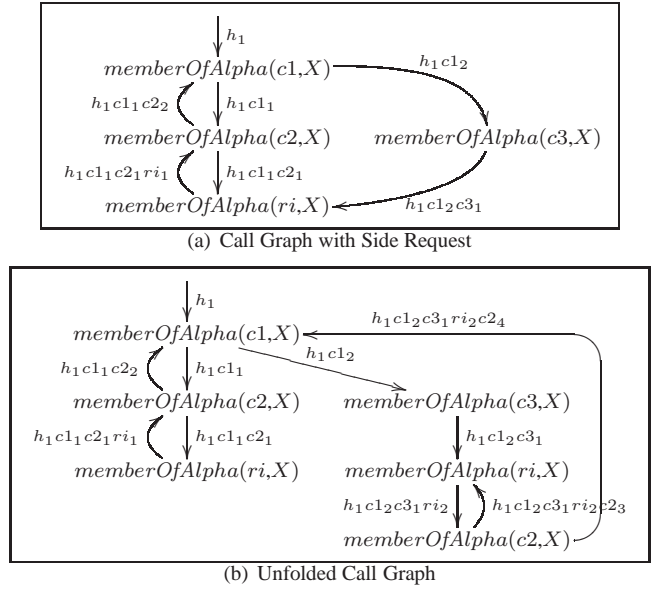
\begin{figure}[!t]
	\centering
\subfigure[Call Graph with Side Request]{
\fbox{
	{\footnotesize
$\xymatrix@R=16pt@C=14pt@M=1pt{
\ar[d]^-{h_1} & \\
membe\ms{rOfA}lpha(c1\mbox{,}X) \ar[d]^-{h_1c1_1} \ar@(r,u)[rd]^-{h_1c1_2} & \\
membe\ms{rOfA}lpha(c2\mbox{,}X) \ar[d]^-{h_1c1_1c2_1} \ar@/^15pt/ [u]^{h_1c1_1c2_2} & membe\ms{rOfA}lpha(c3\mbox{,}X) \ar@(d,l)[ld]^-{h_1c1_2c3_1}\\
membe\ms{rOfA}lpha(ri\mbox{,}X) \ar@/^15pt/[u]^{h_1c1_1c2_1ri_1} &
}$
}}	\label{fig:fold}
}
\subfigure[Unfolded Call Graph]{
\fbox{
	{\footnotesize
$\xymatrix@R=16pt@C=14pt@M=1pt{
\ar[d]^-{h_1} & \\
membe\ms{rOfA}lpha(c1\mbox{,}X) \ar[d]^-{h_1c1_1} \ar[rd]^-{h_1c1_2} & \\
membe\ms{rOfA}lpha(c2\mbox{,}X) \ar[d]^-{h_1c1_1c2_1} \ar@/^15pt/ [u]^{h_1c1_1c2_2} & membe\ms{rOfA}lpha(c3\mbox{,}X) \ar[d]^-{h_1c1_2c3_1}\\
membe\ms{rOfA}lpha(ri\mbox{,}X) \ar@/^15pt/[u]^{h_1c1_1c2_1ri_1} & membe\ms{rOfA}lpha(ri\mbox{,}X) \ar[d]_{h_1c1_2c3_1ri_2} &\\
& membe\ms{rOfA}lpha(c2\mbox{,}X) \ar@/_10pt/[u]_-{h_1c1_2c3_1ri_2c2_3} \ar `r[u] `u[uuu] [uuul]_-{h_1c1_2c3_1ri_2c2_4}
%\ar@/_60pt/[uuul]_-{g_1a_2d_1c_2b_4} 
& 
}$
}}
	
\label{fig:unfold}
}
	%\vspace{-0.4cm}
	\caption{Call Graphs for Side Requests}
	\label{fig:side-graph}
\end{figure}

\noindent The new (compact) call graph is shown in Fig.~\ref{fig:fold}. 
Now, the evaluation of goal {\it member-OfAlpha(c$3$,X)} by \emph{c$3$} leads to a request for goal {\it memberOfAlpha(ri,X)}, which is involved in loop $h_1c1_1$.
\emph{ri} can identify that the request originates from the evaluation of a goal in the same SCC as {\it memberOfAlpha(ri,X)}, since the request identifier $h_1c1_2c3_1$ is side of the identifier $h_1c1_1c2_1$ of the initial request for {\it memberOfAlpha(ri,X)} (i.e., $h_1c1_1c2_1\orderingSide h_1c1_2c3_1$).
However, it cannot identify the loop in which the goal evaluated by \emph{c$3$} is involved. 
This is because loop notifications are only transmitted to higher goals, and thus \emph{ri} is not aware of loop $h_1$.
%Following the ordering associated to the request identifiers, w
We refer to the request from \emph{c$3$} as \emph{side request}, and we call {\it memberOfAlpha(c$3$,X)} a \emph{side goal}.

The main problem with side requests is that it is difficult to determine when they should be responded to.
For example, if \emph{ri} sends answers to \emph{c$3$} at every iteration of loop $h_1c1_1$, \emph{c$1$} would not know when to stop waiting for answers from \emph{c$3$} (since \emph{c$1$} does not know on which goals {\it memberOfAlpha(c$3$,X)} depends). 
On the other hand, \emph{ri} cannot wait until a fixpoint is reached for loop $h_1c1_1$, since only \emph{c$2$} (the principal handling the coordinator) is aware of that.
%In other words, the main problem with side requests is that it is difficult to determine when they should be responded to: even though it is clear that the two goals involved in the side request participate in at least one common loop (if this was not the case, $G_1$ would be disposed by the time the side request is received), it may not be possible to identify (all) the common loop(s). 
To enable the detection of termination, however, a side request should be responded to only when a fixpoint is computed for all the loops lower than the loop in which the side goal is involved.

A simple yet effective solution to this problem is to treat a side request for a goal as a ``new'' request (i.e., a request for a goal that has not yet been evaluated) and to reevaluate the goal. 
Accordingly, when a side request is received, GEM creates a new partial derivation tree for the goal and proceeds with its evaluation. 
This corresponds to inspecting multiple times some branches of the call graph of the program (Fig.~\ref{fig:unfold}); however, it allows us to obtain a call graph formed only by linear loops which, as shown previously in this section, can be successfully evaluated by GEM.
Notice that, despite in the unfolded graph in Fig.~\ref{fig:unfold} some nodes and edges are duplicated with respect to the folded graph in Fig.~\ref{fig:fold}, the flow of answers among goals is equivalent.
This can be easily seen by the fact that edges connect the same nodes in both call graphs.
In Section~\ref{sec:gem-properties} (Theorem~\ref{thm:termination}) we show that a call graph is always finite.

Even though possible in theory, we expect side requests not to occur frequently in the evaluation of a policy.
For instance, no side request was present in any of the example policies in the literature. 
Therefore, we believe the overhead imposed by the proposed solution to be negligible in practice.
Nevertheless, we point out that the size of the unfolded graph is in the worst case exponential with respect to the number of nodes in the original graph.
The reevaluation of side requests, in fact, resembles the approach adopted by SLD resolution, which reevaluates every goal encountered during a computation.
However, thanks to our ability to detect loops, the number of goals evaluated by GEM during a computation is never higher than the number of goals that would be evaluated using SLD resolution.

%In \cite{TZE-TR-10} we discuss an alternative solution that preserves soundness, completeness, and termination of the algorithm without the need of reevaluating side requests.
Since we treat side requests in the same way as new requests, the partial order $\orderingSide$ is not exploited by the version of GEM presented here.
An alternative solution that prevents the reevaluation of side requests and thus requires their identification could be achieved by transmitting loop notifications to the requesters of both higher and lower requests, so that all the principals evaluating a goal in an SCC would be aware of the identifiers of all the loops in the SCC.
Furthermore, when a fixpoint for a loop is reached, the principal handling the loop coordinator should start a new loop iteration to inform all the goals in the loop that the fixpoint has been reached.
These two modifications would enable principals that receive a side request to know (a) in which loop the side request is involved, and (b) when to reply to the side request, that is, when all the loops lower than the one in which the side goal is involved have reached a fixpoint.
Given the complexity of the solution, in this paper we present only the ``basic'' version of GEM, which reevaluates side requests; the implementation of the described optimization is subject of future work.

To conclude, we point out that evaluating every higher request for a goal that is not yet completely evaluated is fundamental for enabling loop detection.
Consider, for instance, a global policy consisting only of clauses~1 and~6 on pages~8 and~9 respectively.
%Clearly, the evaluation of any of these goals would lead to a loop.
Assume that hospital $h$ issues at the same time a request for goals \emph{memberOfAlpha(c$1$,X)} and \emph{memberOfAlpha(c$2$,X)} with identifier $\ID_1$ and $\ID_2$ respectively.
The evaluation of the request by principals \ms{c1} and \ms{c2} leads to a higher request for goals \emph{memberOfAlpha(c$2$,X)} and \emph{memberOfAlpha(c$1$,X)} respectively.
When these second higher requests are received, a partial derivation tree for the requested goals already exists.
If the requests were not further processed, the loop identification would not be possible as no lower request would be issued, and the computation would deadlock.
Therefore, both initial requests must be processed independently.
This ``problem'' is common to all the distributed goal evaluation algorithms whose termination detection exploits request identifiers (e.g.,~\cite{ADNO-POLICY-06}).
Even though relatively simple solutions to this problem can be found (e.g., using timestamped requests, where only the oldest is evaluated), in this paper we focus on a ``basic'' solution for distributed goal evaluation, and do not address efficiency-related issues.

%% file: implementation.tex
\subsection{Implementation}
\label{sec:implementation}

Here, we introduce the data structures and procedures used by GEM to evaluate a goal.

\paragraph{Data Structures.}
%\label{sec:structures}
In GEM, principals communicate by exchanging \queryM\ and \answerM\ messages.
We rely on blocking communication, that is, whenever a principal $a$ sends (respectively receives) a request or response message, no other operation is performed by $a$ until the sending (resp. receipt) process is completed.
In addition, we assume that a message sent by principal $a$ to a principal $b$ is always received (once) by principal $b$.

\begin{definition}
A \query\ is a triple $\qTuple{\ID}{$\ReqM$}{G}$, where:
\begin{itemize}
\item $\ID$ is the request identifier; 
\item \ReqM\ is the principal issuing the request, called \emph{requester};
\item $G$ is a goal $\leftarrow p(loc,t_1,\ldots,t_n)$, where $loc$ is a constant.\hfill$\Box$
\end{itemize}
\end{definition}

A \query\ is an enquiry issued by principal \ReqM\ for the evaluation of goal $G$. 
Each \query\ is uniquely identified by an identifier $\ID$ and is sent to the principal defining $G$. 
%Recall that the location parameter of $G$ must be ground when $G$ is evaluated.

\begin{definition}
Let $r=\qTuple{\ID}{$\ReqM$}{G}$ be a request.
A \answer\ to $r$ is a tuple $\langle\ID,$\AnssM$,\AS,$ \LoopsM$\rangle$, where:
\begin{itemize}
\item $\ID$ is the response identifier; 
\item\AnssM\ is a (possibly empty) set of answers of $G$;
\item$\AS\in\{$\actSM,\loopPSM,\dispSM$\}$ is the status of the evaluation of $G$, where $\ID_1$ is a loop identifier;
\item \LoopsM\ is a set of loop identifiers.\hfill$\Box$
\end{itemize}
\end{definition}

A response has the same identifier of the request to which it refers. 
It contains a (possibly empty) set of answers of the requested goal $G$ (\AnssM) together with the status of $G$'s evaluation (\ASM) and information about the loops in which it is involved (\LoopsM). 
\ASM\ is \dispSM\ if $G$ has been completely evaluated, \actSM\ if additional answers of $G$ may be computed, and \loopPSM\ if the response is sent in the context of the evaluation of loop $\ID_1$. 
%Intuitively, $\ID_1$ identifies the coordinator waiting for the answers.

As discussed in Section~\ref{sec:intuition}, GEM computes the answers of a goal by a depth-first evaluation of its partial derivation tree (Definition~\ref{def:partial-derivation-tree}), which may involve the generation of requests for subgoals evaluated at different locations.
In the implementation, we represent a partial derivation tree as a data structure called \emph{evaluation tree}.
Compared to partial derivation trees, an evaluation tree keeps track of the identifier of the request and status of the evaluation of the selected atom of each \emph{node} in the evaluation tree.
%The building blocks of a evaluation tree are called \emph{nodes}.

\begin{definition}
A \node\ is a triple $\nTuple{\ID}{c}{S}$, where:
\begin{itemize}
	\item $\ID$ is the node identifier;
\item $c$ is a clause;
\item $S\in\{\ms{new},\ms{active},\ms{loop}(\IDs),\ms{answer},\ms{disposed}\}$ is the status of the evaluation of the selected atom in $c$, where \IDs\ is a set of loop identifiers.\hfill$\Box$
\end{itemize}
\end{definition}

The status of a node is \newSM\ if no atom from the body of $c$ has yet been selected for evaluation.
It is set to \actSM\ when a body atom is selected, and to \dispSM\ when the selected atom is completely evaluated. 
The status is set to \loopSS\ if the selected atom is involved in some loops, where \IDsM\ is the set of identifiers of those loops, and to \ansSM\ if $c$ is a fact. 
As mentioned in Section~\ref{sec:preliminaries}, we employ the leftmost selection rule.
Thus, the selected atom of $c$ is always the first body atom.

\begin{definition}
The \emph{evaluation tree} of a goal $G =$ $\leftarrow A$ is a tree with the following properties:
\begin{itemize}
	\item the root is node $\nTuple{\ID_0}{A\leftarrow A}{S_0}$;
	\item there is an edge from the root to a node $\nTuple{\ID_1}{(A\leftarrow B_1,\ldots,$ $B_n)\theta}{S_1}$, where $\ID_1\orderingIn\ID_0$, iff there exists a derivation step $(A\leftarrow A) \stackrel{\theta}{\rightarrow} (A\leftarrow B_1,\ldots,B_n)\theta$ in the partial derivation tree of $G$;
	\item there is an edge from node $\langle\ID_2,A\leftarrow B_1,\ldots,B_n,$ $S_2\rangle$ to node $\nTuple{\ID_3}{(A\leftarrow B_2,\ldots,$ $B_n)\theta}{S_3}$, where $\ID_2\orderingIn\ID_0$ and $\ID_3\orderingIn\ID_0$, iff there exists a derivation step $(A\leftarrow B_1,\ldots,B_n) \stackrel{\theta}{\rightarrow} (A\leftarrow B_2,\ldots,B_n)\theta$ in the partial derivation tree of $G$;\hfill$\Box$
%	\item An answer node is a node $\nTuple{\ID'''}{\Goal\leftarrow\Box}{\ansS}$, where $\ID'''\orderingIn\ID$.
\end{itemize}
\end{definition}

When a principal receives a higher request for a goal $G$, it creates a table for $G$. A table contains all the information about the evaluation of $G$.

\begin{definition}
The \emph{table} of a goal $G$, denoted \tabM$(G)$, is a tuple $\langle\QM,\LoopQM,\ActiveLoops,\AnsSet,\Tree\rangle$, where:
\begin{itemize}
	%\item $\TS\in\{\ms{new},\ms{active},\ms{loop}(\IDs),\ms{disposed}\}$ is the status of the evaluation of $G$, where \IDs\ are loop identifiers,
\item \QM\ is a higher request for $G$;
\item \LoopQM\ is a set of lower requests for $G$;
\item \ActiveLoopsM\ is a set of pairs $\ALTuple{\ID}{counter}$ where \ID\ is a loop identifier and $counter$ is an integer value;
%the number of subgoals in \Tree\ which are involved in loop \ID,
\item \AnsSetM\ is a set of pairs $\ASTuple{ans}{\IDs}$ where $ans$ is an answer of $G$ and \IDs\ is a set of request identifiers;
\item \TreeM\ is the evaluation tree of $G$.\hfill$\Box$
\end{itemize}
\end{definition}

The table of a goal $G$ stores the higher request \Q\ for which it has been created, the set of answers computed so far (\AnsSetM), and the evaluation tree of $G$ (\TreeM). Possible lower requests for $G$ are stored in \LoopQM. 
\ActiveLoops\ keeps a counter for each loop in which $G$ is involved.
The counter of a loop \IDM\ indicates the number of subgoals of $G$ which are involved in loop \IDM, i.e., the number of nodes in $\Tree$ with status \loopSSM\ such that $\ID\in\IDs$. 
The counter is decreased whenever an answer of one of these subgoals is received. 
The status of the root node of \Tree\ indicates the status of the evaluation of $G$.
%table has just been created (\newSM), if $G$ is waiting for the answers of some subgoals in \Tree\ (\actSM), if it is processing some loops ($\ms{loop}(\IDs)$), or if the evaluation of $G$ is terminated (\dispSM).
When $G$ is completely evaluated, the fields of its table are erased, but the answers of $G$ are maintained to speed up the evaluation of future requests for $G$.

\paragraph{Procedures.}
%\label{sec:procedures}
To initiate the evaluation of a goal $G$, a principal $a$ generates a unique sequence of characters $\ID_0$ and sends a request $\langle\ID_0,a,G\rangle$ to the principal defining $G$. %\footnote{Here, we assume that standard cryptographic techniques are in place to avoid collision of identifiers.}
A response $\langle\ID_0,\Anss,$ $\dispS,\{\}\rangle$ is returned to $a$ when the evaluation of $G$ terminates.
%Given a request for a goal $G$ defined in the local program $P_G$, 
GEM computes the answers of $G$ (defined in policy $P_G$) using the following procedures:
\begin{itemize}
	\item \newsubgoal: if the request is not a lower request, invokes \createtable\ to initiate the evaluation of $G$. Otherwise, it sends a loop notification to the requester;
	\item \createtable: creates a table for $G$ and initializes its evaluation tree with the applicable clauses in $P_G$;
	\item \newactive: activates a \newSM\ node in the evaluation tree of $G$; %. If the 
	\item \positivereturn: processes the answers received for a subgoal of $G$;
	\item \answerreturn: determines the requesters of $G$ to whom a response must be sent. %\footnote{Notice that the evaluation of $G$ might be requested by several principals in the distributed system.} 
	It is invoked when there are no more nodes in the evaluation tree of $G$ to activate;
	\item \sendanswer: sends the computed answers of $G$ to the requesters of $G$;
	\item \completion: disposes the table of $G$. It is invoked when $G$ is completely evaluated.
\end{itemize}

\begin{figure}[!t]
	\centering
%\fbox{
%	{\small
%\xymatrix@R=20pt@C=12pt{
%\ar[rr]^-{Incoming}_-{request\ for\ G} &&\newsubgoal \ar[d]^-{G\ is\ a\ ``new''\ goal} \ar[rr]^-{Lower\ request\ for\ G} & & \sendanswer\\
%&&\createtable \ar[d]^-{Initiate\ evaluation\ of\ G} & & \\
%&&\newactive \ar[r]^-{No\ more\ nodes}_-{to\ evaluate} & \answerreturn \ar[ruu]^-{Loop\ iteration\ step\ of\ G\ completed} \ar[r]^-{G\ is\ completely}_-{evaluated} & \completion \ar[uu]_-{Final\ response\ for\ G} \\
%\ar[rr]^-{Incoming\ response}_-{for\ a\ subgoal\ of\ G} &&\positivereturn \ar[u]_-{Resume\ evaluation\ of\ G} & & \\
%}
%}}
\includegraphics[width=0.99\linewidth]{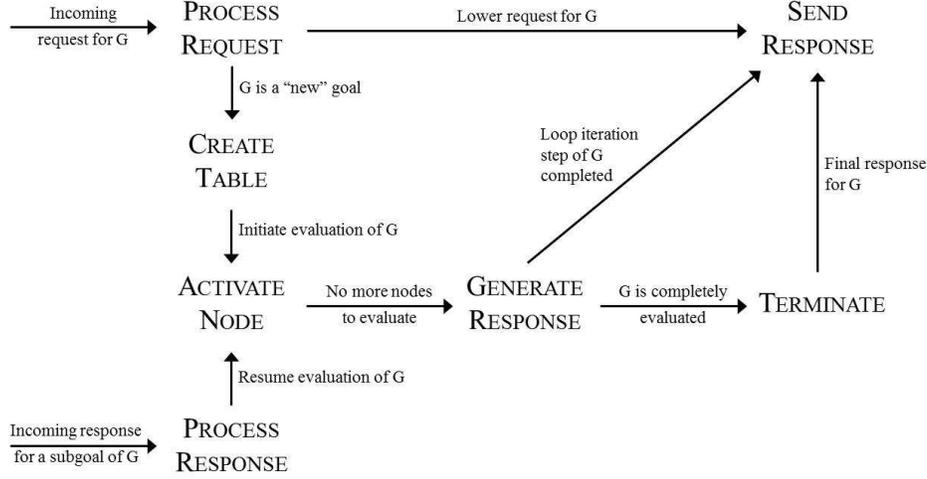}
	\caption{Interaction among the Procedures for the Evaluation of a Goal $G$}
	\label{fig:procedures}
\end{figure}

Each principal in the trust management system runs a listener that waits for incoming requests and responses. 
Whenever a request is received, the listener invokes \newsubgoal. 
Similarly, \positivereturn\ is invoked upon receiving a response to a previously issued request.
The interactions and dependencies among the different procedures are shown in Fig.~\ref{fig:procedures}.

\begin{algorithm}[!t]
{\footnotesize
	\dontprintsemicolon
	\SetInd{0.9em}{0.2em} 
	\SetKwInOut{Input}{input}

	\Input{a request $\langle\ID_0,\Req,G\rangle$}
	\BlankLine

		\uIf{$\exists \tabM(G')=\tTuple{\langle\ID_1,req',G'\rangle}{\ms{LR}}{AG}{AS}{T}$ s.t. $G'$ is a variant of $G$}{
			let $S_{root}$ be the status of the root node of $T$\;
			\uIf{$S_{root} =$ disposed}{
				\sendanswer $(\langle\ID_0,\Req,G'\rangle,\dispS,\{\})$\;
			}\uElseIf{$\ID_0 \orderingIn \ID_1$}{
				$\LoopQ := \LoopQ \cup \{\langle\ID_0,\Req,G'\rangle\}$\;
				\sendanswer $(\langle\ID_0,\Req,G'\rangle,\actS,\{\ID_1\})$\;
			}\uElse{
			let $G''$ be a variable renaming of $G$ s.t. $\not\exists \tab(G'')$\;
			\createtable$(\langle\ID_0,\Req,G''\rangle)$\;
		}
	
		}\uElse{
			\createtable$(\langle\ID_0,\Req,G\rangle)$\;
			}
		
	}
	\caption{\newsubgoal}
	\label{alg:newsubgoal}
\end{algorithm}

\newsubgoal\ (Algorithm~\ref{alg:newsubgoal}) takes as input a request $\langle\ID_0,\Req,G\rangle$ and, if there exists no table for a variant of $G$, invokes procedure \createtable\ to create a table for $G$ (lines~11-12).
If another request for goal $G$ (or a variant of $G$) has been previously received, three situations are possible:
\begin{enumerate}
\item \emph{The request refers to a goal which has been completely evaluated} (lines~3-4).
A response with the answers of $G$ is sent to the requester by invoking \sendanswer. 
\item \emph{The request is a lower request for $G$} (lines~5-7).
This corresponds to the detection of a loop $\ID_1$, where $\ID_1$ is the identifier of $\Q$. The request is added to the set of lower requests $\LoopQ$ and the answers computed so far are sent to the requester together with a notification about loop $\ID_1$, initiating the loop processing phase. 
\item \emph{The request is a side request or originates from a different initial request} (lines~8-10).
We treat the request as a new request; accordingly, a new table for $G$ is created by invoking \createtable. %\footnote{A possible optimization consists of cloning the table of $G'$ which is a variant of $G$, keeping only non-disposed nodes in its tree. This has the advantage of not reevaluating disposed subgoals and not recomputing previously computed answers.} 
\end{enumerate}

\begin{algorithm}[!t]
{\footnotesize
	\dontprintsemicolon
	\SetInd{0.9em}{0.2em} 
	\SetKwInOut{Input}{input}

	\Input{a request $\langle\ID_0,\Req,G=$ $\leftarrow A\rangle$}
	\BlankLine

			create $\tab(G)$\;
			initialize $\tab(G)$ to $\tTuple{\langle\ID_0,\Req,G\rangle}{\{\}}{\{\}}{\{\}}{\langle\ID_0,A\leftarrow A,\newS\rangle}$\;
			\ForEach{clause $H\leftarrow B_1,\ldots,B_n$ applicable to $G$ in the local policy}{
				let $H'\leftarrow B'_1,\ldots,B'_n$ be a variable renaming of the clause s.t. it is variable disjoint from $A$, and $\theta=mgu(A,H')$\;
				let $s$ be a unique sequence of characters\;
				add subnode $\langle\ID_0s,(H'\leftarrow B'_1,\ldots,B'_n)\theta,\newS\rangle$ to the root\;
			}
			\newactive$(G)$\;

	}
	\caption{\createtable}
	\label{alg:createtable}
\end{algorithm}

\createtable\ (Algorithm~\ref{alg:createtable}) inputs a request $\langle\ID_0,\Req,G\rangle$ and creates a table for goal $G$ with \Q\ set to  $\langle\ID_0,\Req,G\rangle$, and \Tree\ initialized with the clauses in the local policy applicable to $G$ (renamed so that they share no variable with $G$) (lines~1-7). 
The identifiers of the subnodes of the root are obtained by concatenating $\ID_0$ with a unique sequence of characters $s$. 
When the initialization of the table of $G$ is terminated, \newactive\ is invoked (line~8).

\begin{algorithm}[!t]
{\footnotesize
	\dontprintsemicolon
	\SetInd{0.9em}{0.2em} 
	\SetKwInOut{Input}{input}

	\Input{a goal $G=$ $\leftarrow A$}
	\BlankLine

	let $\ms{Table}(G)$ be $\tTuple{\ms{HR}}{\ms{LR}}{AG}{AS}{T}$\;
	\uIf{$(\not\exists$ a non-root node $t\in T$ with status $\newS)$ {\bf or} $(\langle A,\IDs\rangle\in AS)$}{
		\answerreturn$(G)$\;	
	}
	\uElse{
		let $S_{root}$ be the status of the root node of $T$\;
		\uIf{$S_{root} =$ new}{
			$S_{root} := \actS$\;
		}
		select the leftmost non-root node $t=\nTuple{\ID_1}{H\leftarrow B_1,\ldots,B_n}{\newS}$ from $T$\;
		\uIf{$n=0$}{
			set the status of $t$ to \ansS\;
			\uIf{$H$ is not subsumed by any answer in $AS$}{
				$AS := AS \cup \{\langle H, \{\}\rangle \}$\;
			}
			\newactive$(G)$\;
		}\uElse{
			%select the leftmost atom $B$ from the body of $\Rule$\;
			\uIf{the location of $B_1$ is not ground}{
				halt with an error message\ \ /* floundering */ \;
			}\uElse{
				set the status of $t$ to \actS\;
				send request $\langle\ID_1,local,B_1\rangle$ to the location of $B_1$\;
			}
		}
	}
	\caption{\newactive}
	\label{alg:newactive}
}
\end{algorithm}

\newactive\ (Algorithm~\ref{alg:newactive}) activates a \newSM\ node from the evaluation tree of a goal $G$. 
First, it sets the status of the root node of the evaluation tree $T$ of goal $G$ to \actSM\ (lines~5-7). 
Then, a node with status \newSM\ is selected from $T$ (line~8). 
If the node's clause is a fact and represents a new answer, it is added to the set of answers $AS$ (with an empty set of recipients), and \newactive\ is invoked again (lines~9-13). 
The answer subsumption check (line~11) is important to avoid sending the same answers of a goal more than once.
If the clause is not a fact, the leftmost body atom $B_1$ of the node's clause is selected for evaluation. 
In case that the location parameter of $B_1$ is not ground, an error is raised and the computation is aborted by \emph{floundering} (lines~15-16). 
%Indeed, the system may make an incorrect decision if the policy is not evaluated in its entirety. 
Otherwise, a request for $B_1$ is sent to the corresponding location; the node identifier is used as request identifier (lines~17-19).
If there are no more nodes with status \newSM, or $G$ is in the set of computed answers $AS$, \answerreturn\ is invoked (lines~2-3).

\begin{algorithm}[!t]
{\footnotesize
	\dontprintsemicolon
	\SetInd{0.9em}{0.2em} 
	\SetKwInOut{Input}{input}
	
	\Input{a request $\langle\ID_0,\Req,G\rangle$, a response status $\AS$, a set of loop identifiers $\Loops$}
	\BlankLine

	let $\ms{Table}(G)$ be $\tTuple{\ms{HR}}{\ms{LR}}{AG}{AS}{T}$\;
	$\Anss := \{\}$\;
	\ForEach{$\langle ans,\IDs\rangle \in AS$ s.t. $\ID_0 \notin \IDs$}{
		$\Anss:= \Anss \cup \{\Ans\}$\;
		$\IDs:= \IDs \cup \{\ID_0\}$\;
	}
%	\uIf{\AnssM\ contains a variant of $G$}{
%		set $\Anss$ to $G$\;
%	}
	send response $\langle\ID_0,\Anss,\AS,\Loops\rangle$ to $Req$\;
	
	\caption{\sendanswer}
	\label{alg:sendanswer}
}
\end{algorithm}

\sendanswer\ (Algorithm~\ref{alg:sendanswer}) inputs a request, a response status, and a set of loop identifiers and sends a response message to the requester, which includes the answers of $G$ that have not been previously sent to that requester (lines~3-7).

\begin{algorithm}[!t]
{\footnotesize
	\dontprintsemicolon
	\SetInd{0.9em}{0.2em} 
	\SetKwInOut{Input}{input}
	
	\Input{a response $\langle\ID_0,\Anss,\AS,\Loops\rangle$}
	\BlankLine

	let $t=\langle\ID_0,H\leftarrow B_1,\ldots,B_n,S_t\rangle$ be the node in the evaluation tree of goal $G=$ $\leftarrow A$ to which the response refers\;
	let $\ms{Table}(G)$ be $\tTuple{\ms{HR}}{\ms{LR}}{AG}{AS}{T}$\;
	let $\nTuple{\ID_1}{A\leftarrow A}{S_{root}}$ be the root node of $T$\;
\uIf{$S_{root} \neq$ disposed}{
	\uIf{$\AS =$ disposed}{
		\uIf{$S_t =$ \loopSSM}{
%			$\forall\IDs'$, dispose all the nodes in $T$ with status $\loopSPS$\;
			dispose all the nodes in $T$ involved in any loop\;
		}
		$S_t := \dispS$\;
	}\uElse{
		\uIf{$S_t =$ \loopSSM}{
			$\IDs := \IDs \cup \Loops$\;
		}\uElseIf{\LoopsM\ $\neq \{\}$}{
			$S_t :=$ \loopLoopsSM\;
		}
		$AG := AG \cup \{\langle\ID_2,0\rangle|\ID_2 \in \Loops$ and $\langle\ID_2,c\rangle \notin AG\}$\;	
		\uIf{$\AS = loop(\ID_3)$}{
			decrease the counter of $\ID_3$ in $AG$ by 1\;
			\uIf{$S_{root} =$ \actSM}{	
				$S_{root} := loop(\{\ID_3\})$\;	
			}%\uElseIf{$\TS(G)=loop(\IDs'')$}{
%				add $\IDs'$ to the set $\IDs''$\;
%			}
		}
	}
	\ForEach{$ans \in$ \AnssM}{
			let $ans'$ be a variable renaming of $ans$ s.t. it is variable disjoint from $B_1$, and $\theta=mgu(B_1,ans')$\;
			%let $\ID_{new}$ be the extension of $\ID$ with a random part\;
			let $s$ be a unique sequence of characters\;
			add subnode $\langle\ID_1s,(H\leftarrow B_2,\ldots,B_n)\theta,\newS\rangle$ of $t$\;				
	}
	\uIf{($S_{root} =$ \actSM) {\bf or} ($S_{root} =$ \loopSSM\ {\bf and} $\forall \ID_4 \in \IDs$, $\langle\ID_4,0\rangle \in AG$)}{
		\newactive$(G)$\;	
	}
}
}
	\caption{\positivereturn}
	\label{alg:positivereturn}
\end{algorithm}

Response messages are processed by \positivereturn\ (Algorithm \ref{alg:positivereturn}). 
The node $t$ to which the response refers is identified by looking at the response identifier (line 1). 
If the status of the response is \dispSM, the selected atom $B_1$ of $t$ is completely evaluated. 
Therefore, $t$ is disposed and, if $B_1$ is in a loop, also all the other nodes in any loop of the SCC are disposed (lines 5-8). 
This is because the termination of a loop is ordered by the principal handling the leader of the SCC once all the goals (and consequently, all the loops) in the SCC are completely evaluated.

%If the answer status is \actSM\ or \loopPSM
Otherwise, the status of $t$ is updated depending on whether the response contains a loop notification, i.e., set \LoopsM\ contains some loop identifier (lines 10-13). In this case, an entry is added to the set of active goals $AG$ for each new loop in \LoopsM\ (line~14).
If the response has been sent in the context of the evaluation of a loop $\ID_3$, the counter of $\ID_3$ in $AG$ is decreased and the status of the table is changed to $loop(\{\ID_3\})$ (lines~15-18).

After updating the node and table status, the set of answers in the response is processed (lines~19-23). In particular, a new subnode of $t$ is created for each answer. The clause of the new node is $(H\leftarrow B_2,\ldots,B_n)\theta$, where $\theta$ is the $mgu$ of $B_1$ and the answer,
and its identifier is obtained by concatenating the identifier $\ID_1$ of the root node of $T$ with a unique sequence of characters $s$. When all answers have been processed, if the principal is not waiting for a response for any subgoal in the evaluation tree of $G$, \newactive\ is invoked to proceed with the evaluation of $G$ (line~25).

\begin{algorithm}[!t]
{\footnotesize
	\dontprintsemicolon
	\SetInd{0.9em}{0.2em} 
	\SetKwInOut{Input}{input}

	\Input{a goal $G=$ $\leftarrow A$}
	\BlankLine
	
	let $\tab(G)$ be $\tTuple{\ms{HR}}{\ms{LR}}{AG}{AS}{T}$\;
	\uIf{$(\not\exists \langle\ID_0,$c$,$\loopSSM$\rangle \in T)$}{ %{\bf or} $(\langle G,\IDs'\rangle \in AS)$}{
		\completion$(G)$\;
	}\uElse{
			let $\nTuple{\ID_1}{A\leftarrow A}{S_{root}}$ be the root node of $T$\;
			\uIf{$G$ is the coordinator of a loop $\ID_1$ {\bf and} $\exists ans \in AS$ s.t. $ans$ has not been sent to some request in LR}{
				%let $\ID''$ be the id of the query in $\Q(G)$\;
				set the counter of $\ID_1$ in $AG$ to the number of subgoals in $T$ involved in loop $\ID_1$\; 
				\uIf{$S_{root} = loop(\IDs_1)$}{
					$S_{root} := loop(\IDs_1\cup\{\ID_1\})$\;
				}\uElse{
					$S_{root} := loop(\{\ID_1\})$\;
				}
				\ForEach{$\langle\ID_2,$\ReqM$,G\rangle \in LR$}{
					\sendanswer$(\langle\ID_2,\Req,G\rangle,loop(\ID_1),\{\})$\;
				}				
			}\uElseIf{$G$ is the leader of the SCC}{
					\completion$(G)$\;
			}\uElse{
				%let $\Q$ be $\langle\ID',\Req,G\rangle$\;
				let \LoopsM\ be the set $\{\ID_3|\langle\ID_3,C\rangle \in AG$ and $\ID_1\orderingIn \ID_3\}$\;
				set the counter of each $\ID_3\in \Loops$ to the number of subgoals in $T$ in loop $\ID_3$\; 
				\uIf{$S_{root} = loop(\IDs_1)$ {\bf and} $\exists \ID_4\in\IDs_1$ s.t. $\ID_1\orderingIn \ID_4$}{
					\sendanswer$(\ms{HR},loop(\ID_4),\Loops)$\;
				}\uElse{%If{($\tab(G)$ is $\actS$) {\bf or} ($\tab(G)$ has status $loop(\ID'')$ {\bf and} $G$ is the coordinator of loop $\ID''$)}{
					\sendanswer$(\ms{HR},\actS,\Loops)$\;
				}
				$S_{root} := \actS$\;
			}
	}
}
	\caption{\answerreturn}
	\label{alg:answerreturn}
\end{algorithm}

\answerreturn\ (Algorithm~\ref{alg:answerreturn}) is invoked when all the clauses in the evaluation tree of a goal $G$ (except for the ones in a loop) have been evaluated. If $G$ is not part of a loop, \completion\ is invoked (lines~2-3). Otherwise, we distinguish three cases:
\begin{enumerate}
	\item If set \ms{LR} is not empty, then goal $G$ is the coordinator of a loop $\ID_1$, where $\ID_1$ is the identifier of the higher request for $G$. If there are new answers of $G$ that have not yet been sent to the lower requests in $LR$, a response with status $loop(\ID_1)$ is sent to each of them (lines~6-14). This corresponds to starting a new loop iteration for loop $\ID_1$. The status of the root node of the evaluation tree $T$ is updated to keep track of the loops that are currently being processed (lines~8-11) and the counter of $\ID_1$ in the set of active goals $AG$ is set to the number of subgoals in $T$ involved in loop $\ID_1$ (i.e., the number of nodes with status \loopSSM\ such that $\ID_1\in\IDs$, line~7).
	This number corresponds to the number of subgoals for which a response in the context of loop $\ID_1$ will be returned. 
	\item If $G$ is the leader of the SCC and no new answers of $G$ have been computed, the loop is terminated by invoking \completion\ (lines~15-16).
	$G$ is the leader of the SCC if the only loop identifier in set $AG$ is the identifier of the higher request for $G$.
	\item Otherwise, a response including the identifier of the loops in which $G$ is involved is sent to the requester of $\Q$ (lines~18-24). The status of the response depends on whether $\Q$ is involved in one of the loops currently being processed (lines~20-23). 
\end{enumerate}

\begin{algorithm}[!t]
{\footnotesize
	\dontprintsemicolon
	\SetInd{0.9em}{0.2em} 
	\SetKwInOut{Input}{input}
	
	\Input{a goal $G$}
	\BlankLine
	
	let $\ms{Table}(G)$ be $\tTuple{\ms{HR}}{\ms{LR}}{AG}{AS}{T}$\;
	dispose all non-answer nodes in $T$\;
	\ForEach{$\langle\ID_0,\Req,G\rangle \in \{\ms{HR}\} \cup \ms{LR}$}{
		\sendanswer$(\langle\ID_0,req,G\rangle,\dispS,\{\})$\;
	}
	$\ms{HR} := null$\;
	$\ms{LR} := AG := \{\}$\;
%	let $S_{root}$ be the status of the root node of $T$\;
%	$S_{root} := \dispS$\;
}
	\caption{\completion}
	\label{alg:completion}
\end{algorithm}

\completion\ (Algorithm~\ref{alg:completion}) is responsible of disposing a table once all the answers of its goal $G$ have been computed.
More precisely, all the table fields are erased except for the set $AS$ of answers of $G$, which are kept in case of future requests for goal $G$.
A response with status \dispSM\ is sent to the requesters of $\Q$ and $\LoopQ$ (lines~3-5).

An example of execution of GEM is in Appendix~B.

%% file: properties.tex
\section{Properties of GEM}
\label{sec:gem-properties}

This section presents the soundness, completeness and termination results of GEM. Moreover, we discuss what information is disclosed by GEM during the
evaluation of a goal.

\subsection{Soundness, Completeness and Termination}
\label{sec:proofs}

Here, we refer to an arbitrary but fixed set $P_1,\ldots,P_n$ of policies, and to the corresponding global policy $P=P_1 \cup \ldots \cup P_n$.
To prove its soundness and completeness, we demonstrate that GEM computes a solution if and only if such a solution can be derived via SLD resolution, which has been proved sound and complete \cite{A-HTCS-90}.
The proofs of the theorems presented in this section are provided in Appendix A.

The following theorem states that each solution computed by GEM can also be derived via SLD resolution using the global policy $P$, and is thus correct. 
Intuitively, this is due to the fact that the solutions generated by the algorithm are obtained using the clause resolution mechanism, which produces correct results.

\begin{theorem}[Soundness]
\label{thm:soundness}
Let $G_1$ be a goal. Let $S$ be the set of tables resulting from running GEM on $G_1$ (w.r.t.\ $P=P_1\cup\ldots\cup P_n$). Let $G_1,\ldots,G_k$ be the goals for which there exists a table in $S$. For each goal $G_i\in\{G_1,\ldots,G_k\}$ let $Sol_i=\{\theta_{i,1},\ldots,\theta_{i,k_i}\}$ be the (possibly empty) set of solutions of $G_i$ generated by the algorithm.
Then, for each $G_i\in\{G_1,\ldots,G_k\}$ and for each $\theta_{i,j}\in Sol_i$ 
there exists an SLD derivation of $P \cup \{G_i\}$ 
with c.a.s. $\sigma$ s.t. $G_i\theta_{i,j}$ is a renaming of $G_i\sigma$.
\end{theorem}

Next, we present the completeness result.

\begin{theorem}[Completeness]
\label{thm:completeness}
Let $G_1$ be a goal. Let $S$ be the set of tables resulting by running GEM on $G_1$ (w.r.t.\ $P=P_1\cup\ldots\cup P_n$). 
Assume that running GEM on $G_1$ (w.r.t.\ $P=P_1\cup\ldots\cup P_n$) did not result in floundering.
If there exists an SLD derivation of $P \cup \{G_1\}$ with c.a.s. $\theta$, 
then there exists a solution $\sigma$ of $G_1$ in $S$ s.t. $G_1\theta$ is a renaming of $G_1\sigma$.
\end{theorem}

Finally, we state that GEM always terminates.

\begin{theorem}[Termination]
\label{thm:termination}
Given a goal $G$ evaluated w.r.t. a finite global policy $P$, GEM terminates.
\end{theorem}

\input{privacy}

%% file: privacy.tex
\subsection{Disclosed Information}
\label{sec:disclosed}

A primary objective of GEM is to preserve the confidentiality of intensional policies. 
Here, we discuss what policy information principals are able to collect during the evaluation of a goal, and classify GEM according to the confidentiality levels defined in Section~\ref{sec:preliminaries}.
%In particular, the question that we address is what can principals infer about the definition of goals evaluated by other principals in the trust management system. 

%Before presenting the answer to this question
First, let us define the following notation.
Let $P$ be a global policy, and $G_a$ and $G_b$ be two goals in $P$ defined by principals $a$ and $b$ respectively.
We say that goal $G_a$ \emph{depends} on goal $G_b$ if there is a path from $G_a$ to $G_b$ in the call graph of the evaluation of $G_a$ with respect to $P$.
Since each edge in the call graph represents a request in GEM, and in trust management each request corresponds to a delegation of authority, if $G_a$ depends on $G_b$ then we say that there is a \emph{chain of trust} from principal $a$ to principal $b$.

We also introduce some notation on request identifiers.
As mentioned in Section~\ref{sec:intuition}, the identifiers in an identifier domain can be defined in several ways (e.g., applying a hash function to the identifier of a higher request).
In this paper, we have considered an identifier domain where identifiers are obtained by concatenating the identifier of a higher request with a sequence of characters. 
Here, we discuss what information is disclosed by GEM during goal evaluation using this identifier domain.
We classify identifiers obtained by concatenation according to two dimensions: \emph{traceability} and \emph{length}.
Given two request identifiers $\ID_1$ and $\ID_2$ for goals $G_1$ and $G_2$ respectively, such that $\ID_2\orderingIn\ID_1$, the traceability dimension refers to the ability of a principal to infer which principals are involved in the evaluation of the goals in the path from $G_1$ to $G_2$.
On the other hand, the length dimension defines the ability to determine the number of goals in the path from $G_1$ to $G_2$.
%In the discussion in this section we refer to the identifier domain proposed in Section~\ref{sec:intuition}.
Let $id_0s_1\cdots s_n$ be a request identifier, where $id_0$ is the identifier of the initial request and each $s_i$ (for $i\in\{1,\ldots,n\}$) is a sequence of characters added by a principal $p_i$ to the identifier $id_0s_1\cdots s_{i-1}$ of a higher request.
For what concerns the traceability dimension, we say that $id_0s_1\cdots s_n$ is a \emph{traceable} identifier if each string $s_i$ uniquely identifies (the location of) principal $p_i$ (as done, for instance, by the identifiers in the example in Section~\ref{sec:intuition}); otherwise, we say that identifier $id_0s_1\cdots s_n$ is \emph{untraceable}.
The length dimension is defined based on the number of characters concatenated by each principal $p_i$ to the request identifier $id_0s_1\cdots s_{i-1}$.
Let $len(s_i)$ denote the number of characters in the string $s_i$.
If $len(s_1) = \ldots = len(s_n)$, then we say that $id_0s_1\cdots s_n$ is a \emph{fixed-length} identifier; otherwise, we say that $id_0s_1\cdots s_n$ is a \emph{variable-length} identifier (we assume that cryptographic techniques are in place to avoid collision of identifiers~\cite{HS-ALP-08}).
Note that a traceable identifier does not necessarily disclose information about the number of goals in the path between two goals; this is because a goal defined by a principal $a$ can have several subgoals defined in $a$'s policy.
Consider, for instance, the traceable request identifier $h\mbox{:}12c1\mbox{:}345$, obtained by concatenating a request identifier with a principal's identifier and a variable-length sequence of digits for each goal evaluated by the principal.
Even though identifier $h\mbox{:}12c1\mbox{:}345$ shows that the principals involved in the computation are hospital $h$ and company $c1$, it does not confer information about the number of goals evaluated by those principals.
Company $c1$, for example, might have evaluated two locally defined goals, concatenating the higher request $h\mbox{:}12$ received from hospital $h$ first with its identifier $c1$ and digit ``$3$'' (separated by a semicolon), and then with the sequence of digits ``$45$''.

We are now ready to present what information a principal $b$ is able to learn about the local policy of a principal $a$ where a goal $G_a$ is defined.
First of all, by requesting the evaluation of $G_a$, $b$ learns the set of answers to the request, i.e., the extensional policy relative to $G_a$; this is necessary for any goal evaluation algorithm. 
As mentioned in Section~\ref{sec:intro}, the confidentiality of extensional policies can be protected, for instance, by relying on hidden credentials~\cite{BHS-CCS-04,FAL-TC-06} or trust negotiation algorithms~\cite{WSJ-DISCEX-00,W-ITRUST-03}.
Here, we are more interested in what $b$ can learn about the intensional policy defining $G_a$. 

By sending a request for $G_a$ (say, with identifier $id_1$), $b$ can learn whether $G_a$ depends on some goal $G_b$ defined in her policy.  
Indeed, if $b$ receives a request for $G_b$ with identifier $id_2$ such that $id_2 \orderingIn id_1$, then $b$ knows that $G_a$ depends on $G_b$.

If $G_a$ depends on a number of goals defined by $b$, then by requesting $G_a$ $b$ learns:
\begin{itemize}
\item what are the goals $G_{b_1},\ldots,G_{b_n}$ defined in her policy on which $G_a$ depends;
%This corresponds to the set of queries $a$ receives with an id
\item for each $G_{b_i} \in \{G_{b_1},\ldots,G_{b_n}\}$, $b$ knows who is the principal $p_i$ that requested $G_{b_i}$; therefore, $b$ learns that $G_a$ depends on a 
goal defined by $p_i$, i.e., that there is a chain of trust from $a$ to $p_i$ (and from $p_i$ to $b$). Principal $b$, however, does not necessarily learn which is the goal defined by $p_i$ on which $G_a$ depends; 
\item depending on how the identifiers are constructed, $b$ might be able to learn additional information about the path from $G_a$ to $G_{b_i}$. 
In particular, if the identifier of the request for $G_{b_i}$ is fixed-length, $b$ is able to infer the number of goals in the path from $G_a$ to $G_{b_i}$. 
Additionally, if the identifier is traceable, $b$ also learns who are the principals defining those goals.
\end{itemize}
Thus, GEM can be classified as E1-I1 according to the classification criteria in Section~\ref{sec:preliminaries}.
In fact, principals learn all the answers of a goal along with some dependencies among the goals involved in an evaluation.

We now illustrate the concepts presented above with an example, using the global policy introduced in Section~\ref{sec:intuition} and the call graph shown in Fig.~\ref{fig:fold} on page~14 (ignore, for now, the request identifiers depicted in the figure).
Assume that the research institute \emph{ri} requests goal {\it memberOfAlpha(c$1$,X)} to \emph{c$1$}.
If the identifiers used in the computation were variable-length and untraceable, \emph{ri} would be able to learn that:
\begin{itemize}
	\item {\it memberOfAlpha(c$1$,X)} depends on goal {\it memberOfAlpha(ri,X)} defined in its policy;
	\item {\it memberOfAlpha(c$1$,X)} depends on \emph{some} goal $G_{C2}$ defined in \emph{c$2$}'s policy and on \emph{some} goal $G_{C3}$ defined by \emph{c$3$}; however, it does not learn which goals they are.
	Furthermore, due to the loop notification received from \emph{c$2$} following the evaluation of goal {\it memberOfAlpha(c$2$,X)}, \emph{ri} learns that {\it memberOfAlpha(c$2$,X)} depends on \emph{some} goal in the path from {\it memberOfAlpha(c$1$,X)} to $G_{C2}$.
\end{itemize}

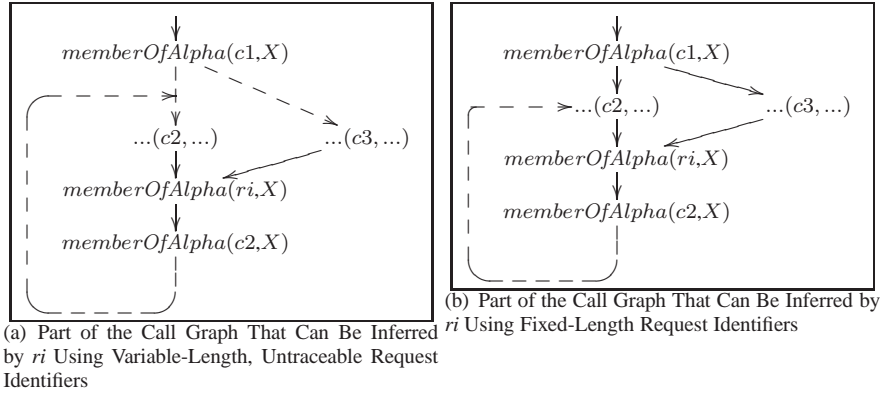
\begin{figure}[!t]
	\centering
		\subfigure[Part of the Call Graph That Can Be Inferred by \emph{ri} Using Variable-Length, Untraceable Request Identifiers]{
	\fbox{
		{\footnotesize
			$\xymatrix@R=10pt@C=11pt@M=1pt{
			& \ar[d] & \\
			& membe\ms{rOfA}lpha(c1\mbox{,}X) \ar@{-->}[dd] \ar@{-->}[rdd] & \\
			& & \\
			& ...(c2,...) \ar[d] & ...(c3,...) \ar[ld] \\
			& membe\ms{rOfA}lpha(ri\mbox{,}X) \ar[d] & \\
			& membe\ms{rOfA}lpha(c2\mbox{,}X)  \ar@{--}[d] \\%\ar@{-->}@/^55pt/[uuu] \\
			& \ar@{-->} `d/8pt[l] `l[lu] `u[uuuur] `/0pt[u] & \\
			}$
		}
	}	
		\label{fig:variable-length}
	}
	\subfigure[Part of the Call Graph That Can Be Inferred by \emph{ri} Using Fixed-Length Request Identifiers]{
	\fbox{
		{\footnotesize
			$\xymatrix@R=10pt@C=11pt@M=1pt{
			& \ar[d] & \\
			& membe\ms{rOfA}lpha(c1\mbox{,}X) \ar[d] \ar[rd] & \\
			\ar@{-->}[r] & ...(c2,...) \ar[d] & ...(c3,...) \ar[ld] \\
			& membe\ms{rOfA}lpha(ri\mbox{,}X) \ar[d] & \\
			& membe\ms{rOfA}lpha(c2\mbox{,}X) \ar@{--}[d] \\ %\ar@{-->}@/^55pt/[uu] & \\
			& \ar@{--} `d/8pt[l] `l[lu] `u/3pt[uuu]  & \\ 
			}$
		}
	}
		\label{fig:fixed-length}
	}
	\caption{Part of the Call Graph That Can Be Inferred by \emph{ri}}
	\label{fig:privacy-graph}
\end{figure} 

Fig.~\ref{fig:variable-length} represents \emph{ri}'s ``view'' of Fig.~\ref{fig:fold}, that is, the part of the call graph that \emph{ri} can infer.
In the graph, we denote with dots (``...'') the predicate symbols and terms that \emph{ri} does not learn; a dashed edge from a goal $G_1$ to a goal $G_2$ indicates that \emph{ri} is able to infer that $G_1$ depends on $G_2$, but not the (number of) goals and principals in the path from $G_1$ to $G_2$. 
Since \emph{ri} does not learn that $G_{C2}$ is actually goal {\it memberOfAlpha(c$2$,X)}, with respect to the call graph in Fig.~\ref{fig:fold}, in Fig.~\ref{fig:variable-length} goal {\it memberOfAlpha(c$2$,X)} is ``duplicated''.
The reason why \emph{ri} is not able to infer that $G_{C2}$ is {\it memberOfAlpha(c$2$,X)}, and even more, does not learn whether {\it memberOfAlpha(c$2$,X)} is in the path from {\it memberOfAlpha(c$1$,X)} to {\it memberOfAlpha(ri,X)} or is a lower goal, is that the only information that \emph{ri} receives in response to the request for {\it memberOfAlpha(c$2$,X)} (besides the answers of the goal) is a notification that {\it memberOfAlpha(c$2$,X)} is in a loop, say with identifier $id_l$.
%With respect to the dependency graph in Figure~\ref{fig:fold}, in Figure~\ref{fig:variable-length} goal {\it memberOf$\alpha$(c$2$,X)} is ``duplicated''; this is because \emph{ri} is not able to infer that $G_{C2}$ is actually {\it memberOf$\alpha$(c$2$,X)}, and even more, \emph{ri} does not learn whether {\it memberOf$\alpha$(c$2$,X)} is in the path from {\it memberOf$\alpha$(c$1$,X)} to {\it memberOf$\alpha$(c$3$,X)} or is a lower goal.
%In fact, the only information that \emph{ri} receives in response to the request for {\it memberOf$\alpha$(c$2$,X)} (further to the answers of the goal) is a notification that {\it memberOf$\alpha$(c$2$,X)} is in a loop, say with identifier $id_l$. 
\emph{ri} can observe that $id_l$ corresponds to a request higher than the request for {\it memberOfAlpha(ri,X)}, i.e., that the loop coordinator is higher than {\it memberOfAlpha(ri,X)}.
However, \emph{ri} cannot infer whether the loop was formed by its own request (in which case \emph{ri} would learn that $G_{C2}$ is {\it memberOfAlpha(c$2$,X)}), or by a request issued by \emph{c$2$} when evaluating {\it memberOfAlpha(c$2$,X)}, or even by the evaluation of a goal on which {\it memberOfAlpha(c$2$,X)} depends.
In other words, because of the variable-length and untraceable nature of identifiers, \emph{ri} does not know the number of goals in the path from {\it memberOfAlpha(ri,X)} to the loop coordinator.
	
In addition to the information above, if the identifiers used in the computations were fixed-length, \emph{ri} would also be able to learn that:
\begin{itemize}
	\item one of the paths from {\it memberOfAlpha(c$1$,X)} to {\it memberOfAlpha(ri,X)} consists of \emph{three} goals: {\it memberOfAlpha(c$1$,X)}, $G_{C2}$, and {\it memberOfAlpha(ri,X)}. 
	Furthermore, \emph{ri} can infer that $G_{C2}$ is the coordinator of loop $id_l$.
	However, \emph{ri} still does not learn whether $G_{C2}$ is {\it memberOfAlpha(c$2$,X)};
	\item the other path from {\it memberOfAlpha(c$1$,X)} to {\it memberOfAlpha(ri,X)} consists of \emph{three} goals: {\it memberOfAlpha(c$1$,X)}, $G_{C3}$, and {\it memberOfAlpha(ri,X)}.
\end{itemize}
The part of the call graph that \emph{ri} can infer in a computation with fixed-length identifiers is shown in Fig.~\ref{fig:fixed-length}.
Note that, in this example, the information that \emph{ri} can infer is the same independently from the traceability of the identifiers.
This is because \emph{ri} already knows all the principals in the paths from {\it memberOfAlpha(c$1$,X)} to {\it memberOfAlpha(ri,X)}: \emph{c$1$} is the principal to whom \emph{ri} sent the initial request, and \emph{c$2$} and \emph{c$3$} are the principals from whom \emph{ri} received the request for {\it memberOfAlpha(ri,X)}.
Even with traceable identifiers, \emph{ri} would not be able to infer more information about the path from {\it memberOfAlpha(c$2$,X)} to $G_{C2}$, as the only information received by \emph{ri} from \emph{c$2$} is the loop identifier.
%Note that the information disclosed by a fixed-length identifier is subsumed by the one disclosed by a traceable identifier.
%In fact, the first conveys ``only'' the number of goals in a path, the second reveals also the principals defining those goals.
%In this example, however, the information that \emph{ri} would be able infer is the same with both types of identifier.
%As shown in Figure~\ref{fig:variable-length}, using variable-length, traceable identifiers prevents \emph{ri} from inferring the number of goals and (some of the) principals in the path from {\it memberOfAlpha(c$1$,X)} to {\it memberOfAlpha(ri,X)} and in the loop formed by \emph{ri}'s request for {\it memberOfAlpha(c2,X)}.

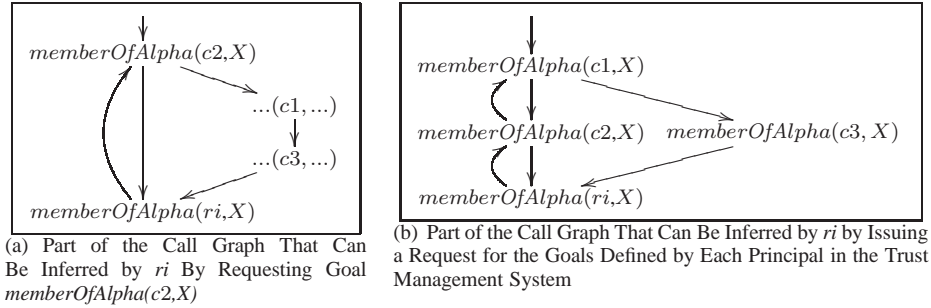
\begin{figure}[!t]
	\centering
	\subfigure[Part of the Call Graph That Can Be Inferred by \emph{ri} By Requesting Goal {\it memberOfAlpha(c$2$,X)}]{
	\fbox{
		{\footnotesize
			$\xymatrix@R=10pt@C=-4pt@M=1pt{
			\ar[d] & \\
			membe\ms{rOfA}lpha(c2\mbox{,}X) \ar[ddd] \ar[rd] & \\
			& ...(c1,...) \ar[d] \\
			& ...(c3,...) \ar[ld] \\
			membe\ms{rOfA}lpha(ri\mbox{,}X) \ar@/^15pt/ [uuu] & \\
			}$
		}
	}
		\label{fig:multiple-requests}
	}
	\hspace{0.2cm}
	\subfigure[Part of the Call Graph That Can Be Inferred by \emph{ri} by Issuing a Request for the Goals Defined by Each Principal in the Trust Management System]{
	\fbox{
		{\footnotesize
			$\xymatrix@R=15pt@C=6pt@M=1pt{
			\ar[d] & \\
			membe\ms{rOfA}lpha(c1\mbox{,}X) \ar[d] \ar[rd] & \\
			membe\ms{rOfA}lpha(c2\mbox{,}X) \ar@/^15pt/ [u] \ar[d] & membe\ms{rOfA}lpha(c3,X) \ar[ld] \\
			membe\ms{rOfA}lpha(ri\mbox{,}X)  \ar@/^15pt/ [u] & \\
			}$
		}
	}
		\label{fig:total-graph}
	}
	\caption{Part of the Call Graph That Can Be Inferred by \emph{ri} by Issuing Multiple Requests Using Fixed-Length Identifiers}
	\label{fig:multiple-graph}
\end{figure} 

A principal might attempt to infer a bigger portion of the call graph by issuing requests for each goal defined by the principals in the trust management system.
For instance, \emph{ri} can infer more information by issuing a request for each goal defined by companies \emph{c$1$}, \emph{c$2$}, and \emph{c$3$}.
In particular, by issuing a request for {\it memberOfAlpha(c$2$,X)}, in a computation with fixed-length and traceable identifiers \emph{ri} would learn that {\it memberOf-Alpha(c$2$,X)} is the goal defined by \emph{c$2$} that depends on {\it memberOfAlpha(ri,X)} (Fig.~\ref{fig:multiple-requests}).
By also issuing a request for {\it memberOfAlpha(c$3$,X)}, \emph{ri} could infer the whole call graph (Fig.~\ref{fig:total-graph}).
Note, however, that the global policy considered here is a relatively simple policy with few goals and principals.
A more complex policy would complicate and sometimes prevent the inference of goal dependencies.
Moreover, some information about the global policy would not be deducible by \emph{ri} when using variable-length and untraceable identifiers.
All the edges in the call graph in Fig.~\ref{fig:total-graph}, for instance, would be dashed edges if variable-length untraceable identifiers were used.
%In fact, in this case it would not be possible for \emph{ri} to learn whether there were additional goals in the path between {\it memberOfAlpha(c$2$,X)} and {\it memberOfAlpha(ri,X)}.
Finally, it is worth noting that even though \emph{ri} might be able to learn the whole call graph, that graph might correspond to different intensional policies~\cite{C-ASP-01}.
For example, \emph{ri} is not able to learn whether {\it memberOfAlpha(c$2$,X)} and {\it memberOfAlpha(c$3$,X)} are connected by disjunction or conjunction in $c1$'s policy.

To conclude, we argue that when using an appropriate identifier domain the knowledge about goal dependencies disclosed by GEM is not sufficient for a principal $b$ to infer the intensional policy relative to a goal $G_a$ defined by a principal $a$.
Principal $b$ always learns whether $G_a$ depends on goals defined in her policy, but most likely not all the goals in the global policy on which $G_a$ depends.
Consider, for instance, clause~2 on page~2.
A principal other than \emph{c$1$} and \emph{mc} cannot learn that {\it memberOfAlpha(c$1$,X)} depends on {\it projectPartner(mc,Y)}.
We believe that the information that $b$ can infer is consistent with the concept of trust management.
In fact, if $G_a$ depends on a goal defined by $b$, then there is a chain of trust from $a$ to $b$; it seems legitimate that the existence of such a chain  may not remain secret to $b$. 
%For example, consider the following global policy:
%
%\vspace{0.2cm} 
%{\small
%%\begin{tabular}{lll}
%${\sf memberOf}\alpha(c1,X) \leftarrow {\sf physician}(c1,X), {\sf memberOf}\alpha(c2,X).$ \\
%\indent${\sf physician}(c1,X) \leftarrow {\sf physician}(c2,X).$ \\
%\indent${\sf physician}(c1,X) \leftarrow {\sf physician}(c3,X).$ \\
%\indent${\sf memberOf}\alpha(c2,X) \leftarrow {\sf memberOf}\alpha(c3,X).$ \\
%%\end{tabular}
%}
%%\vspace{0.2cm}
%
%This policy has the same dependency graph of the global policy presented above (Figure~\ref{fig:privacy-graph1}).
%Therefore, the part of the graph that principal C3 would infer by requesting goal {\it memberOfAlpha(c$1$,X)} to C1 is the same as for the previous example (Figures~\ref{fig:privacy-graph2} and~\ref{fig:privacy-graph3}, depending on how request identifiers are constructed).
%However, there is a big semantic difference between the two global policies. 
%While in the first the set of members of project Alpha is the union of the set of physicians at company C1 and the set of project members at C2, in the latter the set of project members is an intersection of the two.

%% file: evaluation.tex
\section{Practical Evaluation}
\label{sec:evaluation}

We implemented the algorithms presented in Section~\ref{sec:implementation} in Java and conducted several experiments to evaluate the performance of GEM.
In particular, we first tested the implementation with the example policies defined in Section~\ref{sec:intuition} and in Appendix B. 
Then, we modified those policies to assess the scalability of GEM with respect to an increase in the number of principals in the trust management system and the number of clauses in the global policy.

We carried out the experiments by running GEM on four machines located in different area networks.
More precisely, we employed two machines located within the Eindhoven University of Technology (TU/e) network, and two located at the University of Twente (UT).
The two TU/e machines mount an Intel Core 2 Quad 2.4 GHz processor with 3 GB of RAM and a 32 bit Windows operating system.
The UT machines are 32 bit Ubuntu machines with the same processor but 2 GB of RAM.
In each experiment, we have assigned approximately one fourth of the principals (i.e., one fourth of the local policies) in the global policy to each machine.
The exchange of messages between principals on different machines is via HTTP ({\small {\tt javax.servlet.http}} Servlet API).

To present the results of the experiments, we group them into two sets.
The first set of experiments (Section~\ref{sec:experiments1}) studies the performance of GEM for an increasing number of principals, clauses, and loops in the global policy.
The second set (Section~\ref{sec:experiments2}) shows, for some of the global policies in the first set, the effects of increasing the size of the extensional policy (i.e., the number of facts in the global policy).
For each experiment we report the following results:
%\begin{itemize}
%	\item 
the number of principals involved in the computation (denoted by \emph{Princ});
%	\item 
the number of tables created by GEM during the computation (\emph{Tab});
%	\item 
the number of clauses evaluated (\emph{Clauses});
%	\item 
the sum of the computation times on the four machines, expressed in milliseconds (\emph{CTime});
%	\item 
the total time (\emph{TTime}), expressed in milliseconds, given by the \emph{CTime} plus network communication time;
%	\item 
the total memory occupied by GEM on the four machines, expressed in kilobytes (\emph{TMem});
%	\item 
the maximum memory occupied by the tables of the goals created by GEM on the four machines, expressed in kilobytes (\emph{TabMem});
%	\item 
the memory occupied by the tables of the goals created by GEM on the four machines after the tables' disposal, expressed in kilobytes (\emph{EndTabMem});
%	\item 
the number of requests issued during the computation (\emph{Req});
%	\item 
the number of loops identified during the computation (\emph{Loops});
%	\item 
the number of response messages exchanged between principals  (\emph{Resp});
%	\item 
the number of non-empty response messages (i.e., response messages containing at least one answer) exchanged between principals  (\emph{Resp\&Ans});
%	\item 
the total number of answers computed by GEM during the evaluation of the policy (\emph{Ans}).
%\end{itemize}

\subsection{Experiments Set 1: Increasing the Number of Principals, Clauses, and Loops}
\label{sec:experiments1}

In the first set of experiments we conducted three groups of (sub)experiments to evaluate the performance of GEM in response to an increase in: (1) the number of principals and clauses, (2) the number of loops, and (3) both the number of principals, clauses and loops in a global policy.
To evaluate GEM in response to an increase in the number of principals and clauses (experiments group 1), we created six variants of the global policy in Appendix B.
%The call graph of the six variants of the global policy is shown in Figure~\ref{fig:policy-exp1}; in the figure, we use identifiers from 1.0 to 1.5 to denote the variants, where variant 1.0 represents the original policy in Appendix B.
%To keep the figure as simple yet informative as possible, we label the nodes in the graph with the identifier of the principal evaluating the goal they represent rather than with the goal itself, as for the purpose of this section the number of principals involved in a computation is more relevant than the goals they evaluate.
For the second group of experiments, we created six variant of the global policy defined in Section~\ref{sec:intuition}. % to increase the number of loops in the global policy.
%Figure~\ref{fig:policy-exp2} shows the call graph of the six variants of the policy, which are denoted by identifiers 2.0 to 2.5; variant 2.0 corresponds to the original global policy.
Similarly, other six variants of the same policy were created for the experiments in the third group, in order to increase the number of both principals, clauses, and loops. %; the call graph of the six variants, denoted by identifiers 3.0 to 3.5 (where 3.0 corresponds to the original global policy defined in Section~\ref{sec:intuition}), is shown in Figure~\ref{fig:policy-exp3}.
The call graphs of the six variants of the global policies are shown in Fig.~C~1 in the appendix.
Each variant is denoted by an identifier that goes from x.0 to x.5 (where x is either 1, 2, or 3 depending on the experiment group for which they are used), where variant x.0 represents the original policy.
Notice that, for the sake of compactness, Fig.~C~1(b) and~C~1(c) show the folded graph of the global policies, i.e., they do not represent the reevaluation of goals due to side requests.
Since in GEM the computation is based on the unfolded versions of the graph (e.g., the graph in Fig.~\ref{fig:unfold} on page~14 for variants 2.0 and 3.0), the number of lower requests occurring in the actual computation is higher than the one displayed by the figures.
For instance, the number of lower requests for the leader of the SCC goes up to seven in variants (and hence experiments) 2.5 and 3.5.

\begin{table}[!t]
	\centering
{\footnotesize
	\begin{oldtabular}{|c|c|c|c|c|c|c|c|c|c|c|}
	\cline{1-11}
	\multirow{2}{*}{{\bf ID}} & \multirow{2}{*}{{\bf Princ}} & \multirow{2}{*}{{\bf Tab}} & \multirow{2}{*}{{\bf Clauses}} & {\bf TTime (CTime)} & {\bf TMem} & {\bf TabMem} & \multirow{2}{*}{{\bf Req}} & \multirow{2}{*}{{\bf Loops}} & {\bf Resp} & \multirow{2}{*}{{\bf Ans}}\\
	& & & & {\bf in ms} & {\bf in kB} & {\bf (EndTabMem) in kB} & & & {\bf (Resp\&Ans)} & \\\cline{1-11} \noalign{\smallskip}\cline{1-11}%\multicolumn{11}{c}{}\\\cline{1-11}
	1.0 & 4 & 4 & 6 & 286,1 (6,3) & 32 & 18 (15) & 5 & 1 & 9 (6) & 9\\\cline{1-11}
  1.1 &	7 & 7 & 12 & 305,9 (8,5) & 53 & 39 (27) & 9 & 2 & 17 (11) & 26\\\cline{1-11}
	1.2 & 10 & 10 & 18 & 315,2 (11,2) & 78 & 66 (44) & 13 & 3 & 25 (17) & 49\\\cline{1-11}
	1.3 & 13 & 13 & 24 & 322,8 (14,7) & 108 & 97 (63) & 17 & 4 & 33 (22) & 78\\\cline{1-11}
	1.4 & 16 & 16 & 30 & 327,0 (17,0) & 140 & 134 (85) & 21 & 5 & 41 (27) & 113\\\cline{1-11}
	1.5 & 19 & 19 & 36 & 331,7 (19,8) & 180 & 175 (108) & 25 & 6 & 49 (33) & 154\\ \cline{1-11}\noalign{\smallskip}\cline{1-11} %\multicolumn{11}{c}{}\\\cline{1-11}
	2.0 & 4 & 6 & 8 & 830,3 (10,7) & 47 & 37 (32) & 10 & 4 & 31 (17) & 20\\\cline{1-11}
	2.1 & 5 & 10 & 10 & 1037,2 (13,0) & 68 & 61 (50) & 16 & 6 & 48 (19) & 32\\\cline{1-11}
	2.2	& 6 & 15 & 12 & 1283,8 (16,5) & 92 & 91 (74) & 23 & 8 & 72 (30) & 46\\\cline{1-11}
	2.3 & 7 & 21 & 14 & 1459,9 (19,9) & 124 & 122 (100) & 31 & 10 & 95 (34) & 62\\\cline{1-11}
	2.4 & 8 & 28 & 16 & 1673,6 (23,4) & 163 & 153 (127) & 40 & 12 & 125 (47) & 80\\\cline{1-11}
	2.5 & 9 & 36 & 18 & 1840,0 (27,2) & 206 & 189 (159) & 50 & 14 & 154 (53) & 100\\\cline{1-11}\noalign{\smallskip}\cline{1-11} %\multicolumn{11}{c}{}\\\cline{1-11}
	3.0 & 4 & 6 & 8 & 830,3 (10,7) & 47 & 37 (32) & 10 & 4 & 31 (17) & 20\\\cline{1-11}
	3.1 & 7 & 16 & 16 & 1653,6 (23,9) & 151 & 149 (110) & 28 & 12 & 113 (71) & 112\\\cline{1-11}
	3.2 & 10 & 36 & 24 & 3029,0 (45,2) & 454 & 427 (297) & 64 & 28 & 309 (199) & 384\\\cline{1-11}
	3.3 & 13 & 76 & 32 & 5672,4 (78,8) & 1210 & 1133 (752) & 136 & 60 & 773 (535) & 1088\\\cline{1-11}
	3.4 & 16 & 156 & 40 & 10479,7 (134,4) & 3024 & 2827 (1824) & 280 & 124 & 1789 (1276) & 2800\\\cline{1-11}
	3.5 &	19 & 316 & 48 & 21939,5 (310,2) & 7285 & 6862 (4267) & 568 & 252 & 4569 (3477) & 6816\\\cline{1-11}
	\end{oldtabular}
}
	\caption{Performance Evaluation Results: Experiments Set 1}
	\label{tab:results-set1}
\end{table}

Table~\ref{tab:results-set1} presents the results of the three groups of experiments.
Each row in the table shows the results for the variant of the global policy with identifier indicated in column \emph{ID}.
The total time (\emph{TTime}) and the computation time (\emph{CTime}) are the average times for 100 runs of each experiment; the values in all the other columns are constant for every run, as they depend on the structure of the global policy.

Before interpreting the results of the experiments, let us provide some general comments on the relationship between the values in different columns of Table~\ref{tab:results-set1}.
First, in every experiment the number of requests (\emph{Req}) is equal to the number of tables generated (\emph{Tab}) plus the number of loops identified by GEM (\emph{Loops}); this is because a request is either a higher request, which leads to the creation of a table, or a lower request, and thus forms a loop.
Second, the number of response messages (\emph{Resp}) is always at least as large as the number of requests, since to every request is given a response, even if with an empty set of answers.
The number of empty response messages (i.e., response messages with an empty set of answers) can be observed by subtracting the number of response messages containing at least one answer (\emph{Resp\&Ans}) from the total number of response messages \emph{Resp}.
Finally, for some experiments (namely in the computation of variants 2.2 to 2.5 and 3.2 to 3.5), the number of tables generated by GEM is higher than the number of clauses (\emph{Clauses}) and thus the number of goals defined in the global policy.
%Even though this is possible in theory, if for example a lot of requests for goals defined in a global policy occur, in our case the reason is the following: 
This is because while column \emph{Tab} reports the \emph{total} number of tables generated during a computation, column \emph{Clauses} shows the number of \emph{different} clauses being evaluated.
In other words, we do not count twice the clauses used for the evaluation of a goal that is reevaluated because of a side request.
The reason behind this choice is that, on the one hand, we are interested in showing the impact of the number of tables generated and answers computed (\emph{Ans}) during the evaluation of a policy on the memory usage (\emph{TMem} and \emph{TabMem}); on the other hand, considering the number of different clauses gives a better insight on the size of a policy.
%We point out that the number of clauses actually evaluated during a computation is higher than the value reported in column \emph{Clauses}, since some clauses are used in more than one table.

A first interesting outcome of the experiments is that the time and memory results in Table~\ref{tab:results-set1} increase approximately linearly with the number of loops in the global policy (see Fig.~C~2 in the appendix of the paper for a graphical overview).
When the number of loops is low (experiments in groups 1 and 2) the time and memory usage are negligible, but as the number of loops increases considerably (experiments group 3), the \emph{TTime}, \emph{TMem}, and \emph{TabMem} get substantially higher.
This is because an increase in the number of loops (especially if nested, as in the global policy in Fig.~C~1(b) in the appendix of the paper), leads to an increase in the number of response messages, answers, and (since GEM reevaluates side requests) tables generated in a computation.

Another interesting result is represented by the difference between total time and computation time (see Fig.~C~2(a) in the appendix).
In fact, the \emph{TTime} ranges from 16.7 times the \emph{CTime} for policy variant 1.5, up to 80.1 times the \emph{CTime} for variant 2.1.
This implies that most of the \emph{TTime} is devoted to network communication.
%In the graph in Figure~\ref{fig:set1-time}, the number of messages represented on the x-axis is given by the sum of the request (\emph{Req}) and response (\emph{Resp}) messages in Table~\ref{tab:results-set1}; t
Therefore, we can conclude that the larger the number of requests and response messages and the number of answers per response message, the higher the difference between \emph{TTime} and \emph{CTime}.
Furthermore, we point out that in a real distributed system where to each principal corresponds a different machine (possibly in a different area network) the \emph{TTime} would be much higher than the results in Table~\ref{tab:results-set1}, especially when the number of principals increases.

Finally, Table~\ref{tab:results-set1} shows that the number of answers in a computation is always larger than the number of response messages containing at least one answer; in other words, each message in \emph{Resp\&Ans} carries more than one answer on average.
This information, combined with the observation on the difference between \emph{TTime} and \emph{CTime}, %displayed in Figure~\ref{fig:set1-time}, 
suggests that GEM can consistently reduce network overhead with respect to other goal evaluation algorithms which send one message for each computed answer (e.g.,~\cite{ADNO-POLICY-06}), especially when the number of principals and facts in the global policy increases.

\subsection{Experiments Set 2: Increasing the Size of Extensional Policies}
\label{sec:experiments2}

In a real-world policy, we expect the size of the extensional policy (i.e., the number of facts that can be derived from the policy) to substantially exceed the size of the intensional policy (i.e., the number of clauses used to derive new facts).
Consider, for example, the students of a university: while there are only a few rules that define the procedure for becoming a student, the number of students usually goes beyond several thousands.
The goal of the experiments presented in this section is to evaluate the impact on the performance of GEM of an increase in the number of facts in a global policy.
To this end, we considered some of the global policies introduced in Section~\ref{sec:experiments1} and increased the size of their extensional policies by a factor of 10, 50, and 100; in particular, we modified variants 1.0, 1.5, 2.0, 2.5, and 3.2.
We could not perform experiments on variants 3.3 to 3.5 due to the limited memory available on some of the machines used in the experiments.
%\todo{DT: explain why 3.2 and not 3.5. Problem: the reason is performance, I couldn't do more on MY machine...but on 4 machines there wouldn't be problems. Find a better excuse.}

\begin{table}[!t]
	\centering
{\footnotesize
	\begin{oldtabular}{|c|c|c|c|c|c|c|c|c|c|c|}
	\cline{1-11}
	\multirow{2}{*}{{\bf ID}} & \multirow{2}{*}{{\bf Princ}} & \multirow{2}{*}{{\bf Tab}} & \multirow{2}{*}{{\bf Clauses}} & {\bf TTime (CTime)} & {\bf TMem} & {\bf TabMem } & \multirow{2}{*}{{\bf Req}} & \multirow{2}{*}{{\bf Loops}} & {\bf Resp} & \multirow{2}{*}{{\bf Ans}}\\
	& & & & {\bf in ms} & {\bf in kB} & {\bf (EndTabMem) in kB} & & & {\bf (Resp\&Ans)} & \\\cline{1-11}\noalign{\smallskip}\cline{1-11} %\multicolumn{11}{c}{}\\\cline{1-11}
	1.0 & 4 & 4 & 6 & 286,1 (6,3) & 32 & 18 (15) & 5 & 1 & 9 (6) & 9\\\cline{1-11}
  1.0a & 4 & 4 & 24 & 293,7 (13,5) & 81 & 69 (58) & 5 & 1 & 9 (6) & 72\\\cline{1-11}
	1.0b & 4 & 4 & 104 & 334,8 (33,9) & 308 & 300 (255) & 5 & 1 & 9 (6) & 352\\\cline{1-11}
	1.0c & 4 & 4 & 204 & 519,8 (78,8) & 590 & 586 (498) & 5 & 1 & 9 (6) & 702\\\cline{1-11}\noalign{\smallskip}\cline{1-11}
	%\multicolumn{11}{c}{}\\\cline{1-11}
%	1.2 & 10 & 10 & 18 & ? (11,2) & 78 (66) & 13 & 3 & 25 (17) & 49\\\cline{1-10}
%	1.2a & 10 & 10 & 72 & ? (33,2) & 384 (379) & 13 & 3 & 25 (17) & 436\\\cline{1-10}
%	1.2b & 10 & 10 & 612 & ? (547,9) & 3515 (3468) & 13 & 3 & 25 (17) & 4306\\\cline{1-10}
	1.5 & 19 & 19 & 36 & 331,7 (19,8) & 180 & 175 (108) & 25 & 6 & 49 (33) & 154\\\cline{1-11}
	1.5a & 19 & 19 & 144 & 391,6 (80,3) & 1215 & 1210 (790) & 25 & 6 & 49 (33) & 1432\\\cline{1-11}
	1.5b & 19 & 19 & 624 & 1589,4 (817,7) & 5857 & 5764 (3786) & 25 & 6 & 49 (33) & 7112\\\cline{1-11}
	1.5c & 19 & 19 & 1224 & 4356,5 (2986,9) & 11711 & 11541 (7098) & 25 & 6 & 49 (33) & 14212\\\cline{1-11}\noalign{\smallskip}\cline{1-11}
	%\multicolumn{11}{c}{}\\\cline{1-11}
	2.0 & 4 & 6 & 8 & 830,3 (10,7) & 47 & 37 (32) & 10 & 4 & 31 (17) & 20\\\cline{1-11}
	2.0a & 4 & 6 & 26 & 843,7 (23,9) & 192 & 184 (144) & 10 & 4 & 31 (17) & 200\\\cline{1-11}
	2.0b & 4 & 6 & 106 & 948,4 (68,6) & 846 & 844 (651) & 10 & 4 & 31 (17) & 1000\\\cline{1-11}
	2.0c & 4 & 6 & 206 & 1441,2 (161,3) & 1672 & 1658 (1279) & 10 & 4 & 31 (17) & 2000\\\cline{1-11}\noalign{\smallskip}\cline{1-11}
	%\multicolumn{11}{c}{}\\\cline{1-11}
%	2.2	& 6 & 15 & 12 & ? (16,5) & 92	(91) & 23 & 8 & 72 (30) & 46\\\cline{1-10}
%	2.2a & 6 & 15 & 30 & ? (37,3) & 453 (446) & 23 & 8 & 72 (30) & 460\\\cline{1-10}
%	2.2b & 6 & 15 & 210 & ? (369,4) & 4111 (4025) & 23 & 8 & 72 (30) & 4600\\\cline{1-10}
	2.5 & 9 & 36 & 18 & 1840,0 (27,2) & 206 & 189 (159) & 50 & 14 & 154 (53) & 100\\\cline{1-11}
	2.5a & 9 & 36 & 36 & 1871,8 (58,1) & 1017 & 976 (726) & 50 & 14 & 154 (53) & 1000\\\cline{1-11}
	2.5b & 9 & 36 & 116 & 2490,1 (259,1) & 4665 & 4534 (3315) & 50 & 14 & 154 (53) & 5000\\\cline{1-11}
	2.5c & 9 & 36 & 216 & 3875,9 (755,2) & 9225 & 8984 (6538) & 50 & 14 & 154 (53) & 10000\\\cline{1-11}\noalign{\smallskip}\cline{1-11}
	%\multicolumn{11}{c}{}\\\cline{1-11}
	3.2 & 10 & 36 & 24 & 3029,0 (45,2) & 454 & 427 (297) & 64 & 28 & 309 (196) & 384\\\cline{1-11}
	3.2a & 10 & 36 & 78 & 3142,3 (158,1) & 3306 & 3259 (2081) & 64 & 28 & 309 (196) & 3840\\\cline{1-11}
	3.2b & 10 & 36 & 318 & 6277,8 (1789,6) & 16103 & 15954 (10032) & 64 & 28 & 309 (196) & 19200\\\cline{1-11}
	3.2c & 10 & 36 & 618 & 13906,3 (6346,1) & 31938 & 31883 (19968) & 64 & 28 & 309 (196) & 38400\\\cline{1-11}
	\end{oldtabular}
}
	\caption{Performance Evaluation Results: Experiments Set 2}
	\label{tab:results-set2}
\end{table}

Table~\ref{tab:results-set2} shows the results of the second set of experiments.
In the table, suffix ``a'' on a variant's identifier indicates an increase by a factor of 10 of the number of facts in that variant of the global policy, suffix ``b'' indicates an increase by a factor of 50, and suffix ``c'' indicates an increase by a factor of 100.
Note that the number of answers (\emph{Ans}) computed on variants 1.0 and 1.5 grows less than a factor of 10, 50, and 100 because, in order to not modify the    structure of the global policy, the number of facts in the policies of principals \emph{mc$1$} to \emph{mc$6$} in Fig.~C~1(a) in the appendix was not increased; more precisely, those policies  always consists of only two facts. 
Similarly to the previous experiments, \emph{TTime} and \emph{CTime} are the average times for 100 runs of each experiment.

The results in Table~\ref{tab:results-set2} show that memory usage (\emph{TMem} and \emph{TabMem}) and computation time grows faster than the other values for policies with a very large number of computed answers (i.e., variants 1.5c, 2.5c, and 3.2c).
For what concerns memory usage (Fig.~C~3(b) in the appendix of the paper), the extra overhead is due to the accompanying increase of the information that needs to be stored in tables (i.e., clauses, loops, answers).
After the disposal of the tables employed in the computation, there is a decrease of up to 38\% of the memory usage (\emph{EndTabMem}).
This suggests that it is very important to delete as much information as possible from the table of a goal when the goal is completely evaluated, as this leads to a substantial reduction of memory usage.
In this respect, GEM has the advantage of enabling principals to detect when the evaluation of the single goals involved in a computation is completed, and immediately clean up the table of those goals.

Further to the increase in the number of exchanged messages, we believe the extra overhead in \emph{TTime} and \emph{CTime} in computations with a large number of answers (the peaks for variants 1.5c, 2.5c, and 3.2c in Fig.~C~3(a) in the appendix) to be due to the growth of the data structures used by GEM to search, for example, for new answers and clauses to be activated.
In addition, contrarily to experiments set 1, in this set of experiments the \emph{TTime} is not always dominated by network communication time.
In particular, for variant 1.5c the \emph{CTime} is twice the network time, and for variant 1.5b the \emph{CTime} is slightly larger than the network time.
In the remaining experiments the \emph{TTime} ranges from 2.2 times the \emph{CTime} for policy variant 3.2c, up to 77.7 times the \emph{CTime} for variant 2.0.
Moreover, note that the difference between \emph{TTime} and \emph{CTime} always decreases as the number of facts grows.
This is due to the fact that the number of messages exchanged between principals remains constant while the size of the extensional policy is increased.

To conclude, we highlight again how the ``wait'' mechanism that GEM employs to collect a maximum set of answers before sending a response can contribute to reduce the network overhead, especially for global policies with a large extensional policy.
For example, in the experiment on variant 3.2c, GEM sends ``only'' 196 response messages, while other distributed goal evaluation algorithms (e.g.,~\cite{ADNO-POLICY-06}) would send 38400 messages, one for each computed answer.
Intuitively, the latter approach would lead to a network communication time much higher than the 7.5 seconds spent by GEM.

%% file: advanced.tex
\section{Dealing with Negation}
\label{sec:advanced}

GEM is devised to work with \emph{definite} logic programs, i.e., programs without negation.
Negation is used  by some trust management systems (e.g., \cite{CTDEHH-ENTCS-06,CD-IFIPTM-10}) to express non-monotonic constraints, such as separation of duty or ``distrust'' in principals with certain attributes (e.g., employees of a rival company).
Here, we discuss how GEM can be extended to support the use of \emph{negation as failure}.

Negation as failure is an inference rule that derives the truth of a negated body atom $not(B)$ by the failure to derive $B$.
%The problem when allowing the use of negation (as failure) is that the resulting programs may have several minimal models, and therefore no unique stable model~\cite{GL-ICLP-88}.
%In particular, this is true in the presence of \emph{loops through negation} (i.e., loops involving negated atoms).
The problem when allowing the use of negation (as failure) is that in the presence of \emph{loops through negation} (i.e., loops involving negated atoms) a program may have several minimal models~\cite{GL-ICLP-88}.
For instance, program $p\leftarrow not(q)$, $q\leftarrow not(p)$ has two minimal models: $\{p\}$ and $\{q\}$.
Moreover, these two models are not well-founded~\cite{GRS-JACM-91}, as there is no clause in the program demonstrating that $p$ respectively $q$ are true.
%On the other hand, program $p\leftarrow not(p)$ has no model: as there is no clause demonstrating that $p$ is true, $not(p)$ evaluates to true, which implies that $p$ is true (contradiction).
Another undesired consequence of loops through negation is that they may introduce ``inconsistencies'' in a program, as shown at the end of this section.
There are additional consequences of loops through negation~\cite{AB-JLP-94,GRS-JACM-91}, which we do not discuss further as they go beyond the scope of this paper.
%A full treatment of negation in logic programming would require introducing the Clark's completion of a program~\cite{C-LDB-78}, which is beyond the scope of this paper.
In fact, our goal is not to have a full-fledge handling of negation, but to allow the use of negation in policies while preventing loops through negation. 

Loops through negation are a well-studied issue in the logic programming literature. %~\cite{AB-JLP-94}. 
There are three standard solutions to the problems they raise: (a) forbidding the presence of loops through negation, as done by the weakly perfect model semantics~\cite{PP-FI-90}, which is defined only for \emph{weakly stratified programs} (that include \emph{locally stratified}~\cite{P-FDBLP-88} and \emph{stratified} programs~\cite{ABW-FDBLP-88}); (b) using a three-valued semantics~\cite{P-FI-90}, including the truth value {\em undefined} next to {\em true} and {\em false}, as done by the well-founded semantics~\cite{GRS-JACM-91} and Fitting's semantics~\cite{F-JLP-85}, where the semantics of $p$ in the program $p\leftarrow not(p)$ is {\em undefined}; (c) following a multi-model approach, as in stable models~\cite{GL-ICLP-88}.

Our solution follows an approach similar to (a), since solutions (b) and (c) are not suitable for trust management systems.
In fact, relying on a three-valued semantics (b) requires additional mechanisms to determine whether the truth value of the goals involved in a loop through negation is true, false, or \emph{undefined} (e.g., delaying in SLG resolution~\cite{CW-JACM-96}).
%in the context of access control raises the issue of how to interpret the \emph{undefined} answer, as the result of a computation must always lead to a positive or negative decision.
In trust management, however, loops through negation are inherently wrong, as they indicate conflicting policy statements issued by principals among which there is a mutual trust relationship, and thus should not be processed.
Similarly, solution (c) would imply that an access request should be either granted or revoked depending on which (truth) value we ``choose'' to assign to the goals in a computation, which is clearly not a safe approach.
Solution (a) can also not be applied straightforwardly in our context because the definition of weakly stratified program relies on a ``global ordering'' among \emph{all} the (ground) atoms in a global policy.
This would require principals to agree beforehand on the allowed dependencies among (ground) goals; however, in a trust management system principals often do not know each other until their first interaction.
Therefore, rather than forbidding the presence of loops through negation, we prevent their evaluation.
In this respect, the added value of GEM is its ability to detect loops at runtime.
We exploit this feature by introducing an additional runtime check to the algorithm, which causes the computation to flounder if a loop involving a negated goal is detected. 
The check is safe in that if the computation does not flounder, then it always returns a correct answer. 

In summary, GEM can be extended to allow the use of negation in policy statements as follows. Given a clause with a literal $not(B)$ selected for evaluation:
 %\footnote{The computation flounders if a non-ground negative literal is selected for evaluation.}
\begin{enumerate}
	\item if $B$ is not ground, an error is raised and the computation flounders;
	\item if the evaluation of $B$ succeeds with an answer,  then $not(B)$ fails and the clause is disposed;
	\item if $B$ is completely evaluated and has no answers, then $not(B)$ succeeds and a new node is added to the evaluation tree of the goal, removing $not(B)$ from the body;
	\item \emph{if a loop notification for atom $B$ is received, an error is raised and the computation flounders.}
\end{enumerate}
Conditions (1), (2) and (3) are standard when defining negation as failure: (1) is necessary to guarantee correctness \cite{A-HTCS-90}, while (2) and (3) define the semantics of negation. 
Notice that condition (3) captures also the case of \emph{infinite failure}, as done, for instance, by the well-founded semantics \cite{GRS-JACM-91}. 
For example, given a policy composed of clauses $q\leftarrow not(p)$ and $p\leftarrow p$, and a goal $q$, GEM first completely evaluates clause $p\leftarrow p$, detecting the loop and deducing that no answer of $p$ can be derived; consequently, it concludes that $q$ is true.
Condition (4) states that the algorithm flounders if it detects a loop through negation.
The ``floundering message'' is propagated to all the goals involved in the loop similarly to a response message, so that their evaluation is aborted.

Note that floundering due to condition (1) can be avoided by restricting to \emph{well-moded} programs \cite{AM-FAC-94}.
Differently from weakly stratified programs, which require an ordering among all the goals in a global policy, the definition of well-moded program requires each clause \emph{independently} to be well-moded.
Therefore, by requiring local policies to be well-moded, this type of floundering can be prevented.

It is straightforward to demonstrate that the proposed extension of GEM: % has the following properties:
\begin{itemize}
	\item always terminates (for arbitrary global policies and requests), because the only difference with the standard GEM algorithm (which terminates) is the presence of an additional termination condition, and
	\item for non-floundering computations, it is sound and complete with respect to the stable models and well-founded semantics. % \cite{GRS-JACM-91}.
\end{itemize}

We now show how the extended algorithm deals with negation by means of an example.
We consider a scenario inspired by the one introduced in Section~\ref{sec:intro}, where the pharmaceutical company \emph{c$1$} needs to determine the set of principals participating to project \emph{Alpha}.
Project \emph{Alpha} is a multidisciplinary project which requires the collaboration of experts from several fields: physicians, biologists, chemists, etc.
Company \emph{c$1$} already formed a team of qualified chemists to work on the project, and delegates to the partner company \emph{c$2$} the authority of determining the remaining project members.
To avoid interference with the work of its trusted chemists, however, \emph{c$1$} wants to prevent chemists of \emph{c$2$} to take part to the project.
In its definition of project members, \emph{c$2$} includes also the members of project \emph{Alpha} at \emph{c$1$}.
This scenario can be represented by the following policy statements:

\vspace{0.1cm} 
{\small
%\begin{tabular}{lll}
\indent 1. ${\sf memberOfAlpha}(c1\mbox{,}X) \leftarrow {\sf memberOfAlpha}(c2\mbox{,}X), not({\sf chemist}(c2\mbox{,}X)).$ \\
\indent 2. ${\sf memberOfAlpha}(c1\mbox{,}david).$ \\
\indent 3. ${\sf chemist}(c1\mbox{,}david).$ \\
\indent 4. ${\sf memberOfAlpha}(c2\mbox{,}X) \leftarrow {\sf memberOfAlpha}(c1\mbox{,}X).$ \\
\indent 5. ${\sf memberOfAlpha}(c2\mbox{,}alice).$ \\
\indent 6. ${\sf chemist}(c2\mbox{,}alice).$ \\
\indent 7. ${\sf memberOfAlpha}(c2\mbox{,}eric).$ \\
%\indent 8. ${\sf physician}\alpha(c2,eric).$ \\
%\end{tabular}
}
\vspace{-0.3cm} 

%\begin{figure}[!t]
%	\centering
%\subfigure[Dependency Graph with Loops and Negation]{
%\fbox{
%	{\footnotesize
%$\xymatrix@R=20pt@C=15pt{
%& \ar[d] && \\
%& memberOf\alpha(c1,X) \ar[ld] \ar[d]!U|{not} \ar[rd]!U|{not} \ar[rrd]!U|{not} && \\
%memberOf\alpha(c2,X) \ar@/^15pt/[ru] & chemist(c2,alice) & chemist(c2,eric) & chemist(c2,david) \\
%}$
%}}	
%	\label{fig:graph-negation-noloop}
%}
%\subfigure[Part of the Dependency Graph Inspected by GEM Before Floundering]{
%\fbox{
%	{\footnotesize
%$\xymatrix@R=15pt@C=5pt{
%& \ar[d] & \\
%& memberOf\alpha(c1,X) \ar[ld] \ar[d]!U|{not} \ar[rd]!U|{not} & \\ % \ar[rrd]!U|{not} & \\
%memberOf\alpha(c2,X) \ar@/^15pt/[ru] & chemist(c2,alice) & chemist(c2,eric) \ar[d] \\ %& chemist(c2,david) \ar[d] \\
%&& memberOf\alpha(c1,eric) \ar[d] \ar@/_15pt/[u]!U|{not} \\ % & memberOf\alpha(c1,david) \\ %\ar@(ur,l)[ul] \ar[d]
%&& memberOf\alpha(c2,eric) \\ % & \\
%}$
%}}
%\label{fig:graph-negation-loop}
%}
%	%\vspace{-0.4cm}
%	\caption{Dependency Graphs of the Example Global Policy}
%	\label{fig:graph-negation}
%\end{figure}

%The dependency graph of the global policy is shown in Figure~\ref{fig:graph-negation-noloop}; the edges with label $not$ denote a request for the evaluation of a negated goal.
To compute the answers of goal {\it memberOfAlpha(c$1$,X)}, GEM proceeds as follows.
First, clause~1 is evaluated by \emph{c$1$}, leading to a request for goal {\it memberOfAlpha(c$2$,X)} to \emph{c$2$}.
The evaluation of the first applicable clause in \emph{c$2$}'s policy (clause~4) forms a loop, identified by \emph{c$1$}.
The loop processing phase continues at \emph{c$2$}, which identifies the first two answers of {\it memberOfAlpha(c$2$,X)} (i.e., {\it memberOfAlpha(c$2$,alice)} and {\it memberOfAlpha(c$2$,eric)}, clauses~5 and~7 resp.) and sends them to \emph{c$1$}.
For each of these answers, \emph{c$1$} requests to \emph{c$2$} whether the project member is a chemist.
The evaluation of  {\it chemist(c$2$,alice)} succeeds at \emph{c$2$} (clause~6), while {\it chemist(c$2$,eric)} fails; therefore, their negated counterpart in clause~1 fails and succeeds respectively, leading to a new answer at \emph{c$1$}, namely {\it memberOfAlpha(c$1$, eric)}.
This answer, together with the answer derived by evaluating clause~2 (i.e., {\it memberOf-Alpha(c$1$,david)}), is sent by \emph{c$1$} to \emph{c$2$}, starting the second loop iteration.
In this iteration, \emph{c$2$} finds one new answer, {\it memberOfAlpha(c$2$,david)}, which is immediately returned to \emph{c$1$}.
Now, \emph{c$1$} evaluates clause~1 based on the new answer received from \emph{c$2$}.
Since David is not a chemist at \emph{c$2$}, \emph{c$1$} derives again answer {\it memberOfAlpha(c$1$,da-vid)}, which had already been computed in the previous iteration.
Thus, the evaluation of {\it memberOfAlpha(c$1$,X)} terminates with two answers: {\it memberOfAlpha(c$1$,eric)} and {\it memberOfAlpha(c$1$,david)}.

The example above shows that GEM can easily support policies including both loops and negation.
We now show how GEM operates in presence of loops through negation.
Consider the following policy statements complementing the global policy above:

\vspace{0.1cm} 
{\small
%\begin{tabular}{lll}
8. ${\sf chemist}(c2\mbox{,}X) \leftarrow {\sf memberOfAlpha}(c1\mbox{,}X), {\sf chemist}(c3\mbox{,}X).$ \\
\indent 9. ${\sf chemist}(c3\mbox{,}eric).$ \\
%\end{tabular}
}
\vspace{-0.3cm} 

Clause~8 states that all the members of project \emph{Alpha} at \emph{c$1$} that work as chemists at the other partner company \emph{c$3$} are also chemists at \emph{c$2$}.
Note that clause~8 is ``inconsistent'' with clause~1.
In fact, clause~1 defines as members of project \emph{Alpha} principals that \emph{are not} chemists at company \emph{c$2$}; at the same time, clause~8 states that chemists at \emph{c$2$} \emph{are} members of project \emph{Alpha} at \emph{c$1$}.
For this reason, when \emph{c$1$} ascertains that goals {\it memberOfAlpha(c$1$,eric)} and {\it chemist(c$2$,eric)} are in a loop, it raises an error and the computation flounders.
%In the example computation presented above, the requests for goals {\it chemist(c$2$,eric)} and {\it chemist(c$2$,david)} from \emph{c$1$} to \emph{c$2$} would now lead to the evaluation of clause (8), forming a loop (through negation) with goal {\it memberOf\emph{Alpha}(c$1$,X)}.
In fact, in the example computation above, the evaluation of {\it chemist(c$2$,eric)} by \emph{c$2$} would lead to a contradiction: if Eric were not a chemist at \emph{c$2$}, he would be a member of project \emph{Alpha}; however, if Eric were a member of project \emph{Alpha}, he would be a chemist at \emph{c$2$}.

%% file: rw.tex
\section{Related Work}
\label{sec:rw}

Research on goal evaluation has been carried out both in the field of logic programming and trust management. 
In this section we compare our work with existing frameworks focusing on the information disclosed during the evaluation process, based on the classification criteria defined in Section~\ref{sec:system}.
Additionally, we indicate whether the analyzed systems employ a centralized or distributed goal evaluation strategy and discuss the termination detection mechanism they adopt.
Within termination detection, we distinguish between termination of the whole computation initiated by a particular request and termination of the single goals involved in the computation (i.e., detecting when a goal is completely evaluated).
Table~\ref{tab:related} summarizes the results of this analysis. 
In the table, LP denotes the algorithms proposed in the logic programming domain, while TM denotes trust management systems.

\begin{table}[!t]
	\centering
{\small
	\begin{oldtabular}{|c|l|c|c|c|c|}\cline{1-6}
	\multicolumn{2}{|c|}{\multirow{2}{*} {\bf Frameworks}} & \multirow{2}{*}{\bf Evaluation} & {\bf Computation} & {\bf Goal} & \multirow{2}{*}{\bf Classification}\\
	\multicolumn{2}{|c|}{}&& {\bf Termination} & {\bf Termination} &\\\cline{1-6}
	\multirow{7}{*} {\bf LP} & SLG \cite{CW-JACM-96} & centralized & centralized & centralized & E1-I3\\
	\cline{2-6}
	& TP resolution~\cite{SYYZ-TPLP-01} & centralized & centralized & centralized & E1-I3\\
	\cline{2-6}
	& DRA~\cite{GG-ICLP-01} & centralized & centralized & centralized & E1-I3\\
	\cline{2-6}
	& OPTYap \cite{RSC-TPLP-05} & centralized & centralized & centralized & E1-I3\\
	\cline{2-6}
	& Hulin \citeyear{H-VLDB-89} & centralized & centralized & centralized & E1-I3\\
	\cline{2-6}
	& Damasio \citeyear{D-TAPD-00} & distributed & distributed & distributed & E1-I2\\
	\cline{2-6}
	& Hu \citeyear{H-THESIS-97} & distributed & distributed & distributed & E1-I2 \\
	\cline{1-6}
	\cline{1-6}
	\multirow{10}{*} {\bf TM} & RT \cite{LWM-CS-03} & centralized & centralized & centralized & E1-I3\\
	\cline{2-6}
	& Tulip \cite{CE-ICLP-07} & centralized & centralized & centralized & E1-I3\\
	\cline{2-6}
	& SecPAL \cite{BFG-JCS-10} & centralized & centralized & centralized & E1-I3\\
	\cline{2-6}
	& SD3 \cite{JS-PODS-01} & distributed & N/A & N/A & E1-I3\\
	\cline{2-6}
	& Becker et al. \citeyear{BMD-POLICY-09} & distributed & N/A & N/A & E1-I3\\
	\cline{2-6}
	& Cassandra \cite{B-TR-05} & distributed & no & no & E1-I0 \\
	\cline{2-6}
	& \multirow{2}{*}{PeerTrust \cite{ADNO-POLICY-06}} & distributed & distributed & no & E1-I1\\
	\cline{3-6}
	&  & distributed & distributed & distributed & E1-I2\\
	\cline{2-6}
	& MTN \cite{ZW-ESORICS-08} & distributed & distributed & no & E1-I1\\
	\cline{2-6}
	& GEM & distributed & distributed & distributed & E1-I1\\ 
	\cline{1-6}
	\end{oldtabular}
}
	\caption{Comparison Between Goal Evaluation Algorithms}
	\label{tab:related}
\end{table}

%\footnotetext{Dependencies can be deduced only in case of multiple requests to the same principal, as discussed in Section~\ref{sec:privacy}.}

SLG resolution \cite{CW-JACM-96}, TP-resolution~\cite{SYYZ-TPLP-01}, DRA~\cite{GG-ICLP-01}, OPTYap~\cite{RSC-TPLP-05}, and the work by Hulin \citeyear{H-VLDB-89} are centralized tabling systems in which the complete program (i.e., the global policy) is available during the evaluation. 
Therefore, these five systems are classified as E1-I3 according to the classification criteria defined in Section~\ref{sec:system}, that is, they do not preserve the confidentiality of neither extensional nor intensional policies.
%These systems refer to centralized data structures to detect termination. 
SLG identifies loops by observing goal dependencies in the ``call stack'' of the program; termination is detected when no more operations can be applied to the goals in the stack.
SLG resolution is employed in a number of Prolog systems such as, for instance, XSB~\cite{SW-TPLP-12}.
The evaluation strategy employed by GEM is similar to the XSB scheduling strategy called \emph{local evaluation}, which completely evaluates a SCC before returning the answers of the leader to a goal outside the SCC.
Similarly to SLD resolution~\cite{K-IP-74}, in TP-resolution and DRA a goal is evaluated by building a single derivation tree for the goal.
Loops are detected when a subgoal appears more than once in a branch of the tree, and the evaluation of a goal terminates when there are no more nodes in the derivation tree to be evaluated.
OPTYap and Hulin propose a parallel tabled execution strategy to improve the efficiency of goal evaluation.
OPTYap resorts to centralized data structures to identify loops and detect termination.
In \cite{H-VLDB-89}, each process communicates its termination to a global variable, whose access is limited to one process at a time by means of a deadlock mechanism.

Distributed goal evaluation frameworks are presented in \cite{H-THESIS-97} and \cite{D-TAPD-00}.
To detect termination, the work by Hu \citeyear{H-THESIS-97} assumes the presence of global data structures and requires goal dependencies to be propagated among the different principals.
In \cite{D-TAPD-00}, termination detection resorts to a static dependency graph known to all principals and determined at compile time.
Consequently, the confidentiality of (part of) the intensional policies is not preserved, and both algorithms are classified as E1-I2.

In trust management, distributed goal evaluation is a main issue since policies are distributed among principals.  
The trust management frameworks RT \cite{LWM-CS-03} and Tulip \cite{CE-ICLP-07} rely on a centralized goal evaluation strategy, where all the clauses necessary for the evaluation of a goal are collected in a single location.
Similarly, SecPAL \cite{BFG-JCS-10} assumes all the clauses in a global policy to be available to the principal responsible for the evaluation of a goal.
In SD3 \cite{JS-PODS-01}, when queried for a goal, a principal returns to the requester the clauses defining the goal, with the locally defined (body) atoms already evaluated.
Becker et al. \citeyear{BMD-POLICY-09} present an algorithm in which the body atoms of the clauses defining a goal are sent in turn to the principals defining them; each principal evaluates the atom(s) defined in her policy and sends its answers and the remaining atoms to the next principal, until the evaluation fails or all atoms are evaluated.
As a result, policy confidentiality is not preserved by any of these algorithms, which are thus classified as E1-I3. 
Furthermore, neither Jim and Suciu \citeyear{JS-PODS-01} nor Becker et al. \citeyear{BMD-POLICY-09} discuss how termination is detected. 
Cassandra \cite{B-TR-05} employs a distributed evaluation strategy in which no information about intensional policies is disclosed. 
However, it does not detect neither the complete evaluation of single goals, nor the termination of the whole computation. 

PeerTrust \cite{ADNO-POLICY-06} and MTN \cite{ZW-ESORICS-08} detect termination of the computation started by a particular request in a fully distributed way;
this is achieved by ``observing'' when no more messages are exchanged among principals and all goals are quiescent. 
In \cite{ADNO-POLICY-06}, the authors present two solutions: the first, based on the work in~\cite{D-TAPD-00}, is also able to detect the completion of single goals, but requires the dependency graph of the global policy to be known to all principals beforehand.
The second solution, which is also adopted in \cite{ZW-ESORICS-08}, detects termination of the computation without disclosing information about intensional policies.
However, since all request and response messages are tagged with the identifier of the initial request, some information about goal dependencies can be inferred (hence the E1-I1 classification); more precisely, a principal can learn whether a given goal depends on a goal defined in her policy.
In addition, neither PeerTrust nor MTN features a loop identification mechanism.
Consequently, they are not able to detect termination of individual goals, which is required to free the resources used during the computation and to allow the use of negation.
Furthermore, when using negation, the detection of loops through negation allows to preserve the soundness and completeness of the computation with respect to the standard semantics for logic programs.
We enable the identification of loops and the detection of goal termination at the cost of possibly revealing more information about goal dependencies.
In fact, in GEM all the principals involved in a loop are notified about the loop: on the one hand, this enables the principal(s) handling negated goals to terminate the computation with floundering.
On the other hand, this implies that GEM discloses information about the presence of mutual dependencies among goals to more principals than PeerTrust and MTN.
%In Section \ref{sec:advanced} we have shown how GEM can be extended to support negation as failure.
In the example in Section~\ref{sec:disclosed}, for instance, with PeerTrust and MTN the research institute \emph{ri} would not receive any loop notification from company \emph{c$2$}; therefore, \emph{ri} would not learn that there exists a mutual dependency between {\it memberOfAlpha(ri,X)} and {\it memberOfAlpha(c$2$,X)}.

Besides the protection of intensional policies, preserving the confidentiality of extensional policies is also an important requirement of trust management systems, as the answers of a goal might contain sensitive information (e.g., the list of patients of a mental hospital).
%Therefore, the disclosure of extensional policies should also be protected.
Even though none of the existing goal evaluation algorithms satisfies this requirement (see Table~\ref{tab:related}), GEM can be easily adapted to protect the confidentiality of extensional policies.
In particular, by enabling the distributed evaluation of policies, GEM allows principals to discriminate between goals that may be accessed by other principals and goals that may only be used for internal computations, because of their sensitivity.
This distinction is not possible when using an algorithm that relies on a centralized evaluation strategy.
A finer-grained protection of extensional policies can be achieved by integrating GEM with trust negotiation algorithms~\cite{WSJ-DISCEX-00,W-ITRUST-03}.
Trust negotiation algorithms protect the disclosure of extensional policies (i.e., possibly sensitive credentials) by means of \emph{disclosure policies} that specify which credentials a requester must provide to get access to the requested credentials.
Some trust negotiation algorithms deal also with the protection of disclosure policies (e.g.,~\cite{SWY-NDSS-01}); however, they assume that all the credentials of the principals in a trust management system have been already derived when a transaction takes place~\cite{WL-WPES-02}.
GEM, on the other hand, provides a way of deriving those credentials.
Thus, GEM and trust negotiation algorithms can be combined in such a way that a GEM request is evaluated only if the requester satisfies the disclosure policy of that goal, i.e., if she is trustworthy enough to see the answers to the request.
The resulting integrated algorithm enables distributed goal evaluation while preserving the confidentiality of both intensional and extensional policies.
A similar approach is presented in \cite{KM-AAS-08,LMB-ASIACCS-09}.
However, in \cite{KM-AAS-08} the authors do not discuss how to deal with recursive policy statements, while the algorithm presented in \cite{LMB-ASIACCS-09} raises an error in the case that cyclic dependencies are detected, and for this reason is not complete.
MTN~\cite{ZW-ESORICS-08} also applies trust negotiation strategies to distributed goal evaluation, but as discussed in Section~\ref{sec:rw} this algorithm is not able to detect termination of individual goals within a computation.
In~\cite{MBWL-ASIACCS-11}, the authors present a framework to analyze and compare distributed goal evaluation algorithms based on the information about extensional policies that they disclose during a computation.

To conclude, we point out that contrarily to other works on goal evaluation (e.g., \cite{LMW-TISSEC-10}), the distributed evaluation strategy of GEM does not allow to build the complete ``proof'' of a goal.
Building such a proof is in fact similar to constructing the derivation tree of a goal. % having as root the initial goal, as branches the clauses used to evaluate the goal, and as leaf nodes the answers of its subgoals.
Even though cryptographic techniques can be employed to prevent the disclosure of the facts used in the derivation process \cite{LMB-ASIACCS-09}, the construction of such a proof cannot be obtained without disclosing the intensional policies of the principals involved in the evaluation, which is what GEM aims to avoid.
We argue that the approach followed by GEM is consistent with the concept of trust management.
In trust management, in fact, if the policy of a principal $a$ refers to the policy statements of a principal $b$, then $a$ trusts $b$ for the definition and evaluation of those statements.
%In summary, on the one hand a distributed evaluation is the only solution to preserve the confidentiality of intensional policies and, if combined with trust negotiation algorithms or cryptographic techniques, also the confidentiality of extensional policies.
%On the other hand, 
When the proof of a goal is required, the confidentiality requirement should be put aside in favor of a goal evaluation strategy that allows the construction of such a proof (e.g., RT~\cite{LWM-CS-03}).

%% file: conclusions.tex
\section{Conclusions}
\label{sec:conclusions}

In this paper we have presented GEM, a distributed goal evaluation algorithm for trust management systems. 
Differently from many of the existing algorithms, GEM detects the termination of a computation in a completely distributed way without disclosing intensional policies, thereby preserving their confidentiality.
In addition, GEM is able to detect when the single goals within a computation are fully evaluated, by enabling the identification of strongly connected components.
Even though this may lead to the disclosure of some additional information about goal dependencies, it also enables the use of negation (as failure) in policies.
In Section~\ref{sec:disclosed} we show that the information disclosed by GEM is not sufficient to infer the intensional policy of a principal; thus, we believe that the benefits of our solution overcome the drawbacks. 
GEM always terminates and is sound and complete with respect to the standard semantics for logic programs.
As future work, we plan to extend GEM to support constraint rules \cite{LM-PADL-03} and subsumptive tabling. 

Although efficiency is not a primary objective of this paper, GEM can contribute to keep network traffic low.
In fact, in most distributed goal evaluation systems (e.g.,~\cite{ADNO-POLICY-06}) answers are sent as soon as they are computed. 
On the contrary, GEM delays the communication of the answers of a goal until all possible answers have been computed, i.e., until all the branches of the partial derivation tree of the goal have been inspected.
This strategy may delay the identification of the answers of ground goals.
However, it simplifies the termination detection mechanism and we believe reduces the number of messages exchanged by principals during a computation.
The experiments presented in Section~\ref{sec:evaluation} suggest that since the computation time is dominated by network communication, a reduction in the number of messages exchanged between principals leads to a consistently lower computation time.
In addition, since the answers of a goal can be reused for future computations, the proposed solution may reduce the computation time of later evaluations.
%However, it might delay the computation of an answer to a request.

%In this paper we have compared GEM with other trust management frameworks based on their characteristics (Table \ref{tab:related}).
Based on the results of the experiments presented in Section~\ref{sec:evaluation}, we can conclude that GEM performs well both in terms of computation time and memory occupation even for very large global policies.
To confirm this conviction, we have employed GEM in some prototype of real-world distributed systems in the maritime safety and security~\cite{TZE-POLICY-11} and employability~\cite{BEHHTTZ-JTAER-10} domains.
In addition, we are currently designing an advanced version of GEM that implements an ``early loop detection'' strategy to avoid the reevaluation of side requests. 
Finally, since the policy language proposed in this paper can be used to represent the semantics of several existing trust management languages (e.g., RT~\cite{LWM-CS-03} and PeerTrust~\cite{ADNO-POLICY-06}), we point out that GEM can be used to evaluate goals over policies expressed in any of those languages.

%% file: appendix.tex
\begin{appendix}

\input{proofs}

\input{example}

\input{evaluationAppendix}

\end{appendix}

%% file: proofs.tex
\section{Proofs}
\label{sec:proofs-appendix}

As mentioned in Section~3.3, we assume that given a request or response message $M$ sent by a principal $a$ to a principal $b$, one and only one instance of message $M$ is received by $b$.
In other words, we assume no message duplication, and that messages are always received.

We introduce one last definition.

\begin{definition}
Let $S$ be the set of tables resulting from running GEM on a goal $G$ w.r.t. $P=P_1\cup\ldots\cup P_n$.
Let $G_1$ be a goal whose table is in $S$. Let $\theta$ be a solution of $G_1$ using clause $H\leftarrow B_1,\ldots,B_n$. 
Then, by construction $\exists \theta_0,\ldots,\theta_n$ s.t. $\theta_0=mgu(G_1,H)$ and $\theta_j$ is a solution of $B_j\theta_0\cdots\theta_{j-1}$ (with $j\in\{1,\ldots,n\}$). The \emph{ranking} of $\theta$ is defined inductively as follows:
\begin{itemize}
	\item $rank(\theta)=1$ if $n=0$ (i.e., the clause is a fact),
	\item $\ms{rank}(\theta)=1+max(rank(\theta_1),\ldots,rank(\theta_n))$ otherwise, where $rank(\theta_j)$ is the ranking of solution $\theta_j$.\hfill$\Box$
\end{itemize}
\end{definition}

We can now prove the soundness result of GEM.

\vspace{0.3cm}\noindent
{\it Proof of Theorem 1}. We proceed by contradiction and assume that there exists at least a ``wrong'' solution $\theta_{i,j}$ in $Sol_i$, i.e.,\ a solution s.t.\ there is no corresponding SLD derivation of $P \cup \{G_i\}$ with c.a.s. $\sigma$ where $G_i\theta_{i,j}$ is a renaming of $G_i\sigma$ (hypothesis).

Let us choose $\theta_{i,j}$ to be a ``wrong'' solution with minimal ranking ($\ast$). 
Let $G_i=$ $\leftarrow A_i$.
Since $\theta_{i,j}$ is a solution of $G_i$, there exists an evaluation tree of $G_i$ in $S$ created by \createtable\ (lines~2-7) with root $\nTuple{\ID}{A_i\leftarrow A_i}{new}$, a subnode with clause $c = H\leftarrow B_1,\ldots,B_n$ and substitutions $\theta_0,\ldots,\theta_n$ s.t. $\theta_0=mgu(A_i,H)$, and for each $l\in\{1,\ldots,n\}$ there exists:
\begin{itemize}
	\item A node in the evaluation tree of $G_i$ with selected atom $B_l\theta_0\cdots\theta_{l-1}$ (\newactive, lines 8, 18-19).
	\item An evaluation tree of $\leftarrow B_l\theta_0\cdots\theta_{l-1}$ created by \createtable\ (lines~2-7) at the location of $B_l\theta_0\cdots\theta_{l-1}$.
	\item A solution $\theta_l$ of $\leftarrow B_l\theta_0\cdots\theta_{l-1}$; the answer $B_l\theta_0\cdots\theta_l$ is sent to the requester of $\leftarrow B_l\theta_0\cdots\theta_{l-1}$ by \answerreturn\ (lines 12-14 or 20-23) if $\leftarrow B_l\theta_0\cdots\theta_{l-1}$ is involved in a loop, or by \completion\ (lines 3-5) otherwise.
	\item A node with clause $(H \leftarrow B_{l+1},\ldots,B_n)\theta_0\cdots\theta_l$ added to the evaluation tree of $G_i$ by \positivereturn\ (lines 20-23).
\end{itemize}
Then, $\theta_{i,j}=\theta_0\cdots\theta_n$.
If the body of $c$ is empty, then there is a trivial 1-step SLD derivation of $P \cup \{G_i\}$ with c.a.s. $\sigma_i$ (namely the \emph{mgu} of $G_i$ and $c$), therefore contradicting the hypothesis.
So, let us now assume that $n>0$; by construction, for each $l\in\{1,\ldots,n\}$, $rank(\theta_l)$ $<rank(\theta_{i,j})$. So, by the minimality argument ($\ast$), for each $l\in\{1,\ldots,n\}$ there exists an SLD 
derivation of $P \cup \{\leftarrow B_l\theta_0\cdots\theta_{l-1}\}$ 
with c.a.s. $\sigma_l$ s.t. $B_l\sigma_0\cdots\sigma_{l-1}\sigma_l=B_l\theta_0\cdots\theta_{l-1}\theta_l$. 
But then, by standard logic programming results (given the presence of clause $c$), there exists a successful SLD derivation of $P \cup \{G_i\}$ with c.a.s. $\sigma$ s.t. $G_i\sigma=G_i\theta_{i,j}$, contradicting the hypothesis.\hfill$\Box$\\

Since GEM employs a ``wait'' mechanism to determine when the answers of a goal should be sent to the requester, both the completeness and termination properties of the algorithm depend on the correctness of this mechanism.
Therefore, before demonstrating that GEM is complete and always terminates, we prove that the ``wait'' mechanism is correctly implemented, i.e., that the answers of a goal are eventually sent to the requester.
This is particularly challenging in the presence of loops.

In the implementation of GEM proposed in Section~3.3, the ``wait'' mechanism for goals involved in a loop consists of loop counters: at each iteration of a loop $\ID$, the answers of a goal $G$ are only sent when the counter of loop $\ID$ in set \ActiveLoopsM\ is $0$ (procedure \positivereturn, line~24).
Since the counter is set to the number $k$ of subgoals of $G$ which are involved in loop $\ID$ (\answerreturn, lines 7 and 19), at each iteration of loop $\ID$ the principal evaluating $G$ should thus receive $k$ response messages.
In order to prove this, we first show that GEM correctly keeps track of the loops in which the subgoals of $G$ are involved.

\begin{proposition}
\label{pro:loop-notification}
Let $G_1,\ldots,G_m$ be the goals involved in a loop $\ID_1$.
Let $G_i,G_j\in\{G_1,\ldots,G_m\}$ be two goals s.t. $G_j$ is a subgoal of $G_i$.
Then, the node in the evaluation tree of $G_i$ with selected atom $G_j$ has status \loopSS, where $\ID_1\in\IDs$.
\end{proposition}

\vspace{0.3cm}\noindent{\it Proof of Proposition 1}.
Let $G_1$ be the coordinator of loop $\ID_1$.
Let $G_1,\ldots,G_k$ be a subset of $G_1,\ldots,G_m$ s.t. for each $i\in\{2,\ldots,k\}$ goal $G_i$ is a subgoal of $G_{i-1}$, and $G_1$ is a subgoal of $G_k$.
The node in the evaluation tree of $G_1,\ldots,G_k$ with selected atom $G_2,\ldots,G_k,G_1$ respectively has status \loopSS, where $\ID_1\in\IDs$, because of the following observations:
\begin{itemize}
	\item The identifiers $\ID_2,\ldots,\ID_k$ of the requests for goals $G_1,\ldots,G_k$ and the identifier $\ID_{k+1}$ of the request for goal $G_1$ are constructed by procedures \createtable\ (lines~5-6) and \positivereturn\ (lines 21-22) in such a way that $\ID_j\orderingIn \ID_1$, for each $j\in\{2,\ldots,k+1\}$, and thus the lower request $\ID_{k+1}$ for $G_1$ can be identified.
	\item Upon receiving the lower request $\ID_{k+1}$, the principal evaluating $G_1$ returns a response $\langle\ID_{k+1},$\AnssM$_{k+1},S_{k+1}$,$\{\ID_1\}\rangle$ to the principal evaluating $G_k$ (procedure \newsubgoal, lines 5-7).
	\item The status of the node in the evaluation tree of $G_k$ with selected atom $G_1$ is set to $loop(\{\ID_1\})$ (\positivereturn, lines 12-13).
	\item A counter for loop $\ID_1$ is added to set \ActiveLoopsM\ in the table of goal $G_k$ (\positivereturn, line 14).
	\item For each $i\in\{2,\ldots,k\}$, the principal evaluating goal $G_i$ sends to the principal evaluating goal $G_{i-1}$ a response of the form $\langle\ID_i,$\AnssM$_i,S_i$,\IDs$\rangle$, where $\IDs$ is the set of all loops in \ActiveLoopsM\ whose identifier is higher than $\ID_i$ (\answerreturn, lines 18, 21, and 23), and thus $\ID_1\in\IDs$.
	\item The status of the node in the evaluation tree of goal $G_{i-1}$ with selected atom $G_i$ is set to \loopSSM, where $\ID_1\in\IDs$ (\positivereturn, lines 12-13).
	\hfill$\Box$
\end{itemize}

\begin{corol}
\label{cor:counters}
Let $G$ be a goal involved in a loop $\ID$. 
Let $k$ be the number of nodes in the evaluation tree of $G$ with status \loopSSM\ s.t. $\ID\in\IDs$.
When a response is sent to the requester of the higher request for $G$ (or lower request, if $G$ is the loop coordinator), the counter of loop $\ID$ in set \ActiveLoopsM\ in the table of $G$ is set to $k$.
\end{corol}

At each loop iteration, the counters of the loops in which a goal $G$ is involved are set to the number of subgoals of $G$ involved in those loops by procedure \answerreturn, lines 7 and 19.
Hence, we now need to show that at each iteration of a loop $\ID$ the number of response messages with status \ms{loop(\ID)} received by the principal evaluating $G$ is equal to the number of subgoals of $G$ involved in loop $\ID$, i.e., that counters correctly keep track of the number of response messages received by the principal evaluating $G$ at each iteration of loop $\ID$.

Informally, the correctness of counters stems from the fact that at each loop iteration step for a goal $G$ there is \emph{only one} choice of loop identifier to include in the response to the requester of a higher request for $G$.
%This allows us to exactly determine the number of responses containing a given loop identifier that will be received at each loop iteration.
This is because of the following considerations:
\begin{enumerate}

\item Let $G$ be a goal involved in one or more loops. 
	In the loop processing phase, the loop identifier included by the principal evaluating $G$ in the response sent to the requester of a higher request for $G$ is taken from the status of the root node of the evaluation tree of $G$ (procedure \answerreturn, lines 20-21).

\item After a response for $G$ is sent by \answerreturn\ (lines 20-23), the status of the root node of the evaluation tree of $G$ is set to \actS\ (line 24).

\item If $G$ is a non-coordinator goal, then there can be \emph{at most one} loop identifier per time in the status of the root node of its evaluation tree. Therefore, when sending a response for $G$, procedure \answerreturn\ has only one choice of loop identifier to include in the response status.
The reason why a non-coordinator goal can have at most one loop identifier in the status of the root of its evaluation tree is the following. 
The only point where the status of the root node of a non-coordinator goal $G$ is modified to take into account the loop being processed is on line 18 of procedure \positivereturn, and the check on line 17 updates the status only in case it is currently set to \actS. 
We point out that when the response for a subgoal of $G$ is processed by \positivereturn\ the status of the root of the evaluation tree of $G$ is always \actS, due to point (2) above and the fact that GEM only processes one goal at a time (which is due to condition on line 24 of \positivereturn), and thus no response will be received by the principal evaluating $G$ in the context of a loop unless a response for $G$ was previously sent. 

\item if $G$ is the coordinator of a loop $id_l$, then there can be \emph{at most two} loop identifiers per time in the status of the root node of its evaluation tree: one for loop $id_l$, and \emph{at most one} for a higher loop $id_h$. 
Remember that as loop identifiers we use the identifier of the higher request for the coordinator; hence, in this case $id_l$ is the identifier of the higher request for $G$. 
Given the condition on line 20 of \answerreturn, only $id_h$ can be included in the status of a response for $G$ sent to the requester of a higher request. 
In fact, $id_h$ (denoted $id_4$ in the procedure) is the only identifier in the status of the root node of the evaluation tree of $G$ that is higher than $id_l$ (denoted $id_1$ in the procedure), i.e., higher than the identifier of the higher request for $G$. 
Therefore, when sending a response for $G$, \answerreturn\ has only one choice of loop identifier to include in the response status, namely $id_h$.\\
Technically, a coordinator can have at most two loop identifiers in the status of the root node of its evaluation tree because of the following. 
Similarly to non-coordinators, due to condition on line 17 of \positivereturn\ only \emph{one} loop identifier can be added to the root's status on line 18 of \positivereturn.
This occurs when $G$ receives a response from one of its subgoals in the context of a higher loop $id_h$. 
A second loop identifier (the identifier of loop $id_l$) can be added to the root status on lines 8-9 of \answerreturn\ if the response received by the principal evaluating $G$ in the context of loop $id_h$ leads to new answers of $G$, which need to be sent to the goals involved in loop $id_l$. 
No more than two loops at a time will be processed by the principal evaluating $G$ (i.e., $id_l$ and at most one higher loop $id_h$) because of the following reasons. 
Upon receiving a response in the context of a higher loop $id_h$:
\begin{itemize}
\item a response for $G$ in the context of loop $id_h$ will not be sent to a higher goal until a fixpoint for the loop $id_l$ of which $G$ is the coordinator is reached, during which time the status of the root node of the evaluation tree of $G$ is \ms{loop(\{id_h,id_l\})}, and 
\item due to the condition on line 24 of \positivereturn\ no responses for higher goals can be received by the principal evaluating $G$ until a response for $G$ in the context of loop $id_h$ is sent upwards. 
In fact, the counter of loop $id_h$ in the table of higher goals cannot be $0$, because no response for $G$ in the context of loop $id_h$ was sent upwards yet. 
When a response for $G$ is sent upwards, the status of the root node of its evaluation tree becomes \actS\ again (see point (2)).
\end{itemize}

\end{enumerate}

Formally, the correctness of counters is demonstrated by the following Proposition.

\begin{proposition}
\label{pro:one-loops}
Let $G$ be a goal and $G_1,\ldots,G_k$ be the subgoals of $G$ s.t. $G,G_1,\ldots,G_k$ are involved in a loop $\ID_l$.
At each iteration of loop $\ID_l$, the principal evaluating $G$ receives $k$ response messages, one for each subgoal $G_i\in\{G_1,\ldots,G_k\}$.
\end{proposition}

\vspace{0.3cm}\noindent\emph{Proof of Proposition 2}.
Let $G,G_1,\ldots,G_k,\ldots,G_m$ be all the goals involved in loop $\ID_l$, where $m \geq k$.
Let $\ID,\ID_1,\ldots,\ID_m$ be the identifiers of the requests for goals $G,G_1,\ldots,G_m$ respectively.
The proof is by induction on the number $\ell$ of goals $G_j\in\{G_1,\ldots,G_m\}$ s.t. $\ID_j\orderingIn\ID$, that is, the number of goals whose request identifier is lower than the identifier $\ID$ of the request for $G$.
\begin{description}
	\item[Base case:] $\ell=1$. Then, also $k=1$.
	Let $G_j\in\{G_1,\ldots,G_m\}$ be the only goal s.t. $\ID_j\orderingIn\ID$.
	It is straightforward to see that goal $G_j$ is (a variant of) the coordinator of loop $\ID_l$, and $\ID_j$ denotes the lower request for $G_j$.
%	This is because the identifier $\ID_j$ of the lower request for the coordinator is lower than the identifier $\ID$ of the request for $G$, and $G_j$ is the only goal with request identifier lower than $\ID$.
	When there are no more nodes with status \newSM\ in the evaluation tree of $G_j$, i.e., when all the branches of the evaluation tree of $G_j$ have been evaluated, procedure \answerreturn\ is invoked by \newactive\ (lines 2-3).
	By procedure \answerreturn\ (lines 12-14), at each loop iteration one and only one response to the request for the coordinator $G_j$ is sent by GEM to the principal evaluating $G$.
	Thus, at each iteration of loop $\ID_l$ the principal evaluating $G$ receives $k=1$ response messages. Q.e.d.
	\vspace*{0.1cm}
	
	\item[Inductive case:] %assume that at each iteration of loop $\ID_l$ the principal evaluating $G$ receives $k$ response messages for $G$ having up to $\ell-1$ goals $G_j\in\{G_1,\ldots,G_m\}$ s.t. $\ID_j\orderingIn\ID$ ($\ast$).
	Now, assume that $G$ has $\ell$ such goals $G_j\in\{G_1,\ldots,G_m\}$ s.t. $\ID_j\orderingIn\ID$, where $\ell>1$.
	In this case, each subgoal $G_i\in\{G_1,\ldots,G_k\}$ of $G$ is either the coordinator of loop $\ID_l$ or a goal with at most $\ell-k$ subgoals $G_p\in\{G_1,\ldots,G_m\}$ s.t. $\ID_p\orderingIn\ID_i$.
	If $G_i$ is the coordinator of loop $\ID_l$, by the same reasoning done in the base case, one and only one response to the request for $G_i$ is sent by GEM to the principal evaluating $G$ at each loop iteration.\\
	On the other hand, if $G_i$ is not the loop coordinator, there exist at most $\ell-k$ goals $G_p\in\{G_1,\ldots,G_m\}$ s.t. $\ID_p\orderingIn\ID_i$.
%	Let $t$ be the number of subgoals of $G_i$ involved in loop $\ID_l$.
%	By Proposition~\ref{pro:loop-notification}, there are $t$ nodes in the evaluation tree of $G_i$ with status \loopSSM\ s.t. $\ID_l\in\IDs$.
%	Then:
%	\begin{itemize}
%		\item The counter of loop $\ID_l$ in the table of goal $G_i$ is set to the number $t$ of nodes in the evaluation tree of $G_i$ with status \loopSSM\ s.t. $\ID_l\in\IDs$ (i.e., the number of subgoals of $G_i$ involved in loop $\ID_l$, \answerreturn, line 19).
%		\item Since $t\leq\ell-1$, 
	Let $t$ be the number of subgoals of $G_i$ involved in loop $\ID_l$.
	Since $\ell-k\leq\ell-1$, by the inductive hypothesis ($\ast$) the principal evaluating $G_i$ receives $t$ response messages at each iteration of loop $\ID_l$.
%	\end{itemize}
	By Corollary~1, at each loop iteration the counter of loop $\ID_l$ in the table of goal $G_i$ is set to $t$, and is decreased by $1$ every time a response to the requests for its subgoals involved in the loop is received (procedure \positivereturn, lines 15-16).
	Therefore, after $t$ response messages, the counter of loop $\ID_l$ in the table of $G_i$ is $0$, and procedure \positivereturn\ (lines 24-25) resumes the evaluation of goal $G_i$.
	When there are no more nodes with status \newSM\ in the evaluation tree of $G_i$, procedure \newactive\ (lines 2-3) invokes \answerreturn.
	By procedure \answerreturn\ (lines 20-21), one and only one response to the request for $G_i$ is sent by GEM to the principal evaluating $G$ at each iteration of loop $\ID_l$.
	Therefore, at each iteration of loop $\ID_l$ the principal evaluating goal $G$ receives $k$ response messages.\hfill$\Box$
\end{description}

Finally, we show that procedure \completion\ is eventually invoked for any goal in a computation.

\begin{proposition}
\label{pro:termination}
Let $G_1$ be a goal.
Procedure \completion\ is eventually called for $G_1$.
\end{proposition}

\vspace{0.3cm}\noindent\emph{Proof of Proposition 3}.
The proof is divided into two parts.
First, we show that \completion\ is eventually called for a goal $G_1$ that is not involved in a loop.
Then, we show that it is always invoked also if $G_1$ is involved in one or more loops.\\
The first part of the proof is straightforward, and is given by the fact that the number of answers of goal $G_1$ is finite.
This is because of the following observations:
\begin{enumerate}
	\item The global policy $P$ is finite, and the terms in $P$ that are not variables are constants defined in $P$; thus, the Herbrand model of $P$ is finite.
	\item Let $P_{G_1}\in P$ be the policy where goal $G_1$ is defined.
	The answers of $G_1$ are computed by GEM through the clauses in $P_{G_1}$ applicable to $G_1$ (procedure \createtable, lines~3-7).
	Each clause can be either a fact or have the form $H\leftarrow B_1,\ldots,B_m$, such that $B_1,\ldots,B_m$ are defined in a policy in $P$.
	By (1), both the number of facts in $P_{G_1}$ and the number of answers of subgoals $B_1,\ldots,B_n$ are finite.
\end{enumerate}
Thus, the number of answers of goal $G_1$ is finite.
When all the answers of $G_1$ have been computed and all the nodes in the partial tree of $G_1$ have been evaluated, procedure \newactive\ (lines~2-3) invokes \answerreturn, which in turn (lines~2-3) invokes \completion.\\
Consider now the case in which $G_1$ is part of an SCC consisting of loops $\ID_1,\ldots,\ID_k$, s.t. $\ID_k\orderingIn\ldots\orderingIn\ID_1$.
Let $G_1,\ldots,G_m$ be all the goals involved in loops $\ID_1,\ldots,\ID_k$ (where $m\geq k$), and goal $G_{c_i}\in\{G_1,\ldots,G_m\}$ be the coordinator of loop $\ID_i\in\{\ID_1,\ldots,\ID_k\}$.
%Let $G_1,\ldots,G_m$ (with $m\geq k$) be the goals in the SCC which includes loop $\ID$, and $G_l\in\{G_1,\ldots,G_m\}$ be the leader of the SCC.
Because the number of answers of each goal $G_1,\ldots,G_m$ is finite, we have that:
\begin{itemize}
	\item At each iteration of loop $\ID_i$, if new answers of the loop coordinator $G_{c_i}$ are derived, they are sent to the requesters of the lower requests for $G_{c_i}$, starting a new iteration of loop $\ID_i$ (procedure \answerreturn, lines 6-14).
	On the contrary, if no answer of $G_{c_i}$ is computed, the answers of $G_{c_i}$ are sent to the requester of the higher request for $G_{c_i}$ (\answerreturn, lines~20-23). 
		The loops higher than $\ID_i$ in the SCC are then processed.
	\item At each iteration of loop $\ID_1$, if new answers of the leader $G_{c_1}$ are derived, they are sent to the requesters of the lower requests for $G_{c_1}$, starting a new iteration of loop $\ID_1$ (procedure \answerreturn, lines 6-14).
	Notice that this might cause a fixpoint for the loops lower than $\ID_1$ in the SCC to be recomputed.
	On the contrary, if no answer of $G_{c_1}$ is computed, the answers of $G_{c_1}$ are sent to the requester of the higher request for $G_{c_1}$, and a response with status \dispSM\ is sent to the requesters of the lower requests for $G_{c_1}$ (\answerreturn, lines~15-16 and \completion, lines 3-5).
	\item For each goal $G_j\in\{G_1,\ldots,G_m\}$, all the nodes in the evaluation tree of $G_j$ are disposed (\positivereturn, lines~5-8); then, procedure \newactive\ is invoked, which immediately invokes \answerreturn\ (lines~2-3).
	\item The principal evaluating goal $G_j$ sends a response with status \dispSM\ to the requester of the higher request for $G_j$.
	 If $G_j$ is a loop coordinator, the principal evaluating $G_j$ also sends a response with status \dispSM\ to the requesters of the lower requests for $G_j$ (\answerreturn, lines~2-3 and \completion, lines 3-5).
\end{itemize} 
Therefore, procedure \completion\ is always invoked for goal $G_1$.\hfill$\Box$\\

Proposition~\ref{pro:termination} implies that the table of a goal involved in a computation is always disposed.
In fact, the disposal of the table of a goal is carried out by procedure \completion\ (lines 2, 6, and 7).
Consider, for instance, the following variation of the global policy introduced in Section~3.1, where the research insitute \ms{ri} refers to goal \ms{memberOfAlpha}($c1$,$X$) instead of \ms{memberOfAlpha}($c2$,$X$):

\vspace{0.3cm} 
{\small
%\begin{tabular}{lll}
${\sf memberOfAlpha}(c1\mbox{,}X) \leftarrow {\sf memberOfAlpha}(c2\mbox{,}X).$ \\
\indent ${\sf memberOfAlpha}(c2\mbox{,}X) \leftarrow {\sf memberOfAlpha}(ri\mbox{,}X).$ \\
\indent ${\sf memberOfAlpha}(ri\mbox{,}X) \leftarrow {\sf memberOfAlpha}(c1\mbox{,}X).$ \\
%\end{tabular}
}
%\vspace{0.2cm} 

First of all, let us recall that the termination of the evaluation of the goals involved in a loop is commanded by the leader of the SCC (goal \ms{memberOfAlpha}($c1$,$X$) in the example policy). 
When no new answer of the leader is computed by \ms{c1} during a loop iteration, procedure \completion\ is invoked (lines 15-16 of \answerreturn), which disposes the table of the goal and sends a response with status \dispS\ both to the requesters of the higher and lower requests for \ms{memberOfAlpha}($c1$,$X$) (lines 3-5). 
When \ms{ri} receives the response, it disposes all the nodes in the evaluation tree of \ms{memberOfAlpha}($ri$,$X$) involved in a loop (lines 5-8 of \positivereturn), which in this case corresponds to disposing all the non-root nodes. 
At this point, the status of the root node of the evaluation tree of \ms{memberOfAlpha}($ri$,$X$) is \actS\ (see point 2 of the discussion preceding Proposition~\ref{pro:one-loops}).
Therefore, the condition on line 24 of \positivereturn\ is satisfied, and procedure \newactive\ is invoked for \ms{memberOfAlpha}($c1$,$X$). 
Since all the non-root nodes in the evaluation tree of \ms{memberOfAlpha}($c1$,$X$) have status \dispS, \answerreturn\ is invoked (lines 2-3 of \newactive), which in turn (lines 2-3) invokes procedure \completion. 
\completion\ disposes the table of goal \ms{memberOfAlpha}($ri$,$X$) and sends a response with status \dispS\ to \ms{c2}. 
Similarly to \ms{memberOfAlpha}($ri$,$X$), \positivereturn\ disposes all the nodes in the evaluation tree of goal \ms{memberOfAlpha}($c2$,$X$), and a response with status \dispS\ is sent by \ms{c2} to \ms{c1} by procedure \completion. 
Since the root of the evaluation tree of \ms{memberOfAlpha}($c1$,$X$) had already been disposed, in this case the response message is ignored by \ms{c1} (line 4 of \positivereturn).

Next, we prove the completeness and termination results.

\vspace{0.3cm}\noindent
{\it Proof of Theorem 2}. We proceed by contradiction, and assume that $S$ is missing a solution of $G_1$. That is, there exists a successful SLD derivation of $P \cup \{G_1\}$ with c.a.s. $\theta$ and there is no solution $\sigma$ of $G_1$ generated by the algorithm s.t. $G_1\theta=G_1\sigma$ (hypothesis).\\
%Hypothesis: 
This implies that there exist a (maximal) set of goals $G_1,\ldots,G_k$ in $S$ s.t. for each $i\in\{1,\ldots,k\}$ there is a non-empty maximal set of substitutions $\{\theta_{i,1},\ldots,\theta_{i,m_i}\}$ s.t.:
\begin{enumerate}[(a)]
	\item $G_i$ is a goal in $S$.
	\item $\theta_{i,1},\ldots,\theta_{i,m_i}$ are correct solutions of $G_i$ according to SLD resolution: for each $\theta_{i,j}$ there exists a successful SLD derivation of $P \cup \{G_i\}$ with c.a.s. $\theta_{i,j}$ (up to renaming).
	\item The algorithm does not generate the answers $G_i\theta_{i,1},\ldots,G_i\theta_{i,m_i}$ (up to renaming).
\end{enumerate}
The set $G_1,\ldots,G_k$ is not empty as it contains at least $G_1$ (the finiteness of the construction is demonstrated in the proof of Proposition~3).

For each $i,j$, let $der_{i,j}$ be the SLD derivation of $P \cup \{G_i\}$ with c.a.s. $\theta_{i,j}$ of minimal length.
Let us choose integers $p,q$ in such a way that $der_{p,q}$ has minimal length among the derivations in the set $\{der_{i,j}\}$. The fact that $der_{p,q}$ has minimal length implies that for any goal $G'$ in $S$, the following holds: if there exists an SLD derivation of $P \cup \{G'\}$ of length smaller than $\emph{len}(der_{p,q})$ with c.a.s. $\theta'$, then the algorithm generates a solution $\vartheta'$ for which $G'\theta'$ is a renaming of $G'\vartheta' $ ($\ast$).

Let $c$ be the clause used in the first step of the derivation $der_{p,q}$.
If $c$ is a fact, we immediately have a contradiction: since $G_p$ is a goal in $S$, 
this means that there exists an evaluation tree of $G_p = \leftarrow A_p$ created by \createtable\ (lines~2-7) with root node $\nTuple{\ID}{A_p\leftarrow A_p}{new}$ and a node with clause $c$ as subnode of the root node.
Therefore, the algorithm will compute a c.a.s. equivalent to $\theta_{p,q}$ (\newactive), contradicting the hypothesis. % (H).

If $c$ is a rule $H\leftarrow B_1,\ldots,B_n$, and $\sigma_0= mgu(G_p,H)$, then by hypothesis there exist SLD derivations $der_{B_1},\ldots,der_{B_n}$, and substitutions $\sigma_1,\ldots,\sigma_n$ s.t. $H\sigma_0\cdots\sigma_n = G_p\theta_{p,q}$, and for each $i\in\{1,\ldots,n\}$:
\begin{itemize}
	\item $der_{B_i}$ is an SLD derivation of $P \cup \{\leftarrow B_i\sigma_0\cdots\sigma_{i-1}\}$.
	\item The c.a.s. of $der_{B_i}$ is $\sigma_i$, and  $len(der_{B_i}) < len(der_{p,q})$. ($\ast\ast$)
\end{itemize}
Since $G_p$ is a goal in $S$, there exists an evaluation tree of $G_p$ created by \createtable\ (lines~2-7) with root node $\nTuple{\ID}{A_p\leftarrow A_p}{new}$ and a node with clause $c$ as subnode of the root node. Then, it is easy to see that for each $i\in\{1,\ldots,n\}$:
\begin{itemize}
	\item There exists a node in the evaluation tree of $G_p$ with selected atom $B_i\sigma_0\cdots\sigma_{i-1}$ (\newactive, lines 8, 18-19).
	\item There exists an evaluation tree of $\leftarrow B_i\sigma_0\cdots\sigma_{i-1}$ created by \createtable\ (lines~2-7) at the location of $B_i\sigma_0\cdots\sigma_{i-1}$.
	\item Since $len(der_{B_i})<len(der_{p,q})$, by ($\ast$) and ($\ast\ast$) the algorithm computes a solution equivalent to $\sigma_i$ of the goal $\leftarrow B_i\sigma_0\cdots\sigma_{i-1}$.
	\item By Propositions~2 and~3, the answer $B_i\sigma_0\cdots\sigma_i$ is sent to the requester of $\leftarrow B_i\sigma_0\cdots\sigma_{i-1}$ by \answerreturn\ (lines 12-14 or 20-23) if $\leftarrow B_i\sigma_0\cdots\sigma_{i-1}$ is involved in a loop, or by \completion\ (lines 3-5) otherwise.
	\item There exists a node with clause $(H \leftarrow B_{i+1},\ldots,B_n)\sigma_0\cdots\sigma_i$ added to the evaluation tree of $G_p$ by \positivereturn\ (lines 20-23).
\end{itemize}
Therefore, $\sigma_1\cdots\sigma_n$ is (equivalent to) a solution of the evaluation tree of $G_p$, contradicting (a), (b), and (c).\hfill$\Box$

\vspace{0.3cm}\noindent
{\it Proof of Theorem 3}.
We assume that nodes (i.e., goals) in the call graph of $P$ inherit the identifier (and the associated ordering) of the request for which they are created. 
Termination follows from two observations: (i) the call graph of $P$ is finite, and (ii) the number of response messages exchanged by the principals involved in the evaluation of $G$ is finite.

\noindent The call graph of $P$ is finite (i) for the following reasons:
\begin{enumerate}
\item The set of goals over predicates in $P$ (up to renaming) is finite. This is because terms that are not variables are constants in $P$.
\item There is no infinite path in the call graph of $P$ composed of nodes $\ID_1,\ldots,\ID_n$ s.t. $\ID_n\orderingIn\ldots\orderingIn\ID_1$.
	This is because of (1) and because the algorithm never creates a new node with identifier $\ID_i$ for a goal if a node with identifier $\ID_j$ already exists for a variant of that goal and $\ID_i\orderingIn\ID_j$.
\item The outdegree of each node in the call graph of $P$ is finite. This is because the number of atoms in the body of each clause in $P$ is finite.
\end{enumerate}
The number of response messages is finite (ii) because:
\begin{enumerate}
\item The number of answers of each goal defined in $P$ is finite (see the proof of Proposition~3).
\item The (possibly empty) set of answers of a goal are transmitted only when a table for the goal
  is first created (and a node representing the goal is added to the call graph of $P$) or new answers of its subgoals are received.
\item For any nodes $id_1$ and $id_2$, a set of answers that flows from
  $id_2$ to $id_1$ in response to a request $id_2$ never contains answers previously communicated in response to request $id_2$ (\sendanswer, lines 3-4).
\item An empty set of answers may flow from $id_2$ to $id_1$ only if $id_2\orderingIn id_1$ (\answerreturn, lines 20-23, and \completion, lines 3-5), or $id_1$ identifies a lower request and a loop $id_2$ has just been identified (\newsubgoal, lines 5-7).
\item There is no infinite path composed of nodes $\ID_n,\ldots,\ID_1$ in the call graph of $P$ through which the answers flow s.t. $\ID_n\orderingIn\ldots\orderingIn\ID_1$.
\item By Proposition~3, procedure \completion\ is eventually invoked for any goal.
\hfill$\Box$
%There is no path in the goal dependency graph of $P$ that forms an infinite increasing sequence of nodes w.r.t. the ID ordering.
\end{enumerate}

%% file: example.tex
\section{Example}
\label{sec:example}

In this section we show how GEM computes the answers of a goal using the procedures presented in Section~3.3.
As an example global policy, we use a fragment of the policy introduced in Section~1.
In particular, we consider the following policy statements:

\vspace{0.2cm}
{\small
1. ${\sf memberOfAlpha}(c1\mbox{,}X) \leftarrow {\sf projectPartner}(mc\mbox{,}Y), {\sf memberOfAlpha}(Y\mbox{,}X).$\\
\indent 2. ${\sf projectPartner}(mc\mbox{,}c2).$\\
\indent 3. ${\sf projectPartner}(mc\mbox{,}c3).$\\
\indent 4. ${\sf memberOfAlpha}(c2\mbox{,}X) \leftarrow {\sf memberOfAlpha}(c1\mbox{,}X).$ \\
\indent 5. ${\sf memberOfAlpha}(c2\mbox{,}alice).$\\
\indent 6. ${\sf memberOfAlpha}(c3\mbox{,}bob).$\\
}
%\vspace{0.2cm} 

\begin{figure}[!t]
	\centering
	\fbox{
		{\footnotesize
			$\xymatrix@R=20pt@C=20pt{
			& \ar[d] & \\
			& membe\ms{rOfA}lpha(c1\mbox{,}X) \ar[ld] \ar[d] \ar[rd] & \\
			projectPartner(mc\mbox{,}Y) & membe\ms{rOfA}lpha(c2\mbox{,}X) \ar@/^15pt/ [u] & membe\ms{rOfA}lpha(c3\mbox{,}X)
			}$
		}
	}	
	\caption{Call Graph of the Evaluation of \emph{memberOfAlpha(c$1$,X)} with Respect to the Example Global Policy}
	\label{fig:dep-graph-appendix}
\end{figure}
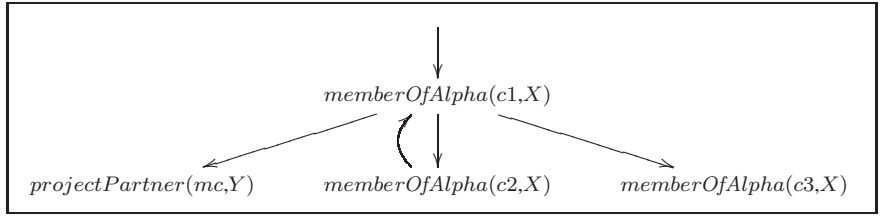  

The call graph of the global policy is shown in Figure~\ref{fig:dep-graph-appendix}.
We illustrate the computation for an initial request {\it (h$_1$,h,memberOfAlpha(c$1$,X))} from hospital $h$ to company \emph{c$1$}.
Table~\ref{tab:callstack} shows the list of all procedure calls made by GEM to produce the response to the initial request.
The first column of the table indicates the order in which the calls are made; the second column denotes the principal and location where each procedure is evaluated.
GEM computes the answers of goal {\it memberOfAlpha(c$1$,X)} by making~53 procedure calls; the number of messages exchanged between different principals, however, is only~14, consisting of~5 request messages and~9 response messages (including the initial request and its response).
Next, we present and discuss some ``screenshots'' showing the status of the computation at various stages.

\begin{table}[!t]	
	\footnotesize
	\begin{oldtabular}{l|c|l}
		{\bf Call} & {\bf Principal} & {\bf Procedure}\\\hline
		1 & c1 & \newsubgoal((h$_1$,h,memberOfAlpha(c1,X)))\\
		2 & c1 & \newactive(memberOfAlpha(c1,X))\\
		3 & mc & \newsubgoal((h$_1$c$1_1$,c1,projectPartner(mc,Y)))\\
		4 & mc & \newactive(projectPartner(mc,Y))\\
		5 & mc & \newactive(projectPartner(mc,Y))\\
		6 & mc & \newactive(projectPartner(mc,Y))\\
		7 & mc & \answerreturn(projectPartner(mc,Y))\\
		8 & mc & \completion(projectPartner(mc,Y))\\
		9 & mc & \sendanswer((h$_1$c$1_1$,c1,projectPartner(mc,Y)),disposed,\{\})\\
		10 & c1 & \positivereturn(h$_1$c$1_1$,\{projectPartner(mc,c2),projectPartner(mc,c3)\},disposed,\{\})\\
		11 & c1 & \newactive(memberOfAlpha(c1,X))\\
		12 & c2 & \newsubgoal((h$_1$c$1_2$,c1,memberOfAlpha(c2,X)))\\
		13 & c2 & \newactive(memberOfAlpha(c2,X))\\
		14 & c1 & \newsubgoal((h$_1$c$1_2$c$2_1$,c2,memberOfAlpha(c1,X)))\\
		15 & c1 & \sendanswer((h$_1$c$1_2$c$2_1$,c2,memberOfAlpha(c1,X)),active,\{h$_1$\})\\
		16 & c2 & \positivereturn(h$_1$c$1_2$c$2_1$,\{\},active,\{h$_1$\})\\
		17 & c2 & \newactive(memberOfAlpha(c2,X))\\
		18 & c2 & \newactive(memberOfAlpha(c2,X))\\
		19 & c2 & \answerreturn(memberOfAlpha(c2,X))\\
		20 & c2 & \sendanswer((h$_1$c$1_2$,c1,memberOfAlpha(c2,X)),active,\{h$_1$\})\\
		21 & c1 & \positivereturn(h$_1$c$1_2$,\{memberOfAlpha(c2,alice)\},active,\{h$_1$\})\\
		22 & c1 & \newactive(memberOfAlpha(c1,X))\\
		23 & c3 & \newsubgoal((h$_1$c$1_3$,c1,memberOfAlpha(c3,X)))\\
		24 & c3 & \newactive(memberOfAlpha(c3,X))\\
		25 & c3 & \newactive(memberOfAlpha(c3,X))\\
		26 & c3 & \answerreturn(memberOfAlpha(c3,X))\\
		27 & c3 & \completion(memberOfAlpha(c3,X))\\
		28 & c3 & \sendanswer((h$_1$c$1_3$,c1,memberOfAlpha(c3,X)),disposed,\{\})\\
		29 & c1 & \positivereturn(h$_1$c$1_3$,\{memberOfAlpha(c3,bob)\},disposed,\{\})\\
		30 & c1 & \newactive(memberOfAlpha(c1,X))\\
		31 & c1 & \newactive(memberOfAlpha(c1,X))\\
		32 & c1 & \newactive(memberOfAlpha(c1,X))\\
		33 & c1 & \answerreturn(memberOfAlpha(c1,X))\\
		34 & c1 & \sendanswer((h$_1$c$1_2$c$2_1$,c2,memberOfAlpha(c1,X)),loop(h$_1$),\{\})\\
		35 & c2 & \positivereturn(h$_1$c$1_2$c$2_1$,\{memberOfAlpha(c1,alice),memberOfAlpha(c1,bob)\},loop(h$_1$),\{\})\\
		36 & c2 & \newactive(memberOfAlpha(c2,X))\\
		37 & c2 & \newactive(memberOfAlpha(c2,X))\\
		38 & c2 & \newactive(memberOfAlpha(c2,X))\\
		39 & c2 & \answerreturn(memberOfAlpha(c2,X))\\
		40 & c2 & \sendanswer((h$_1$c$1_2$,c1,memberOfAlpha(c2,X)),loop(h$_1$),\{h$_1$\})\\
		41 & c1 & \positivereturn(h$_1$c$1_2$,\{memberOfAlpha(c2,bob)\},loop(h$_1$),\{h$_1$\})\\
		42 & c1 & \newactive(memberOfAlpha(c1,X))\\
		43 & c1 & \newactive(memberOfAlpha(c1,X))\\
		44 & c1 & \answerreturn(memberOfAlpha(c1,X))\\
		45 & c1 & \completion(memberOfAlpha(c1,X))\\
		46 & c1 & \sendanswer((h$_1$,h,memberOfAlpha(c1,X)),disposed,\{\})\\
		47 & c1 & \sendanswer((h$_1$c$1_2$c$2_1$,c2,memberOfAlpha(c1,X)),disposed,\{\})\\
		48 & c2 & \positivereturn(h$_1$c$1_2$c$2_1$,\{\},disposed,\{\})\\
		49 & c2 & \newactive(memberOfAlpha(c2,X))\\
		50 & c2 & \answerreturn(memberOfAlpha(c2,X))\\
		51 & c2 & \completion(memberOfAlpha(c2,X))\\
		52 & c2 & \sendanswer((h$_1$c$1_2$,c1,memberOfAlpha(c2,X)),disposed,\{\})\\
		53 & c1 & \positivereturn(h$_1$c$1_2$,\{\},disposed,\{\})\\
		\\\hline
	\end{oldtabular}
	\caption{Procedure Call Stack For the Example Global Policy}
	\label{tab:callstack}
\end{table}

\begin{table}[!t]
	\begin{oldtabular}{l|l}
		\hline
			\multicolumn{2}{l}{{\bf Principal c1}} \\\hline
			HR & (h$_1$,h,$\leftarrow$memberOfAlpha(c1,X))\\
			LR & \{\}\\
			ActiveGoals & \{\}\\
			AnsSet & \{\}\\
			Tree & (h$_1$,memberOfAlpha(c1,X)$\leftarrow$ memberOfAlpha(c1,X),new)\\
			& (h$_1$c$1_1$,memberOfAlpha(c1,X)$\leftarrow$ projectPartner(mc,Y), memberOfAlpha(Y,X),new)\\
		\hline
	\end{oldtabular}
	\caption{Status of the Computation After Procedure Call 1 in Table~\ref{tab:callstack}}
	\label{tab:status1}
\end{table}

When principal \emph{c$1$} receives the request for goal {\it memberOfAlpha(c$1$,X)} from \emph{h}, it calls procedure \newsubgoal\ (Algorithm~1 in Section~3.3) that initializes the table of the goal.
Table~\ref{tab:status1} shows the table of {\it memberOfAlpha(c$1$,X)} resulting from the execution of \newsubgoal\ on the initial request. % {\it (h$_1$,\emph{h},memberOfAlpha(c1,X))}.
The table field \ms{HR} (higher request) is set to the initial request, and the evaluation tree of the goal, \TreeM, is initialized by adding to the root node a subnode representing the only clause in \emph{c$1$}'s local policy applicable to the goal, i.e., clause~1.
The node status is set to \newSM, and the node identifier is obtained by concatenating the request identifier {\it h$_1$} with string {\it c$1_1$}.
To keep the representation more compact, in Table~\ref{tab:status1} and in the other tables presented in this section the evaluation tree of a goal is represented as a list of nodes rather than as the structure defined in Section~3.3.
%The structures of the goals' evaluation trees, however, will be indicated when the tables are presented.

\begin{table}[!t]
	\begin{oldtabular}{l|l}
		\hline
			\multicolumn{2}{l}{{\bf Principal c1}} \\\hline
			HR & (h$_1$,h,$\leftarrow$memberOfAlpha(c1,X))\\
			LR & \{\}\\
			ActiveGoals & \{\}\\
			AnsSet & \{\}\\
			Tree & (h$_1$,memberOfAlpha(c1,X)$\leftarrow$ memberOfAlpha(c1,X),active)\\
			& (h$_1$c$1_1$,memberOfAlpha(c1,X)$\leftarrow$ projectPartner(mc,Y), memberOfAlpha(Y,X),active)
		\\\hline
			\multicolumn{2}{l}{{\bf Principal mc}} \\\hline
			HR & (h$_1$c$1_1$,c1,$\leftarrow$projectPartner(mc,Y))\\
			LR & \{\}\\
			ActiveGoals & \{\}\\
			AnsSet & \{(projectPartner(mc,c2),\{\}),(projectPartner(mc,c3),\{\})\}\\
			Tree & (h$_1$c$1_1$,projectPartner(mc,Y)$\leftarrow$ projectPartner(mc,Y),new)\\
			& (h$_1$c$1_1$mc$_1$,projectPartner(mc,c2),answer)\\
			& (h$_1$c$1_1$mc$_2$,projectPartner(mc,c3),answer)\\
		\hline
	\end{oldtabular}
	\caption{Status of the Computation After Procedure Call 7 in Table~\ref{tab:callstack}}
	\label{tab:status2}
\end{table}

%Clause (1) states that the set of members of project Alpha at \emph{c$1$} is given by the set of project members at all the partner companies certified by \emph{mc}.
%Therefore, 
In order to compute the list of project members, \emph{c$1$} needs to first retrieve from \emph{mc} the list of partner companies in the project, and then for each of these companies the list of its project members.
Table~\ref{tab:status2} shows the status of the computation after goal {\it projectPartner(mc,Y)} has been completely evaluated by \emph{mc} (procedure calls~2 to~7 in Table~\ref{tab:callstack}), i.e., after the set of project partners has been computed.
The request for goal {\it projectPartner(mc,Y)} from \emph{c$1$} to \emph{mc} is generated by the activation of node {\it h$_1$c$1_1$} in \emph{c$1$}'s table (procedure call~2, \newactive), which results in a change of status from \newSM\ to \actSM\ of both the root node and the node itself.
Similarly to \emph{c$1$}, when \emph{mc} receives the request it creates the table of the goal (call~3 in Table~\ref{tab:callstack}), setting \ms{HR} to the higher request and initializing \TreeM\ with clauses~2 and~3 of the global policy presented above.
Two calls to procedure \newactive\ (calls~4 and~5) lead to the identification of two answers of the goal, namely {\it projectPartner(mc,c$2$)} and {\it projectPartner(mc,c$3$)}, which are added to \AnsSetM\ with an empty list of request identifiers.
At the next call to \newactive\ (call~6), the evaluation tree of goal {\it projectPartner(mc,Y)} has no more nodes to activate (i.e., all the branches of the evaluation tree have been inspected) and procedure \answerreturn\ is invoked (call~7).
Since the goal is not involved in any loop, its evaluation is completed and procedure \completion\ is executed next (line~3 of Algorithm~6 in Section~3.3).

\begin{table}[!t]
	\begin{oldtabular}{l|l}
		\hline
			\multicolumn{2}{l}{{\bf Principal c1}} \\\hline
			HR & (h$_1$,h,$\leftarrow$memberOfAlpha(c1,X))\\
			LR & \{(h$_1$c$1_2$c$2_1$,c2,$\leftarrow$memberOfAlpha(c1,X))\}\\
			ActiveGoals & \{\}\\
			AnsSet & \{\}\\
			Tree & (h$_1$,memberOfAlpha(c1,X)$\leftarrow$ memberOfAlpha(c1,X),active)\\
			& (h$_1$c$1_1$,memberOfAlpha(c1,X)$\leftarrow$ projectPartner(mc,Y), memberOfAlpha(Y,X),disposed)\\
			& (h$_1$c$1_2$,memberOfAlpha(c1,X)$\leftarrow$ memberOfAlpha(c2,X),active)\\
			& (h$_1$c$1_3$,memberOfAlpha(c1,X)$\leftarrow$ memberOfAlpha(c3,X),new)\\
			\hline
			\multicolumn{2}{l}{{\bf Principal mc}} \\\hline
			HR & null\\
			LR & \{\}\\
			ActiveGoals & \{\}\\
			AnsSet & \{(projectPartner(mc,c2),\{h$_1$c$1_1$\}),(projectPartner(mc,c3),\{h$_1$c$1_1$\})\}\\
			Tree & (h$_1$c$1_1$,projectPartner(mc,Y)$\leftarrow$ projectPartner(mc,Y),disposed)\\
			& (h$_1$c$1_1$mc$_1$,projectPartner(mc,c2),answer)\\
			& (h$_1$c$1_1$mc$_2$,projectPartner(mc,c3),answer)\\
			\hline
			\multicolumn{2}{l}{{\bf Principal c2}} \\\hline
			HR & (h$_1$c$1_2$,c1,$\leftarrow$memberOfAlpha(c2,X))\\
			LR & \{\}\\
			ActiveGoals & \{\}\\
			AnsSet & \{\}\\
			Tree & (h$_1$c$1_2$,memberOfAlpha(c2,X)$\leftarrow$ memberOfAlpha(c2,X),active)\\
			& (h$_1$c$1_2$c$2_1$,memberOfAlpha(c2,X)$\leftarrow$ memberOfAlpha(c1,X),active)\\
			& (h$_1$c$1_2$c$2_2$,memberOfAlpha(c2,alice),new)\\
		\hline
	\end{oldtabular}
	\caption{Status of the Computation After Procedure Call 15 in Table~\ref{tab:callstack}}
	\label{tab:status3}
\end{table}

As a result of the execution of procedure \completion, the root node of the evaluation tree of goal {\it projectPartner(mc,Y)} is disposed and the answers identified are sent to \emph{c$1$} through procedure \sendanswer\ (procedure call~9, results shown in Table~\ref{tab:status3}).
The response message received by \emph{c$1$} is processed by procedure \positivereturn; the message contains the two answers ({\it projectPartner(mc,c$2$)} and {\it projectPartner(mc,c$3$)}) and an empty set of loop identifiers, and has status \dispSM, indicating that no more answers of goal {\it projectPartner(mc,Y)} will be received.
The evaluation tree of goal {\it memberOfAlpha(c$1$,X)} is updated by adding two subnodes to node {\it h$_1$c$1_1$}, one for each project partner (see \emph{c$1$}'s table in Table~\ref{tab:status3}).

The activation of node {\it h$_1$c$1_2$} by \emph{c$1$} leads to the request for goal {\it memberOfAlpha(c$2$,X)} to \emph{c$2$}.
Accordingly, \emph{c$2$} creates a table for the goal; the evaluation tree of the goal consists of three nodes: the root node and two subnodes, representing clauses~4 and~5 of the global policy, with identifiers {\it h$_1$c$1_2$c$2_1$} and {\it h$_1$c$1_2$c$2_2$} respectively.
The activation of node {\it h$_1$c$1_2$c$2_1$} by \emph{c$2$}, in turn, leads to a request for goal {\it memberOfAlpha(c$1$,X)} to \emph{c$1$}, forming a loop.
The loop is identified by \emph{c$1$} in procedure \newsubgoal\ (call~14 in Table~\ref{tab:callstack}): in fact, the identifier of the higher request for {\it memberOfAlpha(c$1$,X)} ({\it h$_1$}) is a prefix of the identifier of \emph{c$2$}'s request ({\it h$_1$c$1_2$c$2_1$}).
Therefore, the lower request is added by \emph{c$1$} to set \ms{LR}, and a response is sent from \emph{c$1$} to \emph{c$2$} with a notification of loop {\it h$_1$} (call~15).

\begin{table}[!t]
	\begin{oldtabular}{l|l}
		\hline
			\multicolumn{2}{l}{{\bf Principal c1}} \\\hline
			HR & (h$_1$,h,$\leftarrow$memberOfAlpha(c1,X))\\
			LR & \{(h$_1$c$1_2$c$2_1$,c2,$\leftarrow$memberOfAlpha(c1,X))\}\\
			ActiveGoals & \{(h$_1$,1)\}\\
			AnsSet & \{(memberOfAlpha(c1,alice),\{h$_1$c$1_2$c$2_1$\}),(memberOfAlpha(c1,bob),\{h$_1$c$1_2$c$2_1$\})\}\\
			Tree & (h$_1$,memberOfAlpha(c1,X)$\leftarrow$ memberOfAlpha(c1,X),loop(\{h$_1$\}))\\
			& (h$_1$c$1_1$,memberOfAlpha(c1,X)$\leftarrow$ projectPartner(mc,Y), memberOfAlpha(Y,X),disposed)\\
			& (h$_1$c$1_2$,memberOfAlpha(c1,X)$\leftarrow$ memberOfAlpha(c2,X),loop(\{h$_1$\}))\\
			& (h$_1$c$1_3$,memberOfAlpha(c1,X)$\leftarrow$ memberOfAlpha(c3,X),disposed)\\
			& (h$_1$c$1_4$,memberOfAlpha(c1,alice),answer)\\
			& (h$_1$c$1_5$,memberOfAlpha(c1,bob),answer)\\
			\hline
			\multicolumn{2}{l}{{\bf Principal mc}} \\\hline
			HR & null\\
			LR & \{\}\\
			ActiveGoals & \{\}\\
			AnsSet & \{(projectPartner(mc,c2),\{h$_1$c$1_1$\}),(projectPartner(mc,c3),\{h$_1$c$1_1$\})\}\\
			Tree & (h$_1$c$1_1$,projectPartner(mc,Y)$\leftarrow$ projectPartner(mc,Y),disposed)\\
			& (h$_1$c$1_1$mc$_1$,projectPartner(mc,c2),answer)\\
			& (h$_1$c$1_1$mc$_2$,projectPartner(mc,c3),answer)\\
			\hline
			\multicolumn{2}{l}{{\bf Principal c2}} \\\hline
			HR & (h$_1$c$1_2$,c1,$\leftarrow$memberOfAlpha(c2,X))\\
			LR & \{\}\\
			ActiveGoals & \{(h$_1$,1)\}\\
			AnsSet & \{(memberOfAlpha(c2,alice),\{h$_1$c$1_2$\})\}\\
			Tree & (h$_1$c$1_2$,memberOfAlpha(c2,X)$\leftarrow$ memberOfAlpha(c2,X),active)\\
			& (h$_1$c$1_2$c$2_1$,memberOfAlpha(c2,X)$\leftarrow$ memberOfAlpha(c1,X),loop(\{h$_1$\}))\\
			& (h$_1$c$1_2$c$2_2$,memberOfAlpha(c2,alice),answer)\\
			\hline
			\multicolumn{2}{l}{{\bf Principal c3}} \\\hline
			HR & null\\
			LR & \{\}\\
			ActiveGoals & \{\}\\
			AnsSet & \{(memberOfAlpha(c3,bob),\{h$_1$c$1_3$\})\}\\
			Tree & (h$_1$c$1_3$,memberOfAlpha(c3,X)$\leftarrow$ memberOfAlpha(c3,X),disposed)\\
			& (h$_1$c$1_3$c$3_1$,memberOfAlpha(c3,bob),answer)\\
		\hline
	\end{oldtabular}
	\caption{Status of the Computation After Procedure Call 34 in Table~\ref{tab:callstack}}
	\label{tab:status4}
\end{table}

The loop notification sent from \emph{c$1$} to \emph{c$2$} starts the loop processing phase, which involves procedure calls from~16 to~34 in Table~\ref{tab:callstack}.
The results of the loop processing phase are shown in Table~\ref{tab:status4}.
Upon receiving the loop notification, \emph{c$2$} sets the status of the node whose evaluation formed the loop to {\it loop(\{h$_1$\})} and ``freezes'' its evaluation; then, it proceeds with the evaluation of the other nodes of the evaluation tree.
The activation of node {\it h$_1$c$1_2$c$2_2$} (procedure call~17), in particular, leads to the first answer of the goal, i.e., {\it memberOfAlpha(c$2$,alice)}.
Since at this point there are no more nodes to be activated, the computed answer can be sent to \emph{c$1$} with a notification about the loop.
Before sending the answer, \emph{c$2$} sets the counter in \ActiveLoopsM\ to~1 (procedure \answerreturn, call 19) and adds the identifier of \ms{HR} to the set of recipients of answer {\it memberOfAlpha(c$2$,alice)} in \AnsSetM\ (procedure \sendanswer, call~20).

The loop is now processed at \emph{c$1$}.
After adding a subnode to the evaluation tree of goal {\it memberOfAlpha(c$1$,X)} for the answer received from \emph{c$2$}, \emph{c$1$} freezes node {\it h$_1$c$1_2$} and starts the evaluation of node {\it h$_1$c$1_3$} (procedure call~22).
This results in a request from \emph{c$1$} to \emph{c$3$} for the evaluation of goal {\it memberOfAlpha(c$3$,X)}.
The only clause applicable to {\it memberOfAlpha(c$3$,X)} (clause~6 of the global policy) is a fact; therefore, the goal is completely evaluated after one call to procedure \newactive\ (call~24).
The answer of the goal, {\it memberOfAlpha(c$3$,bob)}, is returned to \emph{c$1$} (procedure call~28).
Since the status of the response message is \dispSM, \emph{c$1$} disposes node {\it h$_1$c$1_3$} and adds subnode {\it h$_1$c$1_5$} to it reflecting the answer received from \emph{c$3$} (procedure \positivereturn, call~29).
The next two executions of procedure \newactive\ at \emph{c$1$} lead to the identification of two answers of goal {\it memberOfAlpha(c$1$,X)}, namely {\it memberOfAlpha(c$1$,alice)} and {\it memberOfAlpha(c$1$,bob)}.
Before returning these answers to the requester of \ms{HR} (i.e., \emph{h}), however, all the loops need to be fully processed.
For this reason, \emph{c$1$} sends the two answers to \emph{c$2$} in response to \ms{LR} first; the status of the response message is {\it loop(h$_1$)}, and the status of the root node of the evaluation tree in \emph{c$1$}'s table is changed accordingly (procedure calls~33 and~34 in Table~\ref{tab:callstack}). 

\begin{table}[!t]
	\begin{oldtabular}{l|l}
		\hline
			\multicolumn{2}{l}{{\bf Principal c1}} \\\hline
			HR & null\\
			LR & \{\}\\
			ActiveGoals & \{\}\\
			AnsSet & \{(memberOfAlpha(c1,alice),\{h$_1$c$1_2$c$2_1$,h$_1$\}),(memberOfAlpha(c1,bob),\{h$_1$c$1_2$c$2_1$,h$_1$\})\}\\
			Tree & (h$_1$,memberOfAlpha(c1,X)$\leftarrow$ memberOfAlpha(c1,X),disposed)\\
			& (h$_1$c$1_1$,memberOfAlpha(c1,X)$\leftarrow$ projectPartner(mc,Y), memberOfAlpha(Y,X),disposed)\\
			& (h$_1$c$1_2$,memberOfAlpha(c1,X)$\leftarrow$ memberOfAlpha(c2,X),disposed)\\
			& (h$_1$c$1_3$,memberOfAlpha(c1,X)$\leftarrow$ memberOfAlpha(c3,X),disposed)\\
			& (h$_1$c$1_4$,memberOfAlpha(c1,alice),answer)\\
			& (h$_1$c$1_5$,memberOfAlpha(c1,bob),answer)\\
			& (h$_1$c$1_6$,memberOfAlpha(c1,bob),answer)\\
			\hline
			\multicolumn{2}{l}{{\bf Principal mc}} \\\hline
			HR & null\\
			LR & \{\}\\
			ActiveGoals & \{\}\\
			AnsSet & \{(projectPartner(mc,c2),\{h$_1$c$1_1$\}),(projectPartner(mc,c3),\{h$_1$c$1_1$\})\}\\
			Tree & (h$_1$c$1_1$,projectPartner(mc,Y)$\leftarrow$ projectPartner(mc,Y),disposed)\\
			& (h$_1$c$1_1$mc$_1$,projectPartner(mc,c2),answer)\\
			& (h$_1$c$1_1$mc$_2$,projectPartner(mc,c3),answer)\\
			\hline
			\multicolumn{2}{l}{{\bf Principal c2}} \\\hline
			HR & null\\
			LR & \{\}\\
			ActiveGoals & \{\}\\
			AnsSet & \{(memberOfAlpha(c2,alice),\{h$_1$c$1_2$\}),(memberOfAlpha(c2,bob),\{h$_1$c$1_2$\})\}\\
			Tree & (h$_1$c$1_2$,memberOfAlpha(c2,X)$\leftarrow$ memberOfAlpha(c2,X),disposed)\\
			& (h$_1$c$1_2$c$2_1$,memberOfAlpha(c2,X)$\leftarrow$ memberOfAlpha(c1,X),disposed)\\
			& (h$_1$c$1_2$c$2_2$,memberOfAlpha(c2,alice),answer)\\
			& (h$_1$c$1_2$c$2_3$,memberOfAlpha(c2,alice),answer)\\
			& (h$_1$c$1_2$c$2_4$,memberOfAlpha(c2,bob),answer)\\
			\hline
			\multicolumn{2}{l}{{\bf Principal c3}} \\\hline
			HR & null\\
			LR & \{\}\\
			ActiveGoals & \{\}\\
			AnsSet & \{(memberOfAlpha(c3,bob),\{h$_1$c$1_3$\})\}\\
			Tree & (h$_1$c$1_3$,memberOfAlpha(c3,X)$\leftarrow$ memberOfAlpha(c3,X),disposed)\\
			& (h$_1$c$1_3$c$3_1$,memberOfAlpha(c3,bob),answer)\\
		\hline
	\end{oldtabular}
	\caption{Final Status of the Computation for the Example Global Policy}
	\label{tab:status5}
\end{table}

Now, the second iteration of the loop processing phase starts (procedure calls~35-44).
In this second iteration, \emph{c$2$} identifies a new answer of its goal, i.e., {\it memberOfAlpha(c$2$,bob)}, which is sent back to \emph{c$1$}.
This answer, however, does not lead to new answers at \emph{c$1$}.
Since {\it h$_1$} is the only loop in the SCC (and hence {\it memberOfAlpha(c$1$,X)} is the leader of the SCC), and no new answers of {\it memberOfAlpha(c$1$,X)} have been computed, the loop termination phase can start (line~15 of Algorithm~6 in Section~3.3).
In this phase, \emph{c$1$} sends a response message with status \dispSM\ to both \emph{c$2$} (the other principal in the loop) and \emph{h} (to which also the answers are sent).
Upon receiving this message, \emph{c$2$} disposes all the nodes in the evaluation tree of {\it memberOfAlpha(c$2$,X)} that are involved in some loop (procedure \positivereturn), and forwards the message back to \emph{c$1$} (calls~51 and~52).
\emph{c$1$} simply ignores the message, as the status of the root node of the evaluation tree of {\it memberOfAlpha(c$1$,X)} is already \dispSM\ (line~4 of Algorithm~5 in Section~3.3), and the computation terminates.
Table~\ref{tab:status5} shows the status of the tables of all the goals at the end of the computation.

%% file: evaluationAppendix.tex
\section{Practical Evaluation}
\label{sec:eval-appendix}

%The following figures show the global policies used for the experiments presented in Section~5 of the paper, and provide a graphical overview of the main evaluation results of GEM.

\begin{figure}[!t]
	\centering
	
	\begin{tabular}{ll}
	
	\subfigure[Call Graph of the Global Policies 1.0 to 1.5]{\fbox{
	{\footnotesize
			$\xymatrix@R=10pt@C=7pt{
			&& \ar[d] & &&&&&&\\
			&& c1 \ar[ld] \ar[d] \ar[rd] &&&&&&&\\
			*+[o][F-][d]{1.0}\ar@{--}@<-4pt>[rrrr] & mc1 & c2 \ar@/^6pt/ [u] & c3 \ar[ld] \ar[d] \ar[rd] & &&&&&\\
			*+[o][F-][d]{1.1}\ar@{--}@<-4pt>[rrrrr] && mc2 & c4 \ar@/^6pt/ [u] & c5 \ar[ld] \ar[d] \ar[rd] &&&&&\\
			*+[o][F-][d]{1.2}\ar@{--}@<-4pt>[rrrrrr] &&& mc3 & c6 \ar@/^6pt/ [u] & c7 \ar[ld] \ar[d] \ar[rd] &&&&\\
			*+[o][F-][d]{1.3}\ar@{--}@<-4pt>[rrrrrrr] &&&& mc4 & c8 \ar@/^6pt/ [u] & c9 \ar[ld] \ar[d] \ar[rd] &&&\\
			*+[o][F-][d]{1.4}\ar@{--}@<-4pt>[rrrrrrrr] &&&&& mc5 & c10 \ar@/^6pt/ [u] & c11 \ar[ld] \ar[d] \ar[rd] &&\\
			*+[o][F-][d]{1.5}\ar@{--}@<-4pt>[rrrrrrrrr] &&&&&& mc6 & c12 \ar@/^6pt/ [u] & c13 &
			}$
	}
	\label{fig:policy-exp1}}}
	
	&
	
	\multirow{2}{*}{
		\subfigure[Call Graph of the Global Policies 3.0 to 3.5]{\fbox{
	{\footnotesize
	$\xymatrix@R=13pt@C=5pt{
	&\ar[d] & \\
	&c1 \ar[d] \ar[rd] & \\
	&c2 \ar[d] \ar@/^10pt/[u] & c3 \ar[ld]\\
	*+[o][F-][d]{3.0}\ar@{--}@<-5pt>[rr] & c4 \ar@/^10pt/[u] \ar[d] \ar[rd] & \\
	&c5 \ar[d] \ar@/^10pt/[u] & c6 \ar[ld]\\
	*+[o][F-][d]{3.1}\ar@{--}@<-5pt>[rr]&c7 \ar@/^10pt/[u] \ar[d] \ar[rd] & \\
	&c8 \ar[d] \ar@/^10pt/[u] & c9 \ar[ld]\\
	*+[o][F-][d]{3.2}\ar@{--}@<-5pt>[rr]&c10 \ar@/^10pt/[u] \ar[d] \ar[rd] & \\
	&c11 \ar[d] \ar@/^10pt/[u] & c12 \ar[ld]\\
	*+[o][F-][d]{3.3}\ar@{--}@<-5pt>[rr]&c13 \ar@/^10pt/[u] \ar[d] \ar[rd] & \\
	&c14 \ar[d] \ar@/^10pt/[u] & c15 \ar[ld]\\
	*+[o][F-][d]{3.4}\ar@{--}@<-5pt>[rr]&c16 \ar@/^10pt/[u] \ar[d] \ar[rd] & \\
	&c17 \ar[d] \ar@/^10pt/[u] & c18 \ar[ld]\\
	*+[o][F-][d]{3.5}\ar@{--}@<-5pt>[rr]&c19 \ar@/^10pt/[u] &
	}$
	}
	\label{fig:policy-exp3}}}	
	}
	\\
	
	\subfigure[Call Graph of the Global Policies 2.0 to 2.5]{\fbox{
	{\footnotesize
	$\xymatrix@R=8pt@C=5pt{
	&\ar[d] &&&&&& &\\
	%&c1 \ar[dddd] \ar[rdddddd]^(.8)*[d]+[o][F-]{2.0} \ar[rrddddd]^(.8)*[d]+[o][F-]{2.1} \ar[rrrdddd]^(.8)*[d]+[o][F-]{2.2} \ar[rrrrddd]^(.8)*[d]+[o][F-]{2.3} \ar[rrrrrdd]^(.8)*[d]+[o][F-]{2.4} \ar[rrrrrrd]^(.8)*[d]+[o][F-]{2.5} &&&&&&&\\
	&c1 \ar[dddd] \ar[rdddddd] \ar[rrddddd] \ar[rrrdddd] \ar[rrrrddd] \ar[rrrrrdd] \ar[rrrrrrd] &&&&&&&\\
	&&&&&&& c9 \ar[ld] & *+[o][F-][d]{2.5} \ar@<-4pt>@{-}[l] \\
	&&&&&& c8 \ar[ld] & & *+[o][F-][d]{2.4} \ar@<-4pt>@{--}[ll]\\
	&&&&& c7 \ar[ld] & & & *+[o][F-][d]{2.3} \ar@{--}@<-4pt>[lll]\\
	&c2 \ar[ddd] \ar@/^15pt/[uuuu] &&& c6 \ar[ld] & && & *+[o][F-][d]{2.2} \ar@{--}@<-4pt>[llll]\\
	&&& c5 \ar[ld] & &&&& *+[o][F-][d]{2.1} \ar@{--}@<-4pt>[lllll]\\
	& & c3 \ar[ld] & &&&&& *+[o][F-][d]{2.0} \ar@{--}@<-4pt>[llllll]\\
	&c4 \ar@/^15pt/[uuu] &&&&&& &
	}$
	}
	\label{fig:policy-exp2}}}
	
	&
	
	%\hspace{0.5cm}
	
%	\subfigure[Call Graph of the Global Policies 3.0 to 3.5]{\fbox{
%	{\footnotesize
%	$\xymatrix@R=10pt@C=0pt{
%	&\ar[d] & \\
%	&c1 \ar[d] \ar[rd] & \\
%	&c2 \ar[d] \ar@/^10pt/[u] & c3 \ar[ld]\\
%	*+[o][F-][d]{3.0}\ar@{--}@<-5pt>[rr] & c4 \ar@/^10pt/[u] \ar[d] \ar[rd] & \\
%	&c5 \ar[d] \ar@/^10pt/[u] & c6 \ar[ld]\\
%	*+[o][F-][d]{3.1}\ar@{--}@<-5pt>[rr]&c7 \ar@/^10pt/[u] \ar[d] \ar[rd] & \\
%	&c8 \ar[d] \ar@/^10pt/[u] & c9 \ar[ld]\\
%	*+[o][F-][d]{3.2}\ar@{--}@<-5pt>[rr]&c10 \ar@/^10pt/[u] \ar[d] \ar[rd] & \\
%	&c11 \ar[d] \ar@/^10pt/[u] & c12 \ar[ld]\\
%	*+[o][F-][d]{3.3}\ar@{--}@<-5pt>[rr]&c13 \ar@/^10pt/[u] \ar[d] \ar[rd] & \\
%	&c14 \ar[d] \ar@/^10pt/[u] & c15 \ar[ld]\\
%	*+[o][F-][d]{3.4}\ar@{--}@<-5pt>[rr]&c16 \ar@/^10pt/[u] \ar[d] \ar[rd] & \\
%	&c17 \ar[d] \ar@/^10pt/[u] & c18 \ar[ld]\\
%	*+[o][F-][d]{3.5}\ar@{--}@<-5pt>[rr]&c19 \ar@/^10pt/[u] &
%	}$
%	}
%	\label{fig:policy-exp3}}}	

	\\
	\end{tabular}
	
	\caption{Call Graph of the Global Policies Used in the Experiments Set 1}
	\label{fig:policy-exp}
\end{figure}
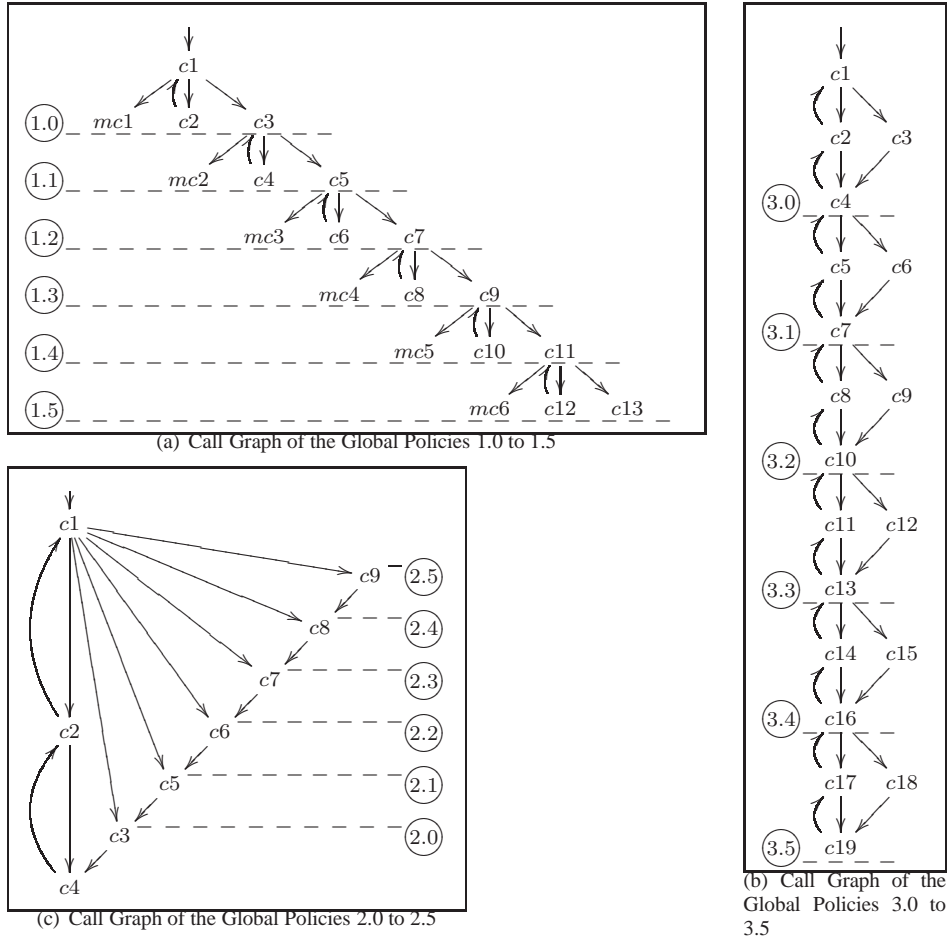

Figure~\ref{fig:policy-exp} shows how the global policies defined in Appendix B and in Section~3.1 have been modified to evaluate the performance of GEM in response to an increase in: (1) the number of principals and clauses (Figure~\ref{fig:policy-exp1}), (2) the number of loops (Figure~\ref{fig:policy-exp2}), and (3) both the number of principals, clauses and loops (Figure~\ref{fig:policy-exp3}) in a global policy.
For each global policy, six variants have been created; in the figures, we use identifiers from x.0 to x.5 (where x is either 1, 2, or 3) to denote the variants, where variant x.0 represents the original policy. % in Appendix B.
To keep the figures as simple yet informative as possible, we label the nodes in the graph with the identifier of the principal evaluating the goal they represent rather than with the goal itself, as for the purpose of the experiments the number of principals involved in a computation is more relevant than the goals they evaluate.

\begin{figure}[!t]
	\centering
	\subfigure[Total and Computation Time for an Increasing Number of Loops in the Computation]{
		\includegraphics[width=0.7\linewidth]{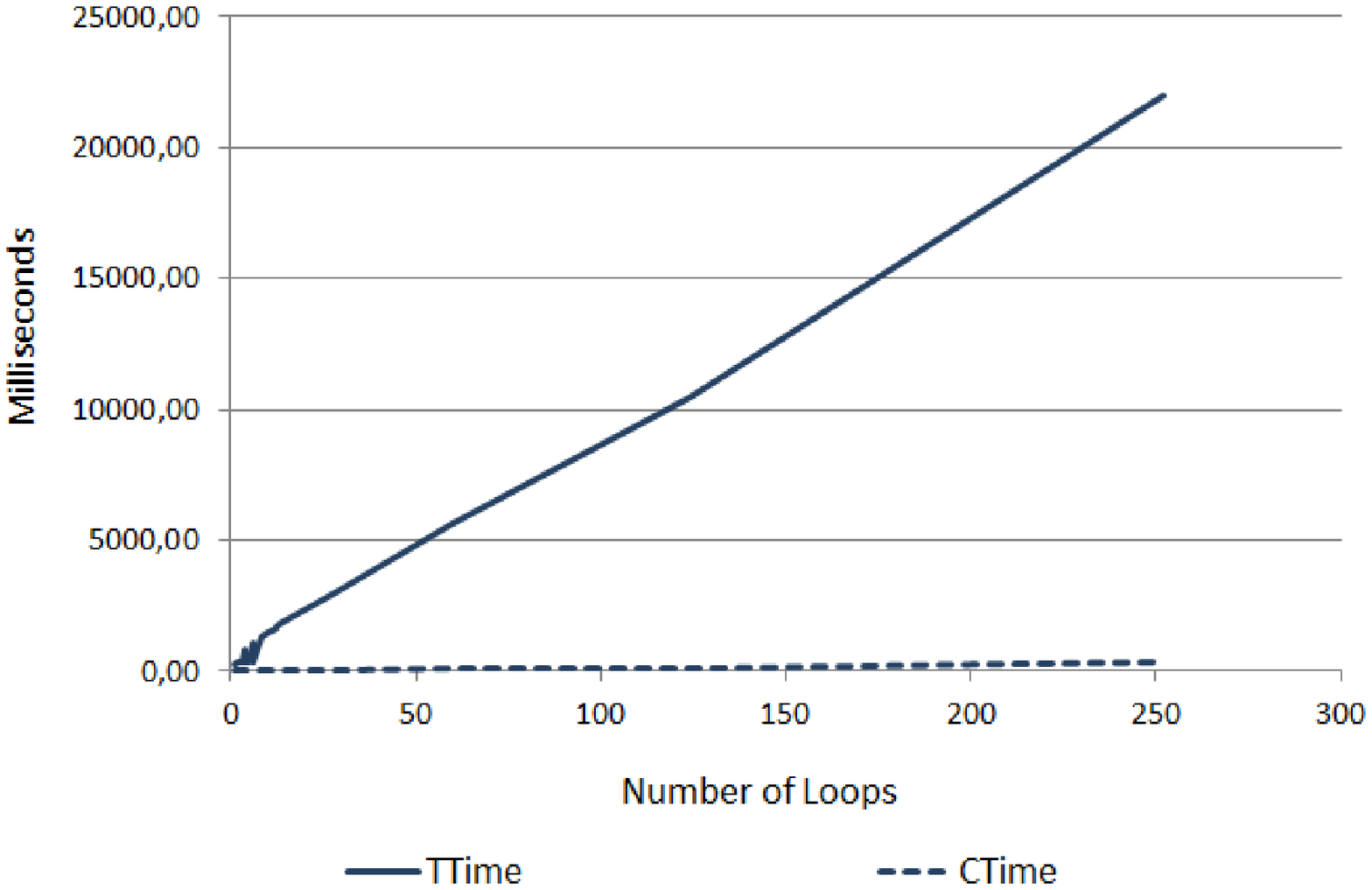}
	\label{fig:set1-time-mem}}
	\subfigure[Total and Tables Memory for an Increasing Number of Loops in the Computation]{
	\includegraphics[width=0.7\linewidth]{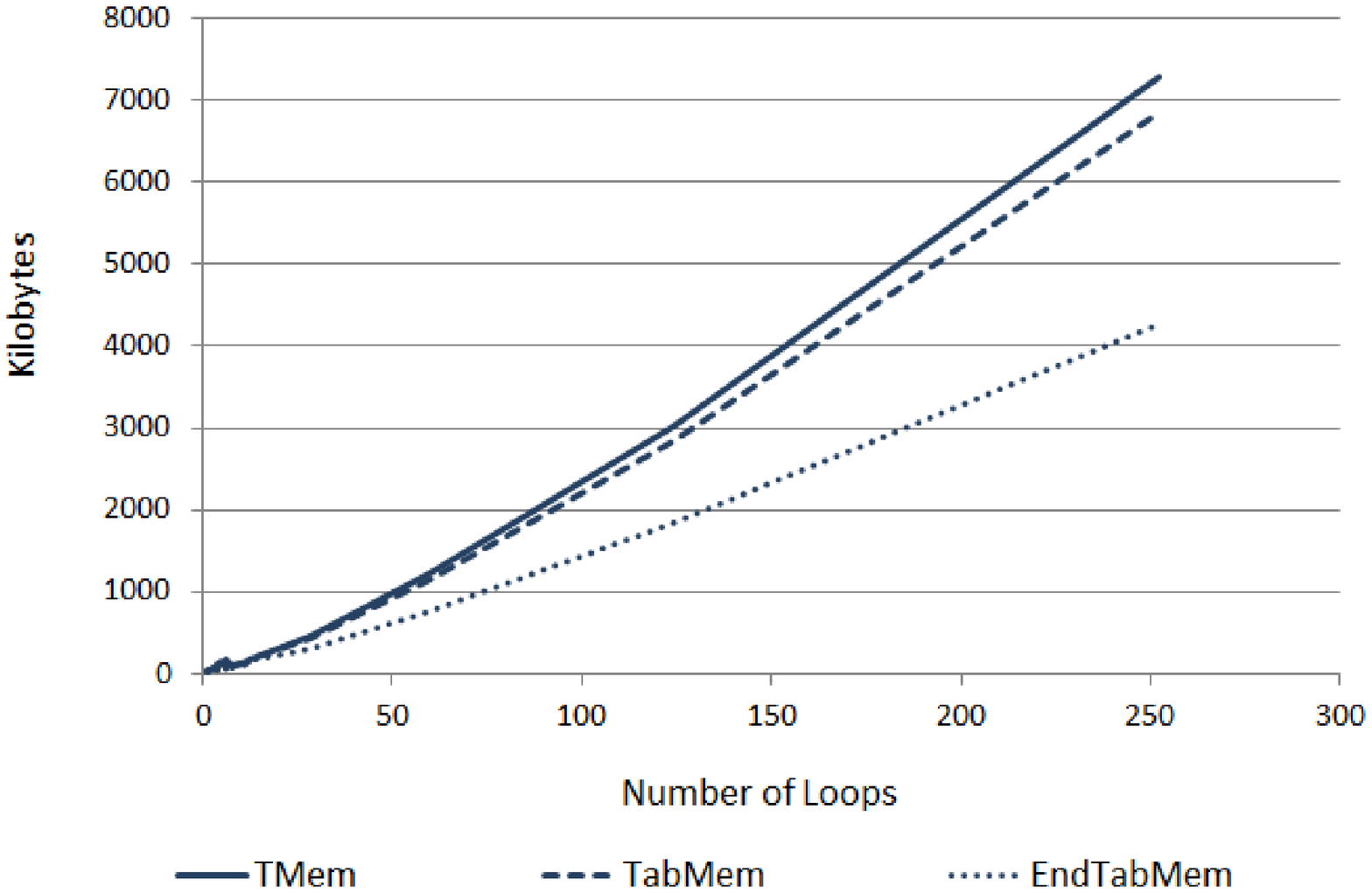}
	\label{fig:set1-time}}
	\caption{Time and Memory Results for Experiments Set 1}
	\label{fig:graphs-set1}
\end{figure}

\begin{figure}[!t]
	\centering
	\subfigure[Time Results with Respect to the Number of Messages Exchanged in the Computation]{
		\includegraphics[width=0.7\linewidth]{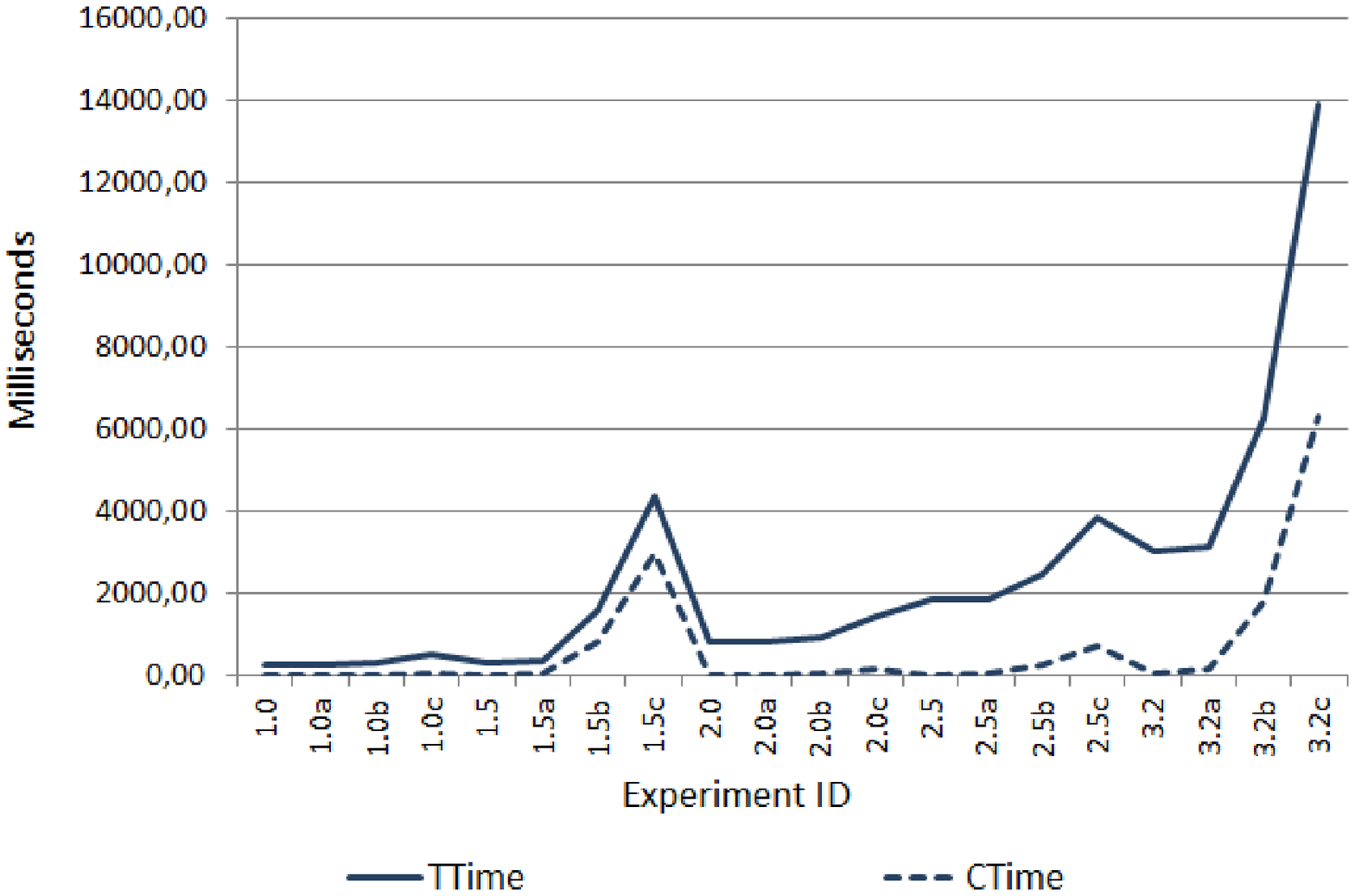}
	\label{fig:set2-time}}
		\subfigure[Memory Results with Respect to the Number of Answers Derived in the Computation]{
	\includegraphics[width=0.7\linewidth]{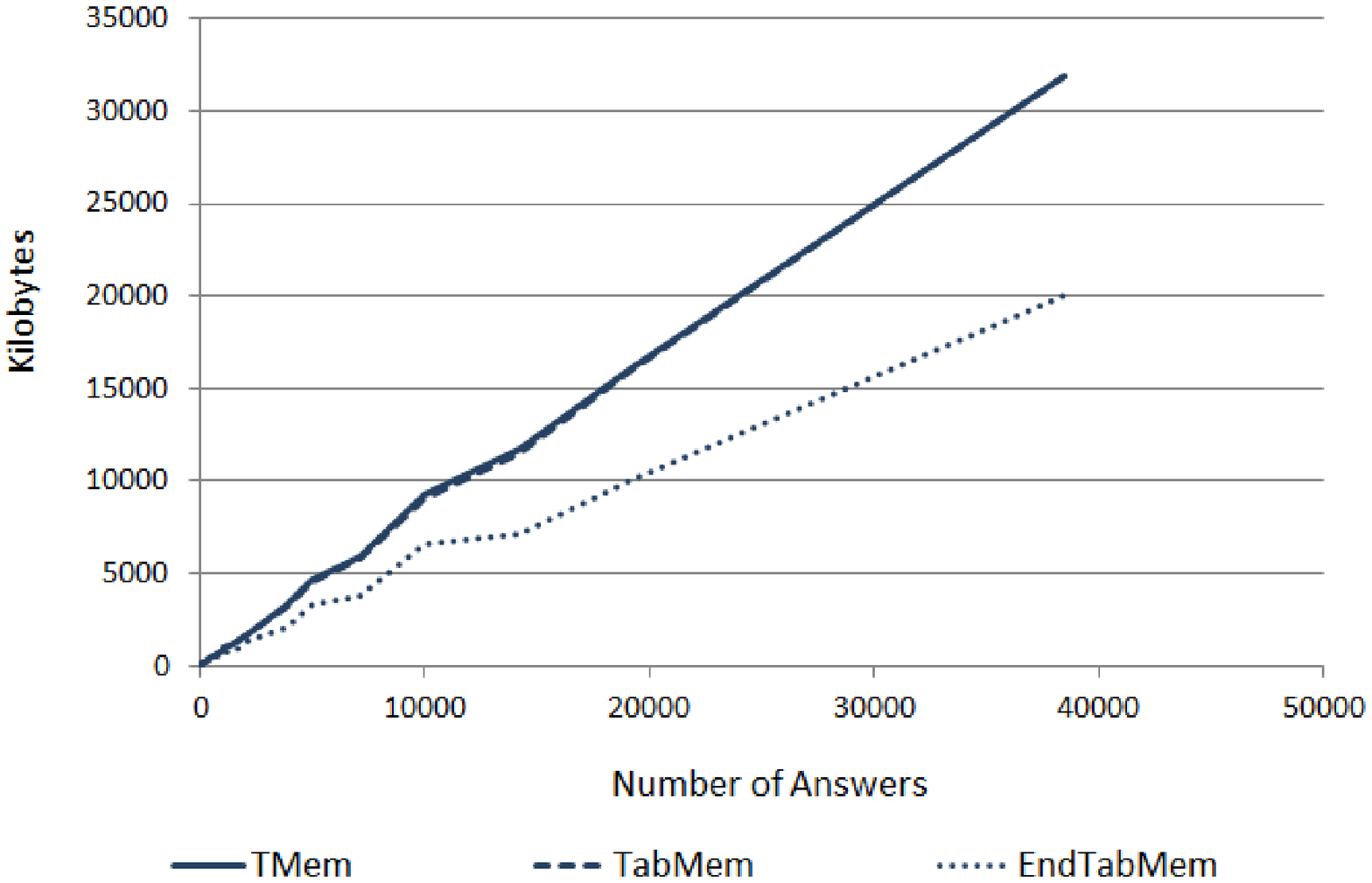}
	\label{fig:set2-mem}}
	\caption{Time and Memory Results for Experiments Set 2}
	\label{fig:graphs-set2}
\end{figure}

Figures~\ref{fig:graphs-set1} and~\ref{fig:graphs-set2} provide a graphical overview of the main evaluation results of GEM, based on the values presented in Tables~1 and~2 in Section~5.

%% file: Journal.bbl
\begin{thebibliography}{}

\bibitem[\protect\citeauthoryear{??}{FOA}{}]{FOAF}
{The Friend of a Friend (FOAF) project}.
\newblock http://www.foaf-project.org/.

\bibitem[\protect\citeauthoryear{Alves, Damasio, Nejdl, and Olmedilla}{Alves
  et~al\mbox{.}}{2006}]{ADNO-POLICY-06}
{\sc Alves, M.}, {\sc Damasio, C.~V.}, {\sc Nejdl, W.}, {\sc and} {\sc
  Olmedilla, D.} 2006.
\newblock {A Distributed Tabling Algorithm for Rule Based Policy Systems}.
\newblock In {\em Proceedings of International Workshop on Policies for
  Distributed Systems and Networks}. IEEE Computer Society, Washington, DC,
  USA, 123--132.

\bibitem[\protect\citeauthoryear{Apt}{Apt}{1990}]{A-HTCS-90}
{\sc Apt, K.~R.} 1990.
\newblock {Logic programming}.
\newblock In {\em Handbook of theoretical computer science (vol. B): formal
  models and semantics}. MIT Press, Cambridge, MA, USA, 493--574.

\bibitem[\protect\citeauthoryear{Apt, Blair, and Walker}{Apt
  et~al\mbox{.}}{1988}]{ABW-FDBLP-88}
{\sc Apt, K.~R.}, {\sc Blair, H.~A.}, {\sc and} {\sc Walker, A.} 1988.
\newblock {\em {Towards a theory of declarative knowledge}}.
\newblock Morgan Kaufmann Publishers Inc., San Francisco, CA, USA, 89--148.

\bibitem[\protect\citeauthoryear{Apt and Bol}{Apt and Bol}{1994}]{AB-JLP-94}
{\sc Apt, K.~R.} {\sc and} {\sc Bol, R.~N.} 1994.
\newblock Logic programming and negation: A survey.
\newblock {\em Journal of Logic Programming\/}~{\em 19/20}, 9--71.

\bibitem[\protect\citeauthoryear{Apt and Marchiori}{Apt and
  Marchiori}{1994}]{AM-FAC-94}
{\sc Apt, K.~R.} {\sc and} {\sc Marchiori, E.} 1994.
\newblock {Reasoning about Prolog programs: from Modes through Types to
  Assertions}.
\newblock {\em Formal Aspects of Computing\/}~{\em 6,\/}~6A, 743--765.

\bibitem[\protect\citeauthoryear{Becker}{Becker}{2005}]{B-TR-05}
{\sc Becker, M.~Y.} 2005.
\newblock {Cassandra: flexible trust management and its application to
  electronic health records}.
\newblock Ph.D. thesis, Computer Laboratory, University of Cambridge, UK.

\bibitem[\protect\citeauthoryear{Becker, Fournet, and Gordon}{Becker
  et~al\mbox{.}}{2010}]{BFG-JCS-10}
{\sc Becker, M.~Y.}, {\sc Fournet, C.}, {\sc and} {\sc Gordon, A.~D.} 2010.
\newblock {SecPAL: Design and semantics of a decentralized authorization
  language}.
\newblock {\em Journal of Computer Security\/}~{\em 18}, 619--665.

\bibitem[\protect\citeauthoryear{Becker, Mackay, and Dillaway}{Becker
  et~al\mbox{.}}{2009}]{BMD-POLICY-09}
{\sc Becker, M.~Y.}, {\sc Mackay, J.~F.}, {\sc and} {\sc Dillaway, B.} 2009.
\newblock {Abductive Authorization Credential Gathering}.
\newblock In {\em Proceedings of the 10th International Conference on Policies
  for Distributed Systems and Networks}. IEEE Press, Piscataway, NJ, USA, 1--8.

\bibitem[\protect\citeauthoryear{Blaze, Feigenbaum, and Lacy}{Blaze
  et~al\mbox{.}}{1996}]{BFL-SP-96}
{\sc Blaze, M.}, {\sc Feigenbaum, J.}, {\sc and} {\sc Lacy, J.} 1996.
\newblock {Decentralized Trust Management}.
\newblock In {\em Proceedings of the IEEE Symposium on Security and Privacy}.
  IEEE Computer Society, 164--173.

\bibitem[\protect\citeauthoryear{B{\"o}hm, Etalle, den Hartog, H{\"u}tter,
  Trabelsi, Trivellato, and Zannone}{B{\"o}hm
  et~al\mbox{.}}{2010}]{BEHHTTZ-JTAER-10}
{\sc B{\"o}hm, K.}, {\sc Etalle, S.}, {\sc den Hartog, J.}, {\sc H{\"u}tter,
  C.}, {\sc Trabelsi, S.}, {\sc Trivellato, D.}, {\sc and} {\sc Zannone, N.}
  2010.
\newblock {Flexible Architecture for Privacy-Aware Trust Management}.
\newblock {\em Journal of Theoretical and Applied Electronic Commerce
  Research\/}~{\em 5}, 77--96.

\bibitem[\protect\citeauthoryear{Bradshaw, Holt, and Seamons}{Bradshaw
  et~al\mbox{.}}{2004}]{BHS-CCS-04}
{\sc Bradshaw, R.~W.}, {\sc Holt, J.~E.}, {\sc and} {\sc Seamons, K.~E.} 2004.
\newblock {Concealing complex policies with hidden credentials}.
\newblock In {\em Proceedings of the 11th Conference on Computer and
  Communications Security}. ACM, New York, NY, USA, 146--157.

\bibitem[\protect\citeauthoryear{Bry}{Bry}{1989}]{B-DOOD-89}
{\sc Bry, F.} 1989.
\newblock {Query Evaluation in Recursive Databases: Bottom-up and Top-down
  Reconciled}.
\newblock In {\em Proceedings of the 1st International Conference on Deductive
  and Object-Oriented Databases}. North-Holland/Elsevier Science Publishers,
  25--44.

\bibitem[\protect\citeauthoryear{Chen and Warren}{Chen and
  Warren}{1996}]{CW-JACM-96}
{\sc Chen, W.} {\sc and} {\sc Warren, D.~S.} 1996.
\newblock {Tabled Evaluation With Delaying for General Logic Programs}.
\newblock {\em Journal of the ACM\/}~{\em 43,\/}~1, 20--74.

\bibitem[\protect\citeauthoryear{Chen}{Chen}{1997}]{C-IJIS-97}
{\sc Chen, Y.} 1997.
\newblock Magic sets and stratified databases.
\newblock {\em International Journal of Intelligent Systems\/}~{\em 12,\/}~3,
  203--231.

\bibitem[\protect\citeauthoryear{Costantini}{Costantini}{2001}]{C-ASP-01}
{\sc Costantini, S.} 2001.
\newblock {Comparing different graph representations of logic programs under
  the Answer Set semantics}.
\newblock In {\em Proceedings of the 1st International Workshop on Answer Set
  Programming}. AAAI Press.

\bibitem[\protect\citeauthoryear{Czenko and Etalle}{Czenko and
  Etalle}{2007}]{CE-ICLP-07}
{\sc Czenko, M.} {\sc and} {\sc Etalle, S.} 2007.
\newblock {Core TuLiP logic programming for trust management}.
\newblock In {\em Proceedings of the 23rd International Conference on Logic
  Programming}. LNCS, vol. 4670. Springer-Verlag, Berlin, Heidelberg, 380--394.

\bibitem[\protect\citeauthoryear{Czenko, Tran, Doumen, Etalle, Hartel, and den
  Hartog}{Czenko et~al\mbox{.}}{2006}]{CTDEHH-ENTCS-06}
{\sc Czenko, M.}, {\sc Tran, H.}, {\sc Doumen, J.}, {\sc Etalle, S.}, {\sc
  Hartel, P.}, {\sc and} {\sc den Hartog, J.} 2006.
\newblock {Nonmonotonic Trust Management for P2P Applications}.
\newblock {\em Electronic Notes in Theoretical Computer Science\/}~{\em
  157,\/}~3 (May), 113--130.

\bibitem[\protect\citeauthoryear{Dam{\'a}sio}{Dam{\'a}sio}{2000}]{D-TAPD-00}
{\sc Dam{\'a}sio, C.~V.} 2000.
\newblock {A Distributed Tabling System}.
\newblock In {\em Proceedings of the 2nd Conference on Tabulation in Parsing
  and Deduction}. 65--75.

\bibitem[\protect\citeauthoryear{Di~Marzo~Serugendo, Foukia, Hassas,
  Karageorgos, Most\'{e}faoui, Rana, Ulieru, Valckenaers, and van
  Aart}{Di~Marzo~Serugendo et~al\mbox{.}}{2004}]{SFHKMRUVA-ESOS-04}
{\sc Di~Marzo~Serugendo, G.}, {\sc Foukia, N.}, {\sc Hassas, S.}, {\sc
  Karageorgos, A.}, {\sc Most\'{e}faoui, S.~K.}, {\sc Rana, O.~F.}, {\sc
  Ulieru, M.}, {\sc Valckenaers, P.}, {\sc and} {\sc van Aart, C.} 2004.
\newblock {Self-Organisation: Paradigms and Applications}.
\newblock {\em Engineering Self-Organising Systems\/}~{\em 2977}, 1--19.

\bibitem[\protect\citeauthoryear{Dong and Dulay}{Dong and
  Dulay}{2010}]{CD-IFIPTM-10}
{\sc Dong, C.} {\sc and} {\sc Dulay, N.} 2010.
\newblock {Shinren: Non-monotonic Trust Management for Distributed Systems}.
\newblock In {\em Proceedings of the 4th International Conference on Trust
  Management}. IFIP, vol. 321. Springer Boston, 125--140.

\bibitem[\protect\citeauthoryear{Ellison, Frantz, Lampson, Rivest, Thomas, and
  Ylonen}{Ellison et~al\mbox{.}}{1999}]{EFLRTY-RFC-99}
{\sc Ellison, C.}, {\sc Frantz, B.}, {\sc Lampson, B.}, {\sc Rivest, R.}, {\sc
  Thomas, B.}, {\sc and} {\sc Ylonen, T.} 1999.
\newblock {SPKI Certificate Theory}.

\bibitem[\protect\citeauthoryear{Fitting}{Fitting}{1985}]{F-JLP-85}
{\sc Fitting, M.} 1985.
\newblock {A Kripke-Kleene-semantics for general logic programs}.
\newblock {\em Journal of Logic Programming\/}~{\em 2,\/}~4, 295--312.

\bibitem[\protect\citeauthoryear{Frikken, Atallah, and Li}{Frikken
  et~al\mbox{.}}{2006}]{FAL-TC-06}
{\sc Frikken, K.}, {\sc Atallah, M.}, {\sc and} {\sc Li, J.} 2006.
\newblock {Attribute-Based Access Control with Hidden Policies and Hidden
  Credentials}.
\newblock {\em IEEE Transactions on Computers\/}~{\em 55}, 1259--1270.

\bibitem[\protect\citeauthoryear{Gelfond and Lifschitz}{Gelfond and
  Lifschitz}{1988}]{GL-ICLP-88}
{\sc Gelfond, M.} {\sc and} {\sc Lifschitz, V.} 1988.
\newblock The stable model semantics for logic programming.
\newblock In {\em Proceedings of the 5th International Conference and Symposium
  on Logic Programming, ICLP'88}. MIT Press, 1070--1080.

\bibitem[\protect\citeauthoryear{Guo and Gupta}{Guo and
  Gupta}{2001}]{GG-ICLP-01}
{\sc Guo, H.-F.} {\sc and} {\sc Gupta, G.} 2001.
\newblock {A Simple Scheme for Implementing Tabled Logic Programming Systems
  Based on Dynamic Reordering of Alternatives}.
\newblock In {\em Proceedings of the 17th International Conference on Logic
  Programming}. Springer-Verlag, London, UK, 181--196.

\bibitem[\protect\citeauthoryear{Hoch and Shamir}{Hoch and
  Shamir}{2008}]{HS-ALP-08}
{\sc Hoch, J.} {\sc and} {\sc Shamir, A.} 2008.
\newblock {On the Strength of the Concatenated Hash Combiner When All the Hash
  Functions Are Weak}.
\newblock In {\em Automata, Languages and Programming}. LNCS, vol. 5126.
  Springer-Verlag, Berlin, Heidelberg, 616--630.

\bibitem[\protect\citeauthoryear{Hu}{Hu}{1997}]{H-THESIS-97}
{\sc Hu, R.} 1997.
\newblock {Efficient tabled evaluation of normal logic programs in a
  distributed environment}.
\newblock Ph.D. thesis, State University of New York at Stony Brook.

\bibitem[\protect\citeauthoryear{Hulin}{Hulin}{1989}]{H-VLDB-89}
{\sc Hulin, G.} 1989.
\newblock Parallel processing of recursive queries in distributed
  architectures.
\newblock In {\em Proceedings of the 15th International Conference on Very
  Large Data Bases}. Morgan Kaufmann Publishers Inc., San Francisco, CA, USA,
  87--96.

\bibitem[\protect\citeauthoryear{Jim and Suciu}{Jim and
  Suciu}{2001}]{JS-PODS-01}
{\sc Jim, T.} {\sc and} {\sc Suciu, D.} 2001.
\newblock {Dynamically distributed query evaluation}.
\newblock In {\em Proceedings of the 20th Symposium on Principles of database
  systems}. ACM, New York, NY, USA, 28--39.

\bibitem[\protect\citeauthoryear{Koshutanski and Massacci}{Koshutanski and
  Massacci}{2008}]{KM-AAS-08}
{\sc Koshutanski, H.} {\sc and} {\sc Massacci, F.} 2008.
\newblock {Interactive access control for autonomic systems: From theory to
  implementation}.
\newblock {\em ACM Transactions on Autonomous and Adaptive Systems\/}~{\em
  3,\/}~3, 1--31.

\bibitem[\protect\citeauthoryear{Kowalski}{Kowalski}{}]{K-IP-74}
{\sc Kowalski, R.}
\newblock Predicate logic as a programming language.
\newblock {\em Information Processing\/}~{\em 74}.

\bibitem[\protect\citeauthoryear{Lee, Minami, and Borisov}{Lee
  et~al\mbox{.}}{2009}]{LMB-ASIACCS-09}
{\sc Lee, A.~J.}, {\sc Minami, K.}, {\sc and} {\sc Borisov, N.} 2009.
\newblock Confidentiality-preserving distributed proofs of conjunctive queries.
\newblock In {\em Proceedings of the 4th International Symposium on
  Information, Computer, and Communications Security}. ACM, New York, NY, USA,
  287--297.

\bibitem[\protect\citeauthoryear{Lee, Minami, and Winslett}{Lee
  et~al\mbox{.}}{2010}]{LMW-TISSEC-10}
{\sc Lee, A.~J.}, {\sc Minami, K.}, {\sc and} {\sc Winslett, M.} 2010.
\newblock {On the Consistency of Distributed Proofs with Hidden Subtrees}.
\newblock {\em ACM Transactions on Information and System Security\/}~{\em
  13,\/}~3, 1--32.

\bibitem[\protect\citeauthoryear{Leuschel, Martens, and Sagonas}{Leuschel
  et~al\mbox{.}}{1998}]{LMS-LOPSTR-98}
{\sc Leuschel, M.}, {\sc Martens, B.}, {\sc and} {\sc Sagonas, K.} 1998.
\newblock {Preserving Termination of Tabled Logic Programs While Unfolding}.
\newblock In {\em Proceedings of the 7th International Workshop on Logic
  Program Synthesis and Transformation}. LNCS, vol. 1463. Springer-Verlag,
  Berlin, 189--205.

\bibitem[\protect\citeauthoryear{Li and Mitchell}{Li and
  Mitchell}{2003}]{LM-PADL-03}
{\sc Li, N.} {\sc and} {\sc Mitchell, J.~C.} 2003.
\newblock {Datalog with Constraints: A Foundation for Trust Management
  Languages}.
\newblock In {\em Proceedings of the 5th International Symposium on Practical
  Aspects of Declarative Languages}. LNCS, vol. 2562. Springer-Verlag, London,
  UK, 58--73.

\bibitem[\protect\citeauthoryear{Li, Winsborough, and Mitchell}{Li
  et~al\mbox{.}}{2003}]{LWM-CS-03}
{\sc Li, N.}, {\sc Winsborough, W.~H.}, {\sc and} {\sc Mitchell, J.~C.} 2003.
\newblock {Distributed credential chain discovery in trust management}.
\newblock {\em Journal of Computer Security\/}~{\em 11,\/}~1, 35--86.

\bibitem[\protect\citeauthoryear{Minami, Borisov, Winslett, and Lee}{Minami
  et~al\mbox{.}}{2011}]{MBWL-ASIACCS-11}
{\sc Minami, K.}, {\sc Borisov, N.}, {\sc Winslett, M.}, {\sc and} {\sc Lee,
  A.~J.} 2011.
\newblock Confidentiality-preserving proof theories for distributed proof
  systems.
\newblock In {\em Proceedings of the 6th Symposium on Information, Computer and
  Communications Security}. ACM, New York, NY, USA, 145--154.

\bibitem[\protect\citeauthoryear{Park}{Park}{1969}]{P-MI-69}
{\sc Park, D.} 1969.
\newblock Fixpoint induction and proofs of program properties.
\newblock {\em Machine Intelligence\/}~{\em 5}, 59--78.

\bibitem[\protect\citeauthoryear{Przymusinska and Przymunsinski}{Przymusinska
  and Przymunsinski}{1990}]{PP-FI-90}
{\sc Przymusinska, H.} {\sc and} {\sc Przymunsinski, T.~C.} 1990.
\newblock {Weakly stratified logic programs}.
\newblock {\em Fundamenta Informaticae\/}~{\em 13,\/}~1, 51--65.

\bibitem[\protect\citeauthoryear{Przymusinski}{Przymusinski}{1990}]{P-FI-90}
{\sc Przymusinski, T.} 1990.
\newblock {The Well-Founded Semantics Coincides with Three-Valued Stable
  Semantics}.
\newblock {\em Fundamenta Informaticae\/}~{\em 13,\/}~4, 445--463.

\bibitem[\protect\citeauthoryear{Przymusinski}{Przymusinski}{1988}]{P-FDBLP-88}
{\sc Przymusinski, T.~C.} 1988.
\newblock {\em {On the declarative semantics of deductive databases and logic
  programs}}.
\newblock Morgan Kaufmann Publishers Inc., San Francisco, CA, USA, 193--216.

\bibitem[\protect\citeauthoryear{Ramakrishnan}{Ramakrishnan}{1991}]{R-JLP-91}
{\sc Ramakrishnan, R.} 1991.
\newblock Magic templates: a spellbinding approach to logic programs.
\newblock {\em Journal of Logic Programming\/}~{\em 11}, 189--216.

\bibitem[\protect\citeauthoryear{Rocha, Silva, and Costa}{Rocha
  et~al\mbox{.}}{2005}]{RSC-TPLP-05}
{\sc Rocha, R.}, {\sc Silva, F.}, {\sc and} {\sc Costa, V.~S.} 2005.
\newblock On applying or-parallelism and tabling to logic programs.
\newblock {\em Theory and Practice of Logic Programming\/}~{\em 5,\/}~1-2,
  161--205.

\bibitem[\protect\citeauthoryear{Seamons, Winslett, and Yu}{Seamons
  et~al\mbox{.}}{2001}]{SWY-NDSS-01}
{\sc Seamons, K.~E.}, {\sc Winslett, M.}, {\sc and} {\sc Yu, T.} 2001.
\newblock {Limiting the Disclosure of Access Control Policies during Automated
  Trust Negotiation}.
\newblock In {\em Proceedings of the Network and Distributed System Security
  Symposium}. The Internet Society.

\bibitem[\protect\citeauthoryear{Shen, Yuan, You, and Zhou}{Shen
  et~al\mbox{.}}{2001}]{SYYZ-TPLP-01}
{\sc Shen, Y.-D.}, {\sc Yuan, L.-Y.}, {\sc You, J.-H.}, {\sc and} {\sc Zhou,
  N.-F.} 2001.
\newblock {Linear tabulated resolution based on Prolog control strategy}.
\newblock {\em Theory and Practice of Logic Programming\/}~{\em 1}, 71--103.

\bibitem[\protect\citeauthoryear{Stine, Kissel, Barker, Lee, and
  Fahlsing}{Stine et~al\mbox{.}}{2008}]{SKBLF-TR-08}
{\sc Stine, K.}, {\sc Kissel, R.}, {\sc Barker, W.~C.}, {\sc Lee, A.}, {\sc
  and} {\sc Fahlsing, J.} 2008.
\newblock {Guide for Mapping Types of Information and Information Systems to
  Security Categories}.
\newblock Special Publication SP 800-60 Rev. 1, National Institute of Standards
  and Technology (NIST).

\bibitem[\protect\citeauthoryear{Swift and Warren}{Swift and
  Warren}{2012}]{SW-TPLP-12}
{\sc Swift, T.} {\sc and} {\sc Warren, D.~S.} 2012.
\newblock {XSB: Extending Prolog with Tabled Logic Programming}.
\newblock {\em Theory and Practice of Logic Programming\/}~{\em 12,\/}~1-2,
  157--187.

\bibitem[\protect\citeauthoryear{Tamaki and Sato}{Tamaki and
  Sato}{1986}]{TS-ICLP-86}
{\sc Tamaki, H.} {\sc and} {\sc Sato, T.} 1986.
\newblock {OLD Resolution with Tabulation}.
\newblock In {\em Proceedings of the 3rd International Conference on Logic
  Programming}. Springer-Verlag, London, UK, 84--98.

\bibitem[\protect\citeauthoryear{Trivellato, Zannone, and Etalle}{Trivellato
  et~al\mbox{.}}{2011}]{TZE-POLICY-11}
{\sc Trivellato, D.}, {\sc Zannone, N.}, {\sc and} {\sc Etalle, S.} 2011.
\newblock {A Security Framework for Systems of Systems}.
\newblock In {\em Proceedings of the 12th International Conference on Policies
  for Distributed Systems and Networks}. IEEE Computer Society, Piscataway, NJ,
  USA.

\bibitem[\protect\citeauthoryear{{Van Gelder}, Ross, and Schlipf}{{Van Gelder}
  et~al\mbox{.}}{1991}]{GRS-JACM-91}
{\sc {Van Gelder}, A.}, {\sc Ross, K.~A.}, {\sc and} {\sc Schlipf, J.~S.} 1991.
\newblock {The well-founded semantics for general logic programs}.
\newblock {\em Journal of the ACM\/}~{\em 38}, 619--649.

\bibitem[\protect\citeauthoryear{Vieille}{Vieille}{1987}]{V-ICLP-87}
{\sc Vieille, L.} 1987.
\newblock {A Database-Complete Proof Procedure Based on SLD-Resolution}.
\newblock In {\em Proceedings of the 4th International Conference on Logic
  Programming}. MIT Press, 74--103.

\bibitem[\protect\citeauthoryear{Winsborough and Li}{Winsborough and
  Li}{2002}]{WL-WPES-02}
{\sc Winsborough, W.~H.} {\sc and} {\sc Li, N.} 2002.
\newblock {Protecting sensitive attributes in automated trust negotiation}.
\newblock In {\em Proceedings of Workshop on Privacy in the Electronic
  Society}. ACM, New York, NY, USA, 41--51.

\bibitem[\protect\citeauthoryear{Winsborough, Seamons, and Jones}{Winsborough
  et~al\mbox{.}}{2000}]{WSJ-DISCEX-00}
{\sc Winsborough, W.~H.}, {\sc Seamons, K.~E.}, {\sc and} {\sc Jones, V.~E.}
  2000.
\newblock {Automated Trust Negotiation}.
\newblock In {\em Proceedings of the DARPA Information Survivability Conference
  and Exposition}. Vol.~1. IEEE Computer Society, Los Alamitos, CA, USA,
  88--102.

\bibitem[\protect\citeauthoryear{Winslett}{Winslett}{2003}]{W-ITRUST-03}
{\sc Winslett, M.} 2003.
\newblock {An introduction to trust negotiation}.
\newblock In {\em Proceedings of International Conference on Trust Management}.
  LNCS, vol. 2692. Springer-Verlag, Berlin, Heidelberg, 275--283.

\bibitem[\protect\citeauthoryear{Yu and Winslett}{Yu and
  Winslett}{2003}]{YW-SP-03}
{\sc Yu, T.} {\sc and} {\sc Winslett, M.} 2003.
\newblock {A Unified Scheme for Resource Protection in Automated Trust
  Negotiation}.
\newblock In {\em Proceedings of the IEEE Symposium on Security and Privacy}.
  IEEE Computer Society, Washington, DC, USA, 110--122.

\bibitem[\protect\citeauthoryear{Zhang and Winslett}{Zhang and
  Winslett}{2008}]{ZW-ESORICS-08}
{\sc Zhang, C.~C.} {\sc and} {\sc Winslett, M.} 2008.
\newblock {Distributed Authorization by Multiparty Trust Negotiation}.
\newblock In {\em Proceedings of the 13th European Symposium on Research in
  Computer Security}. LNCS, vol. 5283. Springer-Verlag, Berlin, Heidelberg,
  282--299.

\bibitem[\protect\citeauthoryear{Zhou and Sato}{Zhou and
  Sato}{2003}]{ZS-PPDP-03}
{\sc Zhou, N.-F.} {\sc and} {\sc Sato, T.} 2003.
\newblock Efficient fixpoint computation in linear tabling.
\newblock In {\em Proceedings of the 5th International conference on Principles
  and Practice of Declaritive Programming}. ACM, New York, NY, USA, 275--283.

\end{thebibliography}
